\documentclass[12pt]{article}
\usepackage{graphicx} 
\usepackage[utf8]{inputenc} \usepackage{amsmath, amsthm, amssymb,amsfonts,mathtools,upgreek,xcolor,dsfont} \usepackage{fullpage}
\usepackage{hyperref}
\numberwithin{equation}{section}
\allowdisplaybreaks

\begin{document}

\begin{titlepage}
	\thispagestyle{empty}
	
	\vspace{35pt}
	
	\begin{center}
	    { \LARGE{ \bf${\cal N}$=4 Supergravity with Local Scaling Symmetry in Four Dimensions}}
		
		\vspace{50pt}
		
		{Nikolaos~Liatsos}
		
		\vspace{25pt}

		{\it Physics Division, National Technical University of Athens\\
		15780 Zografou Campus, Athens, Greece}

            \vspace{15pt} 

		 Email: n.liatsos.s@gmail.com

		\vspace{40pt}
		
		{ABSTRACT}
	\end{center}
We construct the most general four-dimensional ${\cal N}=4$ supergravity coupled to an arbitrary number $n$ of vector multiplets in which the global scaling symmetry is gauged, in addition to a subgroup of $\text{SL}(2,\mathbb{R}) \times \text{SO}(6,n)$. The various gaugings are parametrized by an embedding tensor built out of $2 \binom{n+6}{3}+4(n+6)$ parameters that satisfy a specific set of quadratic consistency constraints, to which we provide explicit solutions. We also derive the local supersymmetry transformation rules and the equations of motion for the four-dimensional ${\cal N}=4$ matter-coupled supergravity with local scaling symmetry. Such supergravity theories do not possess an action, since the scaling symmetry is only an on-shell symmetry of the corresponding ungauged theories. 

\vspace{10pt}

\end{titlepage}

\baselineskip 6 mm

\tableofcontents

\section{Introduction}

${\cal N}=4$ supergravity theories in four spacetime dimensions are of particular interest, since they exhibit the maximum amount of supersymmetry that is compatible with a consistent coupling of the supergravity multiplet to matter multiplets. The first instances of four-dimensional pure ${\cal N}=4$ supergravities were constructed more than 40 years ago in \cite{Das:1977uy, Cremmer:1977tc, Cremmer:1977tt, Freedman:1978ra} and the coupling of ${\cal N}=4$ supergravity to vector multiplets, as well as some of its gaugings, were analyzed a few years later in \cite{deRoo:1984zyh, Bergshoeff:1985ms, deRoo:1985np, deRoo:1985jh, Perret:1987nk,Perret:1988jq}. More recently,  various gauged ${\cal N}=4$ supergravity models originating from  orientifold compactifications of type IIA or IIB supergravity \cite{Frey:2002hf, Kachru:2002he} were studied in detail \cite{DAuria:2002qje,DAuria:2003nhg,Berg:2003ri,Angelantonj:2003rq,Angelantonj:2003up,Villadoro:2004ci,Derendinger:2004jn,Villadoro:2005cu,DallAgata:2009wsi}. 

A systematic parametrization of all the consistent gaugings of four-dimensional ${\cal N}=4$ matter-coupled supergravity is provided in \cite{Schon:2006kz} by means of an appropriately constrained embedding tensor that encodes the embedding of the gauge group into the on-shell global symmetry group of the ungauged theory, namely SL(2,$\mathbb{R}$) $\times$ SO(6,$n$), where $n$ is the number of vector multiplets. The embedding tensor formalism was introduced in \cite{Cordaro:1998tx,Nicolai:2000sc,Nicolai:2001sv} and further developed in \cite{deWit:2002vt,deWit:2004nw,deWit:2005ub,deWit:2007kvg} (see \cite{Samtleben:2008pe,Trigiante:2016mnt,DallAgata:2021uvl} for reviews). The full Lagrangian for the gauged $D=4$, ${\cal N}=4$ matter-coupled supergravity in an arbitrary symplectic frame is given in \cite{DallAgata:2023ahj}, where it was also shown that the supertrace of the squared mass eigenvalues vanishes for any Minkowski vacuum of such a theory that completely breaks ${\cal N}=4$ supersymmetry. This non-trivial result establishes the absence of quadratic divergences in the 1-loop corrections to the scalar potential for this class of vacua. 

The objective of this work is the construction of all possible gaugings of the four-dimensional ${\cal N}=4$ supergravity coupled to an arbitrary number $n$ of vector multiplets that include a gauging of the global scaling symmetry $\mathbb{R}^{+}$ of the equations of motion of the ungauged theory. This symmetry, which is also often referred to as \textit{trombone} symmetry, generalizes the invariance of Einstein's equations of general relativity under Weyl rescalings of the metric:
\begin{equation}
    \label{Weylresc}
    g_{\mu \nu} \rightarrow \Omega^2  g_{\mu \nu} \, , 
\end{equation}
where $\Omega$ is a real constant. Thus, the gauge groups under consideration in this paper will be direct products of a subgroup of $\text{SL}(2,\mathbb{R}) \times \text{SO}(6,n)$ and the scaling symmetry $\mathbb{R}^{+}$. 

The earliest instance of a supergravity theory with local scaling symmetry is the massive ten-dimensional IIA theory constructed in \cite{Howe:1997qt,Lavrinenko:1997qa} by a generalized dimensional reduction \cite{Scherk:1979zr} of eleven-dimensional supergravity. This theory is distinct from Romans' massive IIA supergravity \cite{Romans:1985tz}. Furthermore, supergravities with local trombone symmetry in nine and six spacetime dimensions with a higher-dimensional origin were constructed in \cite{Bergshoeff:2002nv} and \cite{Kerimo:2003am,Kerimo:2004md} respectively. All the aforementioned theories do not possess an action. A general framework for the construction of trombone gauged supergravity theories that makes use of the embedding tensor formalism was established in \cite{LeDiffon:2008sh}, which also provides explicitly the algebraic structures of the embedding tensors that parametrize the consistent gaugings of the maximal supergravities in various spacetime dimensions that involve the on-shell scaling symmetry of the corresponding ungauged theories as well as the quadratic constraints satisfied by these tensors. Furthermore, the authors of \cite{LeDiffon:2008sh} derived the equations of motion that characterize the three-dimensional maximal supergravity with local scaling symmetry, while the corresponding equations for the trombone gauged maximal $({\cal N}=8)$ supergravity in four dimensions were specified in \cite{LeDiffon:2011wt}. In both cases, it is clear that the field equations cannot be obtained from an action. 

As far as half-maximal $({\cal N}=4)$ matter-coupled supergravity in four spacetime dimensions is concerned, the standard gaugings thereof, which describe gauge groups that are entirely embedded into $\text{SL}(2,\mathbb{R}) \times \text{SO}(6,n)$ and do not include the scaling symmetry, are parametrized by an embedding tensor $\Theta$ that can be expressed in terms of two real constant $\text{SL}(2,\mathbb{R}) \times \text{SO}(6,n)$ tensors $\xi_{\alpha M}$ and $f_{\alpha MNP}=f_{\alpha[MNP]}$ \cite{Schon:2006kz}. From the general analysis of \cite{LeDiffon:2008sh} it follows that in this case, the gauging of the on-shell scaling symmetry of the ungauged theory translates into the introduction of additional components $\theta_{\alpha M}$ in the embedding tensor, which transform in the fundamental representation of $\text{SL}(2,\mathbb{R}) \times \text{SO}(6,n)$. Hence, the simultaneous gauging of a subgroup of $\text{SL}(2,\mathbb{R}) \times \text{SO}(6,n)$ and the trombone symmetry can be described by a deformed embedding tensor $\hat{\Theta}$ carrying in total $2 \binom{n+6}{3}+4(n+6)$ parameters.

In this paper, we work out
the generic algebraic consistency constraints on the embedding tensor of any trombone gauged supergravity that are put forward in \cite{LeDiffon:2008sh} for the specific case of four-dimensional ${\cal N}=4$ supergravity coupled to $n$ vector multiplets. This allows us to specify the explicit parametrization of the embedding tensor $\hat{\Theta}$ in terms of the irreducible $\text{SL}(2,\mathbb{R}) \times \text{SO}(6,n)$ tensors $f_{\alpha MNP}$, $\xi_{\alpha M}$ and $\theta_{\alpha M}$ as well as the quadratic constraints on these three tensors which guarantee consistency of the gaugings of $D=4$, ${\cal N}=4$ supergravity that involve the scaling symmetry. We also confirm that these quadratic constraints are sufficient for compatibility of the theory with ${\cal N}=4$ supersymmetry by an explicit derivation of the associated equations of motion. Once more it is apparent that these field equations cannot be reproduced by the variation of an action, which is explained by the fact that the theory under study results from  the gauging of the trombone symmetry of the equations of motion of the ungauged four-dimensional ${\cal N}=4$ supergravity, which, however, is not a symmetry of the action thereof. We should point out that a preliminary discussion of the gaugings of $D=4, \, {\cal N}=4$ supergravity that involve the scaling symmetry can be found in \cite{Prins:2013yra}, whose author has derived an expression for the embedding tensor that parametrizes the aforementioned gaugings as well as a set of quadratic consistency constraints on the irreducible components of this tensor. However, our results cast doubt on the validity of these constraints.  

The rest of this article is organized as follows. In section \ref{sec:Ingred}, we review the field content of the four-dimensional ${\cal N}=4$ supergravity coupled to $n$ vector multiplets and the structure of the coset space $\frac{\text{SL(2,$\mathbb{R}$)}}{\text{SO(2)}} \times \frac{\text{SO(6,$n$)}}{\text{SO(6)} \times \text{SO($n$)}}$ parametrized by the scalar fields of the theory.  We also provide a definition of the relevant symplectic frames. In section \ref{sec:Emb}, we determine the algebraic structure of the embedding tensor that parametrizes the gaugings of $D=4$, ${\cal N}=4$ supergravity that include the trombone symmetry. We also provide the quadratic consistency constraints satisfied by the irreducible components of this embedding tensor and an explicit general solution to these constraints. In section \ref{sec:susy_rules}, we give the local supersymmetry transformation rules of four-dimensional ${\cal N}=4$ supergravity with local scaling symmetry and verify the closure of their algebra on the bosonic fields of the theory. In section \ref{sec:eom}, we derive the full set of equations of motion that characterize the trombone gauged half-maximal supergravity in four dimensions. In section \ref{sec:mass}, we specify the conditions satisfied by solutions to these field equations with constant scalar and vanishing vector and fermionic fields as well as the mass matrices of the fluctuations of the various fields of the theory around such solutions. Furthermore, in appendix \ref{sec:Bianchi}, we present the rheonomic approach underlying the derivation of the local supersymmetry transformations and the fermionic field equations provided in the main text. In appendix \ref{solapp}, we explain in detail the procedure for the solution of the quadratic constraints on the embedding tensor. Finally, in appendix \ref{sec:Tid}, we derive the quadratic constraints satisfied by the fermion shift matrices of the theory ($T$-identities) by appropriately dressing the quadratic constraints on the embedding tensor with the representatives of the scalar coset manifold. 

\section{The Ingredients of $D=4$, ${\cal N}=4$ Supergravity}
\label{sec:Ingred}
In order to establish the notations, we first recall the field content of four-dimensional ${\cal N}=4$ matter-coupled Poincar\'{e} supergravity. The ${\cal N}=4$ supergravity multiplet contains the graviton $g_{\mu\nu}$, four gravitini $\psi_\mu^i$, $i=1,\dots,4$, six vector fields $A_\mu^{ij}=-A_\mu^{ji}$, four spin-1/2 fermions $\chi_i$ (dilatini) and a complex scalar $\tau$ parameterizing the coset manifold $\frac{\text{SL(2,$\mathbb{R}$)}}{\text{SO(2)}}$. This multiplet can be coupled to $n$ vector multiplets, which contain $n$ vector fields $A_\mu^{\underline{a}}$, $\underline{a} = 1, \ldots, n$, $4n$ gaugini $\lambda^{\underline{a}i}$, and $6n$ real scalar fields $\phi^{\underline{a} \underline{m}}$, $\underline{m}=1,\dots,6$, which parameterize the coset manifold $\frac{\text{SO(6,$n$)}}{\text{SO(6)}\times\text{SO($n$)}}$. Overall, the scalar $\sigma$-model is described by the coset space \cite{deRoo:1984zyh,Bergshoeff:1985ms,deRoo:1985jh}
\begin{equation}
    \label{Mscalar}
    \mathcal{M} =\frac{\text{SL(2,$\mathbb{R}$)}}{\text{SO(2)}} \times \frac{\text{SO(6,$n$)}}{\text{SO(6)} \times \text{SO($n$)}}\, .
\end{equation}

\subsection{Scalar Coset Space Representatives}
The SL(2,$\mathbb{R}$)/SO(2) factor of the scalar manifold \eqref{Mscalar} can be represented by a complex SL(2,$\mathbb{R}$) vector $\mathcal{V}_\alpha$ \cite{Schon:2006kz}, where $\alpha=+,-$, which satisfies the constraint 
\begin{equation}
\label{VV*-V*V=e}
    \mathcal{V}_\alpha \mathcal{V}_\beta^* - \mathcal{V}_\alpha^* \mathcal{V}_\beta = -2 i \epsilon_{\alpha \beta} \, ,
\end{equation}
where $\epsilon_{\alpha \beta} = - \epsilon_{\beta \alpha}$ and $\epsilon_{+-}=1$. We will raise and lower SL(2,$\mathbb{R}$) indices according to  the convention
\begin{equation}
    \mathcal{V}^\alpha =  \mathcal{V}_\beta \epsilon^{\beta \alpha}, \qquad \mathcal{V}_\alpha =\epsilon_{\alpha \beta  } \mathcal{V}^\beta ,
\end{equation}
where $\epsilon^{\alpha \beta} = - \epsilon^{\beta \alpha}$ with $\epsilon^{+-}=1$, such that $\epsilon^{\alpha \gamma} \epsilon_{\beta \gamma} = \delta^{\alpha}_\beta$. 

It is also useful to introduce the positive definite symmetric matrix 
\begin{equation}
    \label{Mab}
    M_{\alpha \beta} = \text{Re} ({\cal V}_{\alpha} {\cal V}_{\beta}^*) \, , 
\end{equation}
which satisfies 
\begin{equation}
    M^{\alpha \beta} M_{\beta \gamma} = \delta^{\alpha}_{\gamma}\,.
\end{equation}  
Furthermore, ${\cal V}_\alpha$ carries SO(2)$\cong$U(1) charge +1.

On the other hand, the coset space SO(6,$n$)/(SO(6)$\times$SO($n$)) parametrized by the $6n$ real scalars of the $n$ vector multiplets can be described by means of a coset representative ${L_M}^{\underline{M}}=({L_M}^{\underline{m}},{L_M}^{\underline{a}})$, where $M=1,\dots,n+6$ is a vector index of SO(6,$n$), $\underline{m}=1,\dots,6$ and $\underline{a}=1,\dots,n$ are indices of the fundamental representations of  SO(6) and SO($n$) respectively, while $\underline{M}$ is an SO(6)$\times$SO($n$) index, which decomposes as $\underline{M}=(\underline{m},\underline{a})$. The matrix $L$ is an element of SO(6,$n$), which means that
\begin{equation}
\label{LSO6n}
   \eta_{MN} = \eta_{\underline{M} \underline{N}} {L_M}^{\underline{M}} {L_N}^{\underline{N}} = {L_M}^{\underline{M}} L_{N \underline{M}} ={L_M}^{\underline{m}} L_{N \underline{m}} + {L_M}^{\underline{a}} L_{N \underline{a}} \,,
\end{equation}
where $\eta_{MN}=\eta_{\underline{M} \underline{N}} = \text{diag}(-1,-1,-1,-1,-1,-1,\underbrace{1,\dots,1}_{n \,\text{entries}})$. The constant matrices $\eta_{MN}$ and $\eta_{\underline{M}\underline{N}}$ and their inverses, $\eta^{MN}$ and $\eta^{\underline{M}\underline{N}}$, can be used as metrics to raise and lower the corresponding indices. 

As for the scalar sector of the ${\cal N}=4$ supergravity multiplet, it is useful to introduce the positive definite symmetric matrix $M=LL^T$, with elements 
\begin{equation}
    \label{MMN}
     M_{MN} = -   {L_M}^{\underline{m}} L_{N \underline{m}} + {L_M}^{\underline{a}} L_{N \underline{a}} \, , 
\end{equation}
which satisfies 
   \begin{equation}
\label{MM}
    M^{MN} M_{NP} = \delta^M_P.
\end{equation}

We can trade the matrix elements ${L_M}^{\underline{m}}$ for the antisymmetric SU(4) tensors $L_M{}^{ij}=-L_M{}^{ji}$ defined by 
\begin{equation}
     {L_M}^{ij} = {\Gamma_{\underline{m}}}^{ij} {L_M}^{\underline{m}}, 
\end{equation}
where $\Gamma_{\underline{m}}{}^{i j }$ are six antisymmetric 4$\times$4 matrices that realize the isomorphism between the fundamental representation of SO(6) and the two-fold antisymmetric representation of SU(4). We refer the reader to appendix A of \cite{DallAgata:2023ahj} for the explicit expressions for these matrices.
In terms of $L_M{}^{ij}$, which satisfy the pseudo-reality condition 
\begin{equation}
    \label{pseudor}
  L_{Mij} = (L_M{}^{ij})^* = \frac{1}{2} \epsilon_{ijkl} L_M{}^{kl} ,
\end{equation}
equation \eqref{LSO6n} is written as 
\begin{equation}
    \label{LSO6SU4}
    \eta_{MN} = - {L_M}^{ij} L_{Nij} + {L_M}^{\underline{a}} L_{N \underline{a}} \, .
\end{equation}

\subsection{Fermionic Fields}

The fermionic fields of $D=4$, ${\cal N}=4$ matter-coupled supergravity transform in representations of the isotropy group  $H=\text{SO}(2) \times \text{SO}(6) \times \text{SO}(n)$ of the coset space \eqref{Mscalar}, parametrized by the scalar fields.  More precisely, the gravitini, the dilatini and the gaugini transform in the fundamental representation of SU(4), which is the universal cover of SO(6), while the gaugini alone transform in the fundamental representation of SO($n$) as well. Furthermore, the fermions $\psi^i_\mu$, $\chi^i$ and $\lambda^{\underline{a}i}$ carry SO(2) charges $-\frac{1}{2}$, $+\frac{3}{2}$ and $+\frac{1}{2}$ respectively and have the following chirality properties\footnote{We use the gamma matrix and spinor conventions of  \cite{DallAgata:2021uvl}, which are also summarized in appendix A of \cite{DallAgata:2023ahj}.}:
\begin{equation}
    \label{LH}
    \gamma_5 \psi^i_\mu = \psi^i_\mu, \quad \gamma_5 \chi^i = - \chi^i, \quad \gamma_5 \lambda^{\underline{a}i} = \lambda^{\underline{a}i},
\end{equation}
while their charge conjugates $\psi_{i \mu} = (\psi^i_\mu)^c$, $\chi_i=(\chi^i)^c$ and $\lambda_i^{\underline{a}} = (\lambda^{\underline{a}i})^c$ have opposite SO(2) charges and chiralities. 

\subsection{Symplectic Frames}

The Lagrangian that describes the ungauged theory involves $n+6$ abelian vector fields $A^{\Lambda}_{\mu}$, which are referred to as electric vectors and combine with their magnetic duals, $A_{\Lambda \mu}$, into an SL(2,$\mathbb{R}$) $\times$ SO(6,$n$) vector $A^{M \alpha}_\mu$,  which is also a symplectic vector of Sp($2(6+n)$,$\mathbb{R}$)$\supset$ SL(2,$\mathbb{R}$) $\times$ SO(6,$n$). Following \cite{Schon:2006kz}, we introduce a composite SL(2,$\mathbb{R}$) $\times$ SO(6,$n$) index $\mathcal{M}=M \alpha$ and an antisymmetric symplectic form ${\mathbb{C}}^{\mathcal{M}\mathcal{N}}$ defined by 
\begin{equation}
\label{sympform}
    {\mathbb{C}}^{\mathcal{M}\mathcal{N}} = {\mathbb{C}}^{M \alpha N \beta} \equiv \eta^{MN} \epsilon^{\alpha \beta}.
\end{equation}
Every electric/magnetic split $A^{\cal M}_\mu = A^{M \alpha}_\mu=(A^{\Lambda}_\mu,A_{\Lambda \mu})$ such that \eqref{sympform} decomposes as
\begin{equation}
    \label{CCinv}
    \mathbb{C}^{\mathcal{M} \mathcal{N}} = 
    \begin{pmatrix}
    \mathbb{C}^{\Lambda \Sigma} &   {\mathbb{C}^{\Lambda}}_\Sigma \\
    {\mathbb{C}_{\Lambda}}^\Sigma &  \mathbb{C}_{\Lambda \Sigma} \\ 
    \end{pmatrix}=
    \begin{pmatrix}
     0 & \delta^{\Lambda}_{\Sigma} \\
     - \delta_{\Lambda}^{\Sigma} & 0  
    \end{pmatrix}  
\end{equation}
defines a symplectic frame and any two symplectic frames are related by a symplectic rotation that is an element of Sp($2(6+n)$,$\mathbb{R}$). In \cite{DallAgata:2023ahj}, the choice of the symplectic frame was conveniently parametrized by means of projectors ${\Pi^{\Lambda}}_{\mathcal{M}}$ and $\Pi_{\Lambda \mathcal{M}}$ that extract the electric and magnetic components of a symplectic vector $V^{\mathcal{M}}=(V^{\Lambda}, V_{\Lambda})$ respectively, according to
\begin{equation}
    \label{projup}
    V^{\Lambda} = {\Pi^{\Lambda}}_{\mathcal{M}} V^{\mathcal{M}}, \qquad V_{\Lambda} = \Pi_{\Lambda \mathcal{M}} V^{\mathcal{M}}.
\end{equation}
The properties that must be satisfied by these projectors are provided in the same reference. 

\section{Gauging the Scaling Symmetry}
\label{sec:Emb}
In this section, we will present the structure of the consistent gaugings of half-maximal supergravity in four spacetime dimensions that involve the scaling symmetry. We first recall that the on-shell global symmetry group of the ungauged $D=4$, ${\cal N}=4$ supergravity coupled to $n$ vector multiplets is 
\begin{equation}
    \label{Udual}
    G=\text{SL}(2,\mathbb{R})\times \text{SO}(6,n) \times \mathbb{R}^{+} \, ,
\end{equation}
where $\mathbb{R}^{+}$ denotes the scaling (or trombone) symmetry of the equations of motion, under which the various fields transform as 
\begin{align}
    \delta g_{\mu \nu}  & = 2 \lambda g_{\mu \nu}\, , \qquad \delta A^{\cal M}_\mu = \lambda A^{\cal M}_\mu, \qquad \delta \tau = 0 \, , \qquad \delta \phi^{\underline{a}\underline{m}} = 0 \, , \\
    \delta \psi^i_\mu & = \frac{1}{2} \lambda \psi^i_\mu \, , \qquad \delta \chi^i = - \frac{1}{2} \lambda \chi^i , \qquad \delta \lambda^{\underline{a}i} = - \frac{1}{2} \lambda \lambda^{\underline{a}i}, 
\end{align}
where $\lambda$ is an infinitesimal real parameter. We denote the generators of $G$ by $t_{\hat A}=(t_0,t_A)$, where $t_0$ is the generator of the scaling symmetry $\mathbb{R}^{+}$, while $t_A$ are the generators of $\text{SL}(2,\mathbb{R})\times\text{SO}(6,n)$, where the index $A$ labels the adjoint representation of $\text{SL}(2,\mathbb{R})\times\text{SO}(6,n)$, and thus decomposes as $A=([MN],(\alpha \beta))$. 

\subsection{Embedding Tensor}
For the purpose of constructing the most general ${\cal N}=4$ locally supersymmetric theory in four spacetime dimensions whose gauge group is the direct product of the scaling symmetry $\mathbb{R}^{+}$ and a subgroup of $\text{SL}(2,\mathbb{R})\times\text{SO}(6,n)$, we will employ the embedding tensor formalism, which was introduced in \cite{Cordaro:1998tx,Nicolai:2000sc,Nicolai:2001sv} and further developed in \cite{deWit:2002vt,deWit:2004nw,deWit:2005ub,deWit:2007kvg} (see \cite{Samtleben:2008pe,Trigiante:2016mnt,DallAgata:2021uvl} for reviews). 
In this formulation of the gauging procedure, all the information about the embedding of the gauge group into $G$ is encoded in an embedding tensor ${\hat\Theta}_{\cal M}{}^{\hat A}$, by means of which the gauge group generators $X_{\cal M}$ are expressed as linear combinations of the generators of $G$ according to  
\begin{equation}
    X_{\cal M} = 
{\hat\Theta}_{\cal M}{}^{\hat A} t_{\hat A} ={\hat\Theta}_{\cal M}{}^0 t_0 + {\hat\Theta}_{\cal M}{}^A t_A \, .
\end{equation}
We also introduce vector gauge fields $A^{\cal M}_{\mu}=A^{M \alpha}_{\mu}$, which decompose into electric vectors $A^{\Lambda}_{\mu}$ and magnetic vectors $A_{\Lambda \mu}$, and the gauge covariant exterior derivative 
\begin{equation}
\label{hatd}
\hat{d} = d - g A^{\cal M} X_{\cal M} \, , 
\end{equation}
where $g$ is the gauge coupling and $A^{\cal M} = A^{\cal M}_\mu d x^\mu $. The action of \eqref{hatd} on a $p$-form depends on the representation of $G$ carried by the latter. Furthermore, we note that  
\begin{equation}
\label{B}
    {\cal B} \equiv g  {\hat\Theta}_{\cal M}{}^0 A^{\cal M} 
\end{equation}
is the linear combination of the one-form potentials $A^{\cal M}$ that gauges the scaling symmetry. Moreover, from the coupling of the embedding tensor in \eqref{hatd} it follows that the latter has scaling weight $-1$. 

Following \cite{LeDiffon:2008sh}, we consider the following ansatz for the embedding tensor:
\begin{eqnarray}
\begin{aligned}
\label{embans}
         {\hat\Theta}_{\cal M}{}^{NP} &  = \Theta_{\cal M}{}^{NP} + \zeta_1 (t^{NP})_{\cal M}{}^{\cal Q} \theta_{\cal Q} \, , \\
    {\hat\Theta}_{\cal M}{}^{\beta \gamma} & = {\Theta}_{\cal M}{}^{\beta \gamma} + \zeta_2 (t^{\beta \gamma})_{\cal M}{}^{\cal Q} \theta_{\cal Q} \, , \\
    {\hat\Theta}_{\cal M}{}^0 &= \theta_{\cal M}\,,
\end{aligned}    
\end{eqnarray}
where $\zeta_1$ and $\zeta_2$ are real constants and $\Theta_{\cal M}{}^A=(\Theta_{\cal M}{}^{NP},{\Theta}_{\cal M}{}^{\beta \gamma})$ is the embedding tensor that parametrizes the standard gaugings of four-dimensional ${\cal N}=4$ supergravity, which do not involve the trombone symmetry. Consistency of the standard gaugings restricts $\Theta_{\cal M}{}^A$ to the $(\mathbf{2},\mathbf{n+6}) + \left( \mathbf{2}, \mathbf{\binom{n+6}{3}} \right)$ representation of $\text{SL}(2,\mathbb{R})\times\text{SO}(6,n)$, which means that this embedding tensor is built out of two real constant $\text{SL}(2,\mathbb{R})\times\text{SO}(6,n)$ tensors $\xi_{\alpha M}$ and $f_{\alpha MNP}=f_{\alpha[MNP]}$ \cite{Schon:2006kz}. The explicit expressions for its components, $\Theta_{\alpha M}{}^{NP}$ and $\Theta_{\alpha M}{}^{\beta \gamma}$, in terms of $\xi_{\alpha M}$ and $f_{\alpha MNP}$ can be found in \cite{Schon:2006kz,DallAgata:2023ahj}. Also, $(t_A)_{\cal M}{}^{\cal N}$ are the matrix elements of the generators of $\text{SL}(2,\mathbb{R})\times\text{SO}(6,n)$ in the fundamental representation, which are taken to be 
\begin{equation}
\label{gen}
    (t_{PQ})_{M \alpha}{}^{N \beta} = \delta^N_{[P} \eta_{Q]M} \delta^\beta_\alpha, \qquad (t_{\gamma \delta})_{M \alpha}{}^{N \beta} = \delta^\beta_{(\gamma} \epsilon_{\delta) \alpha} \delta_M^N \, .
\end{equation}

We now recall that the non-abelian two-form field strengths $H^{\cal M}$ of the vector gauge fields $A^{\cal M}$ involve Stueckelberg-type terms of the form \cite{deWit:2004nw,deWit:2005hv} 
\begin{equation}
    H^{\cal P} \supset g Z^{\cal P}{}_{{\cal M} {\cal N}} B^{{\cal M} {\cal N}}  , 
\end{equation}
where $B^{{\cal M} {\cal N}} = B^{({\cal M} {\cal N})} $ are two-form gauge fields and the intertwining tensor $Z^{\cal P}{}_{{\cal M} {\cal N}}$ is given by 
\begin{equation}
\label{Z}
    Z^{\cal P}{}_{{\cal M} {\cal N}} \equiv X_{({\cal M} {\cal N})}{}^{\cal P} ,
\end{equation}
where 
\begin{equation}
\label{XMNP}
    X_{{\cal M} {\cal N} }{}^{\cal P} \equiv {\hat\Theta}_{\cal M}{}^{\hat A} (t_{\hat A})_{\cal N}{}^{\cal P} = - \theta_{\cal M} \delta^{\cal P}_{\cal N} + {\hat\Theta}_{\cal M}{}^A (t_A)_{\cal N}{}^{\cal P} 
\end{equation}
are the matrix elements of the gauge group generators in the fundamental representation of $G$. As pointed out in \cite{LeDiffon:2008sh}, the two-form field content of the theory should be the same in the presence and in the absence of a gauging of the scaling symmetry, since it is fixed by supersymmetry. Therefore, the intertwining tensor must project onto the adjoint $(\mathbf{3},\mathbf{1})+\left(\mathbf{1},\mathbf{\frac{1}{2}}\mathbf{(n+6)(n+5)}\right)$ representation of SL(2,$\mathbb{R}$) $\times$ SO(6,$n$) in its lower indices $({\cal M} {\cal N})$. This requirement determines the parameters $\zeta_1$ and $\zeta_2$ in the ansatz \eqref{embans} for the embedding tensor.

More precisely, the two-fold symmetric tensor product of the fundamental $(\mathbf{2},\mathbf{n+6})$ representation of SL(2,$\mathbb{R}$) $\times$ SO(6,$n$) decomposes as 
\begin{align}
    \label{symprod}
    &((\mathbf{2},\mathbf{n+6}) \times (\mathbf{2},\mathbf{n+6}) )_{\text{sym.}} \nonumber \\ 
    &= \left(\mathbf{3}, \mathbf{\frac{1}{2}(n+6)(n+7)-1} \right)  + (\mathbf{3},\mathbf{1})+\left(\mathbf{1},\mathbf{\frac{1}{2}}\mathbf{(n+6)(n+5)}\right)
\end{align}
We require the projection of the intertwining tensor onto the representation $\left(\mathbf{3}, \mathbf{\frac{1}{2}(n+6)(n+7)-1} \right)$ to vanish, i.e. 
\begin{equation}
\label{proj}
    Z^{P \gamma}{}_{(M(\alpha|N)\beta)} - \frac{1}{n+6} \eta_{MN} \eta^{RS} Z^{P \gamma}{}_{R(\alpha|S|\beta)} = 0 \, .
\end{equation}
Substituting the ansatz \eqref{embans} into the above condition and using the explicit expressions \eqref{gen} for the generators of SL(2,$\mathbb{R}$) $\times$ SO(6,$n$), we find $\zeta_1+\zeta_2=-2$, or $\zeta_2 = - 2 - \zeta_1$. Therefore, the components of the embedding tensor ${\hat\Theta}_{\cal M}{}^{\hat{A}}$ read
\begin{align}
    \label{ThaMNPz1}
    {\hat\Theta}_{\alpha M}{}^{NP} & = f_{\alpha M}{}^{NP} + \delta^{[N}_M \xi^{P]}_\alpha - \zeta_1 \delta^{[N}_M \theta^{P]}_\alpha,  \\
    \label{ThaMbcz1}
    {\hat\Theta}_{\alpha M}{}^{\beta \gamma} & = \delta^{(\beta}_\alpha \xi^{\gamma)}_M - (2+ \zeta_1) \delta^{(\beta}_\alpha \theta^{\gamma)}_M \, , \\
    \label{ThaMz1}
    {\hat\Theta}_{\alpha M}{}^0 & = \theta_{\alpha M} \, .
    \end{align} 
Without loss of generality, we can set $\zeta_1=-1$, which leads to 
    \begin{align}
    \label{ThaMNP}
    {\hat\Theta}_{\alpha M}{}^{NP} & = f_{\alpha M}{}^{NP} + \delta^{[N}_M \xi^{P]}_\alpha + \delta^{[N}_M \theta^{P]}_\alpha,  \\
    \label{ThaMbc}
    {\hat\Theta}_{\alpha M}{}^{\beta \gamma} & = \delta^{(\beta}_\alpha \xi^{\gamma)}_M - \delta^{(\beta}_\alpha \theta^{\gamma)}_M \, , \\
    \label{ThaM}
    {\hat\Theta}_{\alpha M}{}^0 & = \theta_{\alpha M} \, .
        \end{align} 
Indeed, for any other value of $\zeta_1$, the parametrization \eqref{ThaMNPz1}-\eqref{ThaMz1} of the embedding tensor can be obtained from \eqref{ThaMNP}-\eqref{ThaM} by the redefinition $\xi_{\alpha M} \rightarrow \xi_{\alpha M} - (\zeta_1 + 1) \theta_{\alpha M}$. In the rest of this paper, the components of the embedding tensor $\hat{\Theta}$ will be given by equations \eqref{ThaMNP}-\eqref{ThaM}, which result in the following expression for the entries of the gauge group generators:        
\begin{align}
\label{Xtens}
    X_{M \alpha N \beta}{}^{P \gamma} = & - \delta^\gamma_\beta {f_{\alpha M N}}^P + \frac{1}{2} \left( \delta^P_M \delta^\gamma_\beta \xi_{\alpha N} - \delta^P_N \delta^\gamma_\alpha \xi_{\beta M} - \eta_{MN} \delta^\gamma_\beta \xi^P_\alpha  + \delta^P_N \epsilon_{\alpha \beta} \xi^\gamma_M\right) \nonumber \\
    & - \delta^P_N \delta^\gamma_\beta \theta_{\alpha M}  + \frac{1}{2} \left( \delta^P_M \delta^\gamma_\beta \theta_{\alpha N}  + \delta^P_N \delta^\gamma_\alpha \theta_{\beta M} - \eta_{MN} \delta^\gamma_\beta \theta^P_\alpha - \delta^P_N  \epsilon_{\alpha \beta} \theta^\gamma_M \right)  .
    \end{align}
Furthermore, the intertwining tensor equals
\begin{equation}
    Z^{\cal P}{}_{{\cal M} {\cal N}} = Z^{{\cal P} A} (t_A)_{{\cal M} {\cal N}} \, , 
\end{equation}
where
\begin{equation}
    Z^{M \alpha NP} = - \frac{1}{2} \Theta^{\alpha M NP} + \frac{3}{2} \eta^{M[N|} \theta^{\alpha|P]}, \qquad Z^{M \alpha \beta \gamma} = \frac{1}{2} \epsilon^{\alpha (\beta} \left( \xi^{\gamma)M} + \theta^{\gamma)M} \right).
\end{equation}

We should mention that a parametrization of the embedding tensor describing the gaugings of $D=4, \, {\cal N}=4$ supergravity that involve the scaling symmetry was first given in \cite{Prins:2013yra}. In this reference, the embedding tensor ${\hat\Theta}_{\cal M}{}^{\hat A}$ is given by \eqref{ThaMNPz1}-\eqref{ThaMz1} with $\zeta_1=2$ and $\theta_{\alpha M}= - \frac{1}{2} \kappa_{\alpha M}$, so it clearly satisfies the linear constraint \eqref{proj}. Equivalently, the expression for $X_{M \alpha N \beta}{}^{P \gamma}$ provided in \cite{Prins:2013yra} can be obtained from \eqref{Xtens} by redefining $\xi_{\alpha M}$ as $\xi_{\alpha M} \rightarrow \xi_{\alpha M} - 3 \theta_{\alpha M}$ and then setting $\theta_{\alpha M}=- \frac{1}{2} \kappa_{\alpha M}$.
\subsection{Quadratic Consistency Constraints}
In order for the gauging to be consistent, the embedding tensor ${\hat\Theta}_{\cal M}{}^{\hat A}$ must also be invariant under the action of the gauge group that it  defines, which translates into the quadratic constraints \cite{LeDiffon:2008sh}
\begin{align}
\label{quad1}
    0 & = {\hat\Theta}_{\cal M}{}^{\hat A} t_{\hat A} \theta_{\cal N} = X_{{\cal M} {\cal N}}{}^{\cal P} \theta_{\cal P} \, , \\ 
    \label{quad2}
       0 & = {\hat\Theta}_{\cal M}{}^{\hat A} t_{\hat A} \Theta_{\cal N}{}^B = X_{{\cal M} {\cal N}}{}^{\cal P} \Theta_{\cal P}{}^B + {\hat\Theta}_{\cal M}{}^A \Theta_{\cal N}{}^C f_{AC}{}^B , 
\end{align}
where ${f_{AB}}^{C}$ are the structure constants of the Lie algebra of $ \text{SL}(2,\mathbb{R}) \times \text{SO}(6,n)$, defined by $[t_A,t_B]={f_{AB}}^C t_C $. From equations \eqref{quad1} and \eqref{quad2} it follows that the gauge group generators satisfy 
\begin{equation}
    \label{closure}
    [X_{\cal M},X_{\cal N}] = - X_{{\cal M} {\cal N}}{}^{\cal P} X_{\cal P} \, ,
\end{equation}
which amounts to the closure of the gauge algebra. Using \eqref{ThaMNP}-\eqref{Xtens} we find that conditions \eqref{quad1} and \eqref{quad2} are equivalent to the following quadratic constraints on the $ \text{SL}(2,\mathbb{R}) \times \text{SO}(6,n)$ tensors $f_{\alpha MNP}$, $\xi_{\alpha M}$ and $\theta_{\alpha M}$:
    \begin{align}
        \label{q2}
         \epsilon^{\alpha \beta} \xi_{\alpha (M|} \theta_{\beta|N)} & = 0 \, ,  \\ 
         \label{q3}
         \epsilon^{\alpha \beta} \left(\theta^P_\alpha f_{\beta MNP} + \xi_{\alpha [M|} \theta_{\beta |N]} - 3 \, \theta_{\alpha M} \theta_{\beta N}  \right) & = 0 \, ,\\
         \label{q4}
         \theta^P_{(\alpha} f_{\beta) MNP} + \xi_{(\alpha[M} \theta_{\beta)N]} & = 0 \, ,\\
         \label{q5}
         \xi^M_{(\alpha} \theta_{\beta)M} + \theta^M_\alpha \theta_{\beta M}  & = 0 \, ,\\ \label{q6}
         \xi^P_{(\alpha} f_{\beta)MNP} - \xi_{(\alpha[M} \theta_{\beta)N]} & = 0 \, , \\ \label{q7}
         \xi^M_{(\alpha} \theta_{\beta)M} + \xi^M_\alpha \xi_{\beta M} & = 0 \, ,\\ \label{q8} \epsilon^{\alpha \beta} \left( \xi^P_\alpha f_{\beta MNP} + \xi_{\alpha M} \xi_{\beta N} - 3 \xi_{\alpha[M|} \theta_{\beta|N]} \right) & = 0 \, ,\\ \label{q9}
         3 f_{\alpha [MN|R} f_{\beta|PQ]}{}^R + 2 \xi_{(\alpha[M} f_{\beta)NPQ]}  + 2 \theta_{(\alpha[M} f_{\beta)NPQ]} & = 0 \, , \\
         \label{q10}
         \epsilon^{\alpha \beta} \theta_{\alpha[M|} f_{\beta|NPQ]} & =0 \, ,\\
         \label{q11}
         \epsilon^{\alpha \beta} ( f_{\alpha MNR} f_{\beta PQ}{}^R - \xi_{\alpha [M|} f_{\beta|N]PQ} + \xi_{\alpha [P|} f_{\beta|Q]MN} + \theta_{\alpha [M|} f_{\beta|N]PQ} - \theta_{\alpha [P|} f_{\beta|Q]MN} & \nonumber\\ + \xi_{\alpha[M|} \xi_{\beta [P} \eta_{Q]N]} 
         - \xi_{\alpha[M|} \theta_{\beta [P} \eta_{Q]N]} + \xi_{\alpha[P|} \theta_{\beta [M} \eta_{N]Q]} - 3 \theta_{\alpha[M|} \theta_{\beta [P} \eta_{Q]N]} ) & = 0 \, .
                              \end{align}
In the absence of a gauging of the scaling symmetry, i.e. for $\theta_{\alpha M}=0$, the above constraints consistently reduce to the quadratic identities $(2.20)$ of \cite{Schon:2006kz}, obeyed by the tensors $f_{\alpha M N P}$ and $\xi_{\alpha M}$ that parametrize the standard gaugings of half-maximal supergravity in four dimensions. Furthermore, given the remark in the last paragraph of the previous subsection, the quadratic constraints \eqref{q2}-\eqref{q11} should reproduce the corresponding constraints in \cite{Prins:2013yra} if one performs the redefinition $\xi_{\alpha M} \rightarrow \xi_{\alpha M} - 3 \theta_{\alpha M}$ and then sets $\theta_{\alpha M} = - \frac{1}{2} \kappa_{\alpha M}$. However, we have found that this is not the case and, in fact, the quadratic constraints on the irreducible components of the embedding tensor ${\hat\Theta}$ provided in \cite{Prins:2013yra} are not consistent with the formula for $X_{{\cal M} {\cal N}}{}^{\cal P}$ therein, since they do not imply the closure relation \eqref{closure}; thus, we respectfully disagree with them. 

\subsection{Solution to the Quadratic Constraints}
\label{sec:solmain}

In order to solve the quadratic constraints \eqref{q2}-\eqref{q11} for any number $n\geq1$ of vector multiplets, we employ a strategy similar to the one followed in \cite{LeDiffon:2008sh,LeDiffon:2011wt} for the solution of the corresponding constraints for the maximal supergravities with local scaling symmetry in various spacetime dimensions. In particular, we decompose the embedding tensor with respect to the subgroup $\text{SO}(1,1)_B \times  \text{SO}(1,1)_A \times \text{SO}(5,n-1)$ of $ \text{SL}(2,\mathbb{R}) \times \text{SO}(6,n)$, where $\text{SO}(1,1)_B$ is a subgroup of SL(2,$\mathbb{R}$), while $\text{SO}(1,1)_A \times \text{SO}(5,n-1)$ is embedded into SO(6,$n$). When $ \text{SO}(6,n)$ is broken to $  \text{SO}(1,1)_A \times \text{SO}(5,n-1)$, an SO(6,$n$) vector $v_M$ decomposes into an SO(5,$n-1$) vector $v_{\hat{m}}$, where $\hat{m}=1,\dots,n+4$ is an index labelling the fundamental representation of SO(5,$n-1$), and two SO(5,$n-1$) singlets $v_{\oplus}$ and $v_{\ominus}$ with $\text{SO}(1,1)_A$ weights 0,$+1$ and $-1$ respectively, according to the branching rule
\begin{equation}
    \label{n+6br}
    \mathbf{n+6} = \mathbf{(n+4)^0} + \mathbf{1^{+1}} + \mathbf{1^{-1}} \, , 
\end{equation}
while the SO(6,$n$)-invariant metric $\eta_{MN}$ decomposes as 
\begin{equation}
    \eta_{MN} = ( \eta_{\oplus \ominus} = \eta_{\ominus \oplus} = 1  ,\eta_{\hat{m} \hat{n}})  ,
\end{equation}
where $\eta_{\hat{m} \hat{n}}$ is the SO(5,$n-1$)-invariant metric. Furthermore, with respect to the subgroup $\text{SO}(1,1)_B$ of SL(2,$\mathbb{R}$), an SL(2,$\mathbb{R}$) vector $v_{\alpha}$ splits into two scalars $v_{+}$ and $v_{-}$ with weights $+1$ and $-1$ respectively under $\text{SO}(1,1)_B$. Therefore, with respect to $\text{SO}(1,1)_B \times  \text{SO}(1,1)_A \times \text{SO}(5,n-1)$, the embedding tensor components $\xi_{\alpha M}$ and $\theta_{\alpha M}$ decompose as
\begin{align}
     \label{xdec}
    \xi_{\alpha M} & = (\xi_{+ \hat{m}}, \xi_{+ \oplus}, \xi_{+ \ominus}, \xi_{- \hat{m}}, \xi_{- \oplus}, \xi_{- \ominus}) , \\
    \label{thdec}
     \theta_{\alpha M} & = (\theta_{+ \hat{m}}, \theta_{+ \oplus}, \theta_{+ \ominus}, \theta_{- \hat{m}}, \theta_{- \oplus}, \theta_{- \ominus}) \, . 
\end{align}
Moreover, the $ \text{SL}(2,\mathbb{R}) \times \text{SO}(6,n)$ tensor $f_{\alpha MNP}$ decomposes as 
\begin{equation}
\label{fdec}
    f_{\alpha M N P} = (f_{+ \hat{m} \hat{n} \hat{p}}, f_{+ \oplus \hat{n} \hat{p}},f_{+ \ominus \hat{n} \hat{p}}, f_{+ \oplus \ominus \hat{p}}, f_{- \hat{m} \hat{n} \hat{p}}, f_{- \oplus \hat{n} \hat{p}},f_{- \ominus \hat{n} \hat{p}}, f_{- \oplus \ominus \hat{p}} ) ,
\end{equation}
given the branching rule of the 3-fold antisymmetric representation of SO(6,$n$) with respect to the subgroup  $\text{SO}(1,1)_A \times \text{SO}(5,n-1)$:
\begin{equation}
    \mathbf{\binom{n+6}{3}} = \mathbf{\binom{n+4}{3}^0} + \mathbf{\binom{n+4}{2}^{+1}} + \mathbf{\binom{n+4}{2}^{-1}} + \mathbf{(n+4)^0}
\end{equation}

Our choice to decompose the embedding tensor with respect to the subgroup $\text{SO}(1,1)_B \times  \text{SO}(1,1)_A \times \text{SO}(5,n-1)$ of $ \text{SL}(2,\mathbb{R}) \times \text{SO}(6,n)$ is motivated by the fact that the respective decomposition of $\theta_{\alpha M}$ contains an $\text{SO}(5,n-1)$ singlet, for instance $\theta_{+ \oplus}$, which, if non-zero, allows one to solve all of the quadratic constraints \eqref{q2}-\eqref{q11}. Indeed, 
in appendix \ref{solapp}, we prove that for non-vanishing $\theta_{\alpha M}$, the general solution to these constraints is parametrized by three real $\text{SO}(5,n-1)$ singlets $\theta_{+ \oplus}$, $\theta_{- \oplus}$, $\xi_{+ \oplus}$, a real $\text{SO}(5,n-1)$ vector $\theta_{+ \hat{m}}$, an antisymmetric rank two  $\text{SO}(5,n-1)$ tensor $f_{+ \oplus \hat{m} \hat{n}}$, whose weights under $\text{SO}(1,1)_B \times \text{SO}(1,1)_A$ are indicated by their relevant indices, and four additional real $\text{SO}(1,1)_B \times  \text{SO}(1,1)_A \times \text{SO}(5,n-1)$ tensors $\zeta_{+ \hat{m}}, \, \zeta_{- \hat{m}}, \, \zeta_{- \oplus \hat{m} \hat{n}}= \zeta_{- \oplus [\hat{m} \hat{n}]}, \,  \zeta_{+ \hat{m} \hat{n} \hat{p}} =\zeta_{+ [\hat{m} \hat{n} \hat{p}]} $, which are eigenvectors of the operator $\delta_{f_{+ \oplus}} \equiv  f_{+ \oplus}{}^{\hat{m} \hat{n}} t_{\hat{m} \hat{n}}$, where $t_{\hat{m} \hat{n}} = t_{[\hat{m} \hat{n}]}$ are the generators of $\text{SO}(5,n-1)$, according to 
\begin{align}
   \label{z+meig} 
   \delta_{f_{+ \oplus}} \zeta_{+ \hat{m}} &  \equiv - f_{+ \oplus \hat{m}}{}^{\hat{n}} \zeta_{+ \hat{n}} = \frac{1}{2} \left( \xi_{+ \oplus} + \theta_{+ \oplus} \right) \zeta_{+ \hat{m}} \, , \\
   \label{z+mnpeig}
   \delta_{f_{+ \oplus}} \zeta_{+ \hat{m} \hat{n} \hat{p}} & \equiv - 3 f_{+ \oplus [\hat{m}}{}^{\hat{q}} \zeta_{+ \hat{n} \hat{p}] \hat{q}} = \frac{1}{2} \left( \xi_{+ \oplus} + \theta_{+ \oplus} \right) \zeta_{+ \hat{m} \hat{n} \hat{p}} \, ,   \\
   \label{z-+mneig}
    \delta_{f_{+ \oplus}}  \zeta_{- \oplus \hat{m} \hat{n}} & \equiv 2 f_{+ \oplus [\hat{m}}{}^{\hat{p}} \zeta_{- \oplus \hat{n}] \hat{p}} = \left( -  \xi_{+ \oplus} + \theta_{+ \oplus} \right) \zeta_{- \oplus \hat{m} \hat{n}} \, , \\
    \label{z-meig}
    \delta_{f_{+ \oplus}} \zeta_{- \hat{m}} & \equiv - f_{+ \oplus \hat{m}}{}^{\hat{n}} \zeta_{- \hat{n}}  = -\frac{1}{2} \left( \xi_{+ \oplus} - 3 \theta_{+ \oplus} \right) \zeta_{- \hat{m}} \, . 
   \end{align}
The remaining irreducible components of the embedding tensor can be expressed in terms of the aforementioned tensors as 
\begin{align}
    \label{th+-1}
    \theta_{+ \ominus} = & -\frac{1}{2 \theta_{+ \oplus}} \theta_{+}^{\hat{m}} \theta_{+ \hat{m}} \, , \\ 
    \label{th--1}
    \theta_{- \ominus} = & - \frac{1}{2 \theta_{+ \oplus}^2} \theta_{+}^{\hat{n}} \theta_{+\hat{n}} \theta_{- \oplus} - \frac{1}{\theta_{+ \oplus}} \zeta_{-}^{\hat{m}} \theta_{+ \hat{m}} \, ,\\
    \label{th-m1}
     \theta_{- \hat{m}} = & \,  \frac{\theta_{- \oplus}}{\theta_{+ \oplus}} \theta_{+ \hat{m}} + \zeta_{- \hat{m}} \, ,
    \\
    \label{xi+-1}
    \xi_{+ \ominus} = & - \frac{1}{2} \frac{\xi_{+ \oplus}}{\theta_{+ \oplus}^2} \theta_{+}^{\hat{m}} \theta_{+ \hat{m}} - \frac{1}{\theta_{+ \oplus}} \zeta_{+}^{\hat{m}} \theta_{+ \hat{m}} \, , \\
    \label{xi-+1}
         \xi_{- \oplus} =& \, \frac{\xi_{+ \oplus}}{\theta_{+ \oplus}} \theta_{- \oplus} \, , \\
    \label{xi--1}
    \xi_{- \ominus} = & - \frac{\theta_{- \oplus}}{\theta_{+ \oplus}^2} \zeta_{+}^{\hat{m}} \theta_{+ \hat{m}} + \frac{\xi_{+ \oplus}}{\theta_{+ \oplus}} \theta_{- \ominus} \, , \\
    \label{xi+m1}
    \xi_{+ \hat{m}} = & \, \frac{\xi_{+ \oplus}}{\theta_{+ \oplus}} \theta_{+ \hat{m}} + \zeta_{+ \hat{m}} \, ,\\
    \label{xi-m1}
     \xi_{- \hat{m}} = & \, \frac{\xi_{+ \oplus} \theta_{- \oplus}}{\theta_{+ \oplus}^2} \theta_{+ \hat{m}} + \frac{\theta_{- \oplus}}{\theta_{+ \oplus}} \zeta_{+ \hat{m}} + \frac{\xi_{+ \oplus}}{\theta_{+ \oplus}} \zeta_{- \hat{m}} \, ,\\ 
     \label{f++-m1}
     f_{+ \oplus \ominus \hat{m}} = & \, \frac{1}{\theta_{+ \oplus}}  f_{+ \oplus \hat{m} \hat{n}} \theta_{+}^{\hat{n}} - \frac{1}{2} \zeta_{+ \hat{m}} \, , \\
     \label{f-+-m1}
      f_{- \oplus \ominus \hat{m}} = & \, \frac{\theta_{- \oplus}}{\theta_{+ \oplus}^2} f_{+ \oplus \hat{m} \hat{n}} \theta_{+}^{\hat{n}} + \frac{1}{\theta_{+ \oplus}} \zeta_{- \oplus \hat{m} \hat{n}} \theta_{+}^{\hat{n}} - \frac{1}{2} \frac{\theta_{- \oplus}}{\theta_{+ \oplus}} \zeta_{+ \hat{m}} + \frac{1}{2} \left( \frac{\xi_{+ \oplus}}{\theta_{+ \oplus}} - 3  \right) \zeta_{-\hat{m}} \, , \\
      \label{f+-mn1}
       f_{+ \ominus \hat{m} \hat{n}} = & - \frac{2}{\theta_{+ \oplus}^2} \theta_{+[\hat{m}} f_{+ \oplus \hat{n}] \hat{p}} \theta_{+}^{\hat{p}} - \frac{1}{2 \theta_{+ \oplus}^2} \theta_{+}^{\hat{p}} \theta_{+ \hat{p}}  f_{+ \oplus \hat{m} \hat{n}} \nonumber \\ & - \frac{1}{\theta_{+ \oplus}} \left(  \zeta_{+ \hat{m} \hat{n} \hat{p}} \theta_{+}^{\hat{p}} + \zeta_{+[\hat{m}} \theta_{+ \hat{n}]} \right) , \\
       \label{f-+mn1}
       f_{- \oplus \hat{m} \hat{n}} = &  \, \frac{\theta_{- \oplus}}{\theta_{+ \oplus}} f_{+ \oplus \hat{m} \hat{n}} + \zeta_{- \oplus \hat{m} \hat{n}} \, , \\
       \label{f--mn1}
        f_{- \ominus \hat{m} \hat{n}} = &  \, \frac{1}{\theta_{+ \oplus}^2} \zeta_{-}^{\hat{p}} \theta_{+\hat{p}} f_{+ \oplus \hat{m} \hat{n}}  - 
       \frac{1}{2} \frac{\theta_{-\oplus}}{\theta_{+ \oplus}^3} \theta_{+}^{\hat{p}} \theta_{+ \hat{p}} f_{+ \oplus \hat{m} \hat{n}} - 2 \frac{\theta_{-\oplus}}{\theta_{+ \oplus}^3} \theta_{+ [\hat{m}} f_{+ \oplus \hat{n}] \hat{p}} \theta_{+}^{\hat{p}} \nonumber \\ & + \frac{2}{\theta_{+ \oplus}^2} \zeta_{-[\hat{m}} f_{+ \oplus \hat{n}] \hat{p}} \theta_{+}^{\hat{p}} - \frac{1}{2 \theta_{+ \oplus}^2}  \theta_{+}^{\hat{p}} \theta_{+ \hat{p}} \zeta_{- \oplus \hat{m} \hat{n}} - \frac{2}{\theta_{+ \oplus}^2} \theta_{+[\hat{m}} \zeta_{- \oplus \hat{n}]\hat{p}} \theta_{+}^{\hat{p}}  \\
    & - \frac{\theta_{- \oplus}}{\theta_{+ \oplus}^2} \zeta_{+ \hat{m} \hat{n} \hat{p}} \theta_{+}^{\hat{p}} - \frac{\theta_{- \oplus}}{\theta_{+ \oplus}^2} \zeta_{+ [\hat{m}} \theta_{+ \hat{n}]} + \frac{\xi_{+\oplus}}{\theta_{+ \oplus}^2} \zeta_{- [\hat{m}} \theta_{+ \hat{n}]} - \frac{3}{\theta_{+ \oplus}} \zeta_{- [\hat{m}} \theta_{+ \hat{n}]} \, , \nonumber \\
    \label{f+mnp1}
    f_{+ \hat{m} \hat{n} \hat{p}} = & \, \frac{3}{\theta_{+ \oplus}} \theta_{+ [\hat{m}} f_{+ \oplus \hat{n} \hat{p}]} + \zeta_{+ \hat{m} \hat{n} \hat{p}} \, ,\\
     \label{f-mnp1}
      f_{- \hat{m} \hat{n} \hat{p}} =  & \, 3 \frac{\theta_{- \oplus}}{\theta_{+ \oplus}^2} \theta_{+ [\hat{m}} f_{+ \oplus \hat{n} \hat{p}]}  + \frac{\theta_{- \oplus}}{\theta_{+ \oplus}} \zeta_{+ \hat{m} \hat{n} \hat{p}} - \frac{3}{\theta_{+ \oplus}} \zeta_{- [\hat{m}} f_{+ \oplus \hat{n} \hat{p}]} + \frac{3}{\theta_{+ \oplus}} \theta_{+ [\hat{m}} \zeta_{- \oplus \hat{n} \hat{p}]} \, .
   \end{align}
Furthermore, the $\zeta$ tensors must satisfy the polynomial constraints:
\begin{align}
    \label{z+mz+m}
    &  \zeta^{\hat{m}}_{+} \zeta_{+ \hat{m}} = 0 \, ,\\
    \label{z-mz-m1} 
     &  \zeta^{\hat{m}}_{-} \zeta_{- \hat{m}} = 0 \, ,\\
      \label{z-mnz+n1}
     & \zeta_{- \oplus \hat{m} \hat{n}} \zeta_{+}^{\hat{n}} = 0 \, ,\\
     \label{z-mnz-n1}
     & \zeta_{- \oplus \hat{m} \hat{n}} \zeta_{-}^{\hat{n}} = 0 \, , \\
     \label{z+mz-n1}
     & \zeta_{+ \hat{m}} \zeta_{- \hat{n}} =  0 \, ,  \\ 
      \label{zmnpzp1}
   &   \zeta_{+ \hat{m} \hat{n} \hat{p}} \zeta_{+}^{\hat{p}} = 0  \, ,\\ 
   \label{zmnpz-p1}
   &   \zeta_{+ \hat{m} \hat{n} \hat{p}} \zeta_{-}^{\hat{p}} = 0 \, ,\\
   \label{3zz+2zz1}
   &  3 \zeta_{+ \hat{r}[\hat{m} \hat{n}} \zeta_{+ \hat{p} \hat{q}]}{}^{\hat r} + 2 \zeta_{+ [\hat{m}} \zeta_{+ \hat{n} \hat{p} \hat{q}]} = 0  \, ,\\
   \label{z-mz+npq1}
    &  \zeta_{-[\hat{m}} \zeta_{+ \hat{n} \hat{p} \hat{q}]} = 0 \, ,\\
    \label{zzeta}
    & \zeta_{- \oplus \hat{p}}{}^{\hat{q}} \zeta_{+ \hat{m} \hat{n} \hat{q}} + \zeta_{+ [\hat{m}} \zeta_{- \oplus \hat{n}] \hat{p}} - \zeta_{+ \hat{p}} \zeta_{- \oplus \hat{m} \hat{n}} - 2 \left( \frac{\xi_{+ \oplus}}{\theta_{+\oplus}} - 1  \right) \zeta_{- [\hat{m}} f_{+ \oplus \hat{n}] \hat{p}} \nonumber \\
    & + \frac{2}{\theta_{+ \oplus}} \zeta_{-[\hat{m}} f_{+ \oplus \hat{n}]}{}^{\hat{q}} f_{+ \oplus \hat{p} \hat{q}} - \frac{1}{2} \left( \frac{\xi_{+ \oplus}^2}{\theta_{+ \oplus}} - 2 \xi_{+ \oplus} - 3 \theta_{+ \oplus} \right) \zeta_{-[\hat{m}} \eta_{\hat{n}] \hat{p}} = 0 \, .
       \end{align}
  It is easier to construct solutions to the system of equations \eqref{z+meig}-\eqref{z-meig} and \eqref{z+mz+m}-\eqref{zzeta} than to the original set of quadratic constraints, \eqref{q2}-\eqref{q11}. A simple solution to the former can be obtained by setting $\zeta_{+ \hat{m}} = \zeta_{- \hat{m}} = \zeta_{- \oplus \hat{m} \hat{n}} = \zeta_{+ \hat{m} \hat{n} \hat{p}} = 0  $, which leaves a non-trivial embedding tensor parametrized by  $\theta_{+ \oplus}$, $\theta_{- \oplus}$, $\xi_{+ \oplus}$, $\theta_{+ \hat{m}}$ and $f_{+ \oplus \hat{m} \hat{n}}$. 

  Interestingly, $\text{SO}(1,1) \times \text{SO}(5,n-1)$, where the $\text{SO}(1,1)$-factor is identified with the diagonal of $\text{SO}(1,1)_B$ and $\text{SO}(1,1)_A$ \cite{Schon:2006kz}, is the global symmetry group of the ungauged five-dimensional ${\cal N}=4$ supergravity coupled to $n-1$ vector multiplets \cite{Awada:1985ep,DallAgata:2001wgl}, which, upon compactification on a circle, yields a four-dimensional ${\cal N}=4$ supergravity coupled to $n$ vector multiplets. Thus, at least a subset of the gaugings of four-dimensional ${\cal N}=4$ supergravity described by the general solution to the quadratic constraints \eqref{q2}-\eqref{q11} given in this subsection must have a higher-dimensional origin as Scherk-Schwarz reductions from five dimensions.  

  Obviously, the breaking of the $\text{SO}(6,n)$ symmetry to $\text{SO}(1,1)_A \times \text{SO}(5,n-1)$ requires the existence of at least one vector multiplet, i.e. $n \geq 1$. Let us now discuss the case of pure $D=4, \, {\cal N} = 4$ supergravity, for which $n=0$. Adding the quadratic constraints \eqref{q5} and \eqref{q7} we obtain
  \begin{equation}
      \label{(xi + th)^2 = 0}
      \eta^{MN} \left( \xi_{\alpha M} + \theta_{\alpha M} \right) \left( \xi_{\beta N} + \theta_{\beta N} \right) = 0 \, . 
  \end{equation}
For $n=0$, $\eta^{MN}  = - \delta^{MN} = -  \text{diag}(1,1,1,1,1,1)$, so equation \eqref{(xi + th)^2 = 0} implies that 
\begin{equation}
    \label{xi=-th}
     \xi_{\alpha M} = - \theta_{\alpha M} \, ,      
\end{equation}
which ensures the validity of the constraints \eqref{q2}, \eqref{q5} and \eqref{q7}. Furthermore, as a result of \eqref{xi=-th}, the rest of the quadratic constraints on the embedding tensor are simplified to 
\begin{align}
    \label{q1xi=-th}
    \epsilon^{\alpha \beta} \left( \theta^P_\alpha f_{\beta MNP} - 4 \theta_{\alpha M} \theta_{\beta N}\right) & = 0 \, , 
    \\
    \label{q2xi=-th}
     \theta^P_{(\alpha} f_{\beta) MNP} & = 0 \, , \\
     \label{q3xi=-th}
     f_{\alpha [MN|R} f_{\beta|PQ]}{}^R & = 0  \, , \\ 
     \label{q4xi=-th}
     \epsilon^{\alpha \beta} \theta_{\alpha[M|} f_{\beta|NPQ]} & =0 \, ,\\
     \label{q5xi=-th}
      \epsilon^{\alpha \beta} \left( f_{\alpha MNR} f_{\beta PQ}{}^R + 2  \theta_{\alpha [M|} f_{\beta|N]PQ} -2 \theta_{\alpha [P|} f_{\beta|Q]MN} \right)&  = 0 \,. 
\end{align}
A non-trivial solution to the conditions \eqref{q1xi=-th}-\eqref{q5xi=-th} can be obtained by assigning arbitrary non-zero values to the components $f_{+123}$, $\theta_{+4}$, $\theta_{+5}$ and $\theta_{+6}$ of the embedding tensor and setting all the other components of $f_{\alpha MNP}$ and $\theta_{\alpha M}$ equal to zero. In this case, the vector fields $A^{1+}_{\mu}$, $A^{2+}_{\mu}$ and $A^{3+}_{\mu}$ gauge an $\text{SO}(3)$ subgroup of $\text{SO}(6)$, while the linear combinations $ \sum_{M=4}^{6} \theta_{+M} A^{M+}_{\mu}$ and $\sum_{M=4}^{6} \theta_{+M} A^{M-}_{\mu}$ of the vector fields gauge the subgroup of $\text{SL}(2,\mathbb{R})$ generated by $\sigma_3$ and $\sigma_1+i \sigma_2$ and the former gauges the scaling symmetry as well. Another simple way of satisfying \eqref{q1xi=-th}-\eqref{q5xi=-th} is by setting $f_{\alpha MNP} = 0$ and $\theta_{\alpha M} = v_{\alpha} A_M$, where $v_{\alpha}$ and $A_M$ are arbitrary real $\text{SL}(2,\mathbb{R})$ and $\text{SO}(6)$ vectors respectively. Therefore, the gauging of the scaling symmetry is consistent for $n=0$ as well,  since there still exist solutions to the quadratic constraints \eqref{q2}-\eqref{q11} with non-vanishing $\theta_{\alpha M} $. 
\subsection{Gauge Covariant Field Strengths}
Under a gauge transformation with infinitesimal parameters $\zeta^{\cal M}(x)$ the one-form gauge fields $A^{\cal M}$ transform as 
\begin{equation}
   \label{dA}
   \delta_{\zeta} A^{\cal M} = \hat{d} \zeta^{\cal M} = d \zeta^{\cal M} + g {X_{{\cal N} {\cal P}}}^{\cal M} A^{\cal N} \zeta^{\cal P}
\end{equation}
and their non-abelian two-form field strengths are defined by 
\cite{deWit:2005hv,deWit:2004nw,deWit:2005ub}
\begin{align}
    \label{HMa}
    H^{M \alpha} = & \, dA^{M \alpha} + \frac{g}{2} X_{N \beta P \gamma}{}^{M \alpha} A^{N \beta} \wedge A^{P \gamma} + g Z^{M \alpha A} B_{A} \nonumber \\
    =& \, dA^{M \alpha} + \frac{g}{2} X_{N \beta P \gamma}{}^{M \alpha} A^{N \beta} \wedge A^{P \gamma} \\
    & - \frac{g}{2} \Theta^{\alpha M}{}_{NP} B^{NP} + \frac{3}{2} g \theta^{\alpha}_N B^{MN} + \frac{g}{2} \left( \xi^M_\beta + \theta^M_\beta \right) B^{\alpha \beta},   \nonumber 
\end{align}
where $B^{MN}=B^{[MN]}$ and $B^{\alpha \beta}=B^{(\alpha \beta)}$ are two-form gauge fields in the adjoint representations of SO(6,$n$) and SL(2,$\mathbb{R}$) respectively. The above field strengths transform covariantly under \eqref{dA}, i.e. 
\begin{equation}
    \label{cov}
    \delta_\zeta  H^{\cal M} = - g {X_{{\cal N} {\cal P}}}^{\cal M} {\zeta}^{\cal N} H^{\cal P},
\end{equation}
provided the two-form gauge fields transform as (see for example \cite{Trigiante:2016mnt})
\begin{align}
      \label{dzBMN}
    \delta_{\zeta} B^{MN} &= \epsilon_{\alpha \beta} \left(-2 \zeta^{[M| \alpha} H^{|N] \beta} + A^{[M| \alpha} \wedge \delta_{\zeta} A^{|N] \beta}\right), \\[2mm]
    \label{dzBab}
    \delta_{\zeta} B^{\alpha \beta} & = \eta_{MN} \left( 
2 \zeta^{M (\alpha|} H^{N |\beta)} - A^{M (\alpha|} \wedge \delta_{\zeta} A^{N |\beta)} \right).
\end{align}
Furthermore, the field strengths \eqref{HMa} are invariant under the following tensor gauge transformations, which are parametrized by one-forms $\Xi^{MN} = \Xi^{[MN]} $ and $\Xi^{\alpha \beta}=\Xi^{(\alpha \beta)}$:
\begin{align}
    \label{dxA}
    \delta_\Xi A^{M \alpha} & = - g Z^{M \alpha  A} \Xi_A = \frac{g}{2} \Theta^{\alpha M}{}_{NP} \Xi^{NP} - \frac{3}{2} g \theta^\alpha_N \Xi^{MN} - \frac{g}{2} \left( \xi^M_\beta  + \theta^M_\beta \right) \Xi^{\alpha \beta}, \\ 
    \label{dXiBMN}
    \delta_{\Xi} B^{MN} & = \hat{d}  {\Xi}^{MN} + \epsilon_{\alpha \beta} A^{[M|\alpha} \wedge \delta_{\Xi} A^{|N]\beta}, \\
    \label{dXiBab}
    \delta_{\Xi} B^{\alpha \beta} & = \hat{d} \Xi^{\alpha \beta} - \eta_{MN} A^{M(\alpha|} \wedge \delta_{\Xi} A^{N |\beta)} , 
    \end{align}
where
\begin{equation}
\label{dXMN}
   \hat{d}  {\Xi}^{MN} \equiv d  {\Xi}^{MN}  -2 g \theta_{\alpha P} A^{P \alpha} \wedge \Xi^{MN} + 2 g {\hat\Theta}_{\alpha P Q}{}^{[M|} A^{P \alpha} \wedge \Xi^{|N]Q} 
\end{equation}
and 
\begin{align}
    \label{dXab}
     \hat{d} {\Xi}^{\alpha \beta} \equiv & \, d {\Xi}^{\alpha \beta} - 2 g \theta_{\gamma M} A^{M \gamma} \wedge \Xi^{\alpha\beta} \nonumber \\ 
     & - g \left(\xi^{(\alpha|M} - \theta^{(
     \alpha|M}\right)A_{M \gamma} \wedge \Xi^{|\beta) \gamma} - g \left(\xi_{\gamma M} - \theta_{\gamma M} \right)A^{M (\alpha} \wedge \Xi^{\beta) \gamma} , 
\end{align}
since the parameters of the tensor gauge transformations have scaling weight $+2$, as do the two-form gauge fields. 

The gauge covariant three-form field strengths of the two-form gauge fields are defined by \cite{deWit:2005hv}
\begin{align}
\label{HMN}
    {\cal H}^{(3) MN} = \, & {\hat d}B^{MN}   + \epsilon_{\alpha \beta} A^{[M|\alpha} \wedge \left( dA^{|N]\beta} + \frac{g}{3} {X_{P \gamma Q \delta}}^{|N]\beta} A^{P \gamma} \wedge A^{Q \delta} \right), \\
\label{Hab}    
{\cal H}^{(3) \alpha \beta} = \, & {\hat d}B^{\alpha \beta} - \eta_{MN} A^{M(\alpha|} \wedge \left(  dA^{N|\beta)} + \frac{g}{3} {X_{P \gamma Q \delta}}^{N|\beta)} A^{P \gamma} \wedge A^{Q \delta} \right) .
\end{align}
In addition, the field strengths \eqref{HMa}, \eqref{HMN} and \eqref{Hab} satisfy the Bianchi identities
\begin{align}
    \label{2Bianchi}
    \hat{d} H^{M \alpha} & = g Z^{M \alpha A} {\cal H}^{(3)}_A  , \\
    \label{3Bianchi}
    Z^{M \alpha A} \hat{d} {\cal H}^{(3)}_A & = X_{N \beta P \gamma}{}^{M \alpha} H^{N \beta} \wedge H^{P \gamma} ,  
\end{align}
where $\hat{d} H^{M \alpha}  \equiv d H^{M \alpha} + g X_{N \beta P \gamma}{}^{M \alpha} A^{N \beta} \wedge H^{P \gamma}$.

\subsection{Scalar Sector}

In this subsection, we discuss the interplay of the scalar sector of $D=4$, ${\cal N}=4$ matter-coupled supergravity with its gaugings that involve the scaling symmetry.

For the coset space parametrized by the complex scalar of the ${\cal N}=4$ supergravity multiplet, we define the gauged SL(2,$\mathbb{R}$)/SO(2) zweibein by 
\begin{equation}
\label{hatPdef}
	\hat{P}=\frac{i}{2} \epsilon^{\alpha \beta} {\cal V}_{\alpha} \hat{d} {\cal V}_\beta
\end{equation} 
and the gauged SO(2) connection by 
\begin{equation}\label{hatcalA}
	\hat{\cal A} = - \frac{1}{2} \epsilon^{\alpha \beta} {\cal V}_{\alpha}  \hat{d} {\cal V}_\beta^* \, ,
\end{equation} 
where  
\begin{equation}
    \hat{d} {\cal V}_\alpha \equiv d {\cal V}_\alpha + \frac{1}{2} g   \left(\xi_{\alpha M} - \theta_{\alpha M}\right) A^{M \beta} {\cal V}_{\beta} + \frac{1}{2} g \left( \xi^{ \beta M} - \theta^{ \beta M} \right) A_{M \alpha} {\cal V}_{\beta} \, .
\end{equation}
The definitions \eqref{hatPdef} and \eqref{hatcalA} follow from the expansion of the gauged Maurer-Cartan left-invariant one-form associated with the coset manifold SL(2,$\mathbb{R}$)/SO(2) along the basis $\{\sigma_1, i \sigma_2, \sigma_3\}$ of the Lie algebra $\mathfrak{sl}(2,\mathbb{R})$, where $i \sigma_2$ spans its compact $\mathfrak{so}(2)$ subalgebra (see \cite{DallAgata:2023ahj} for more details).

The Maurer-Cartan equation satisfied by this form 
implies the Bianchi identity
\begin{equation}
    \label{DhatPhat}
    \hat{D} \hat{P} \equiv d \hat{P} - 2 i \hat{\cal A} \wedge \hat{P} = \frac{i}{2} g \left(\xi_{\alpha M} - \theta_{\alpha M} \right){\cal V}^\alpha  {\cal V}_\beta {H}^{M \beta}  
    \end{equation}
and gives the following expression for the gauged SO(2) curvature:
\begin{equation}
\label{gSO(2)curv}
     \hat{F} \equiv d \hat{\cal A} = \, i \hat{P}^* \wedge \hat{P} + \frac{g}{2} \left( \xi^\alpha_M  - \theta^\alpha_M \right) M_{\alpha \beta} H^{M \beta}  .
\end{equation}
With some algebra, one can also derive the useful identity
\begin{equation}
\label{hatDV=hatPV}
    \hat{D} {\cal V}_{\alpha} \equiv \hat{d}  {\cal V}_{\alpha} - i \hat{\cal A} {\cal V}_{\alpha} = \hat{P} {\cal V}_{\alpha}^* \, .
\end{equation}

On the other hand, for the coset space parametrized by the $6n$ real scalars of the $n$ vector multiplets, we define the gauged SO(6,$n$)/($\text{SU}(4)\times \text{SO}(n)$) vielbein by 
\begin{equation}
    \label{harPaij}
    {\hat P}_{\underline{a}}{}^{ij} = L^{M}{}_{\underline{a}} {\hat d} L_M{}^{ij} , 
\end{equation}
the gauged SU(4)$\simeq$SO(6) connection by 
\begin{equation}
    \label{SU4conn}
        {\hat\omega}^i{}_j  = L^{M ik } \hat{d} L_{Mjk}
\end{equation}
and the gauged SO($n$) connection by 
\begin{equation}
    \label{SOnconn}
    {\hat\omega}_{\underline{a}}{}^{\underline{b}} = L^M{}_{\underline{a}} \hat{d} {L_M}^{\underline{b}}  ,
\end{equation}
where
\begin{equation}
    \label{hatdLM}
     \hat{d} L_M{}^{\underline{M}} \equiv d  L_M{}^{\underline{M}} + g A^{N \alpha} {\hat\Theta}_{\alpha N M}{}^P L_P{}^{\underline{M}} .
\end{equation}

In this case, the relevant gauged Maurer-Cartan equations imply the Bianchi identity 
\begin{equation}
    \label{scalarBianchi3}
    \hat{D} {{\hat{P}}_{\underline{a}}}{}^{ij}  \equiv   d {{\hat{P}}_{\underline{a}}}{}^{ij} + {{\hat{\omega}}_{\underline{a}}}{}^{\underline{b}} \wedge {{\hat{P}}_{\underline{b}}}{}^{ij} - {\hat{\omega}^i}{}_{k} \wedge {{\hat{P}}_{\underline{a}}}{}^{kj} - {\hat{\omega}^j}{}_{k} \wedge {{\hat{P}}_{\underline{a}}}{}^{ik} 
     = g {{\hat\Theta}_{\alpha M }}{}^{N P}   L_{N \underline{a}} {L_P}^{ij}  H^{M \alpha}  , 
\end{equation}
as well as the following expressions for the gauged SU(4) and SO($n$) curvatures $ {{\hat{R}}^i}{}_{j}  $ and ${{\hat{R}}_{\underline{a}}}{}^{ \underline{b}}$ respectively:
\begin{align}
    \label{SU4curv}
     {{\hat{R}}^i}{}_{j} &  \equiv d {\hat{\omega}^i}{}_{j} - {\hat{\omega}^i}{}_{k} \wedge {\hat{\omega}^k}{}_{j} = {\hat{P}}^{\underline{a} ik } \wedge {\hat{P}}_{\underline{a} jk}  + g {{\hat\Theta}_{\alpha M }}{}^{N P}   {L_N}^{ik} L_{Pjk} H^{M \alpha} , \\
    \label{SOncurv}
    {{\hat{R}}_{\underline{a}}}{}^{ \underline{b}}  & \equiv  d {\hat\omega}_{\underline{a}}{}^{\underline{b}} + {\hat\omega}_{\underline{a}}{}^{\underline{c}} \wedge {\hat\omega}_{\underline{c}}{}^{\underline{b}}    = 
 - {\hat{P}}_{\underline{a}ij} \wedge {\hat P}^{\underline{b} i j} + g {{\hat\Theta}_{\alpha M }}{}^{N P}   L_{N \underline{a}} {L_P}^{\underline{b}} H^{M \alpha} .
\end{align}
One can also derive the following useful relations: 
\begin{align}\label{hatDLij}
  \hat{D} {L_M}^{ij} & \equiv  \hat{d} {L_M}^{ij} - {\hat{\omega}^i}{}_{k} {L_M}^{kj}  - {{\hat{\omega}}^j}{}_{k} {L_M}^{ik} =   {L_M}^{\underline{a}} {\hat{P}_{\underline{a}}}{}^{ij} , \\[2mm]
\label{hatDLa}
 \hat{D} {L_M}^{\underline{a}} &\equiv  \hat{d} {L_M}^{\underline{a}} + {\hat{\omega}^{\underline{a}}}{}_{\underline{b}} {L_M}^{\underline{b}} = {L_M}^{ij} {\hat{P}^{\underline{a}}}{}_{ij} \, .  
\end{align}

\section{Supersymmetry Algebra}
\label{sec:susy_rules}
In this section, we provide the local supersymmetry transformation rules of $D=4$, ${\cal N}=4$ matter-coupled supergravity with local scaling symmetry and the algebra that they obey. In the geometric approach \cite{Castellani:1991eu}, the local supersymmetry transformations of the spacetime fields follow from the restrictions to spacetime of the Lie derivatives of the corresponding superfields along a tangent vector that is dual to the gravitino super-one-forms, as explained in appendix \ref{sec:Bianchi}. 

In particular, the ${\cal N}=4$ local supersymmetry transformations of the bosonic fields $e^a_{\mu}$, ${\cal V}_{\alpha}$, $L_{Mij}$, $L_{M \underline{a}}$ and $A^{M \alpha}_{\mu}$ are the same as in the ungauged (or standard gauged) theory and are given by \cite{DallAgata:2023ahj}
\begin{align}
    \label{deag}
    \delta_{\epsilon} e^a_\mu = \, & {\bar\epsilon}^i \gamma^a \psi_{i \mu} + {\bar\epsilon}_i \gamma^a \psi^i_\mu  \, , \\
       \label{dVg}  
       \delta_{\epsilon} {\cal V}_\alpha = \, & {\cal V}_{\alpha}^* {\bar\epsilon}_i \chi^i, \\
       \label{dLMijg}
        \delta_{\epsilon} L_{Mij} = \, & L_{M \underline{a}} ( 2 {\bar\epsilon}_{[i} \lambda^{\underline{a}}_{j]} + \epsilon_{ijkl} {\bar\epsilon}^k \lambda^{\underline{a}l} ) \, , \\
        \label{dLMag}
        \delta_{\epsilon} {L_M}^{\underline{a}} = \, & 2 {L_M}^{ij} {\bar\epsilon}_i \lambda^{\underline{a}}_j + c.c. \, ,\\
    \label{dAMag}
     \delta_{\epsilon} A^{M \alpha}_\mu = \, & ({\cal V}^\alpha)^* {L^M}_{ij} {\bar\epsilon}^i \gamma_\mu \chi^j - {\cal V}^\alpha L^{M \underline{a}} {\bar\epsilon}^i \gamma_\mu \lambda_{\underline{a}i} + 2 {\cal V}^\alpha {L^M}_{ij} {\bar\epsilon}^i \psi^j_\mu + c.c. \, ,
     \end{align}
     while the corresponding transformations of the linear combinations
     \begin{equation}
     \label{BMa}
         B^{M \alpha}_{\mu \nu} \equiv - \frac{1}{2} {{\Theta}^{\alpha M}}_{NP} B^{NP}_{\mu \nu} + \frac{3}{2} \theta^{\alpha}_N B^{MN}_{\mu \nu} + \frac{1}{2} \left(\xi^M_\beta + \theta^M_\beta \right) B^{\alpha \beta}_{\mu \nu}
     \end{equation}
            of the antisymmetric tensor gauge fields, which appear in the gauge covariant field strengths \eqref{HMa} of the vector gauge fields, read
     \begin{align}
    \label{dB}
        \delta_{\epsilon} B^{M\alpha}_{\mu \nu}  = \, & -4 i Z^{ M\alpha NP} {L_N}^{\underline{a}} {L_P}^{ij} {\bar\epsilon}_i \gamma_{\mu \nu} \lambda_{\underline{a}j} + \frac{1}{2} \left( \xi^M_\beta + \theta^M_\beta \right) ({\cal V}^\alpha)^* ({\cal V}^\beta)^* {\bar\epsilon}_i \gamma_{\mu \nu} \chi^i \nonumber \\[2mm]
        & + 4 i  Z^{ M \alpha NP} {L_N}^{\underline{a}} L_{Pij} {\bar\epsilon}^i \gamma_{\mu\nu} \lambda^j_{\underline{a}}   + \frac{1}{2} \left( \xi^M_\beta + \theta^M_\beta \right) {\cal V}^\alpha  {\cal V}^\beta
        {\bar\epsilon}^i \gamma_{\mu\nu} \chi_i \nonumber \\[2mm]
        & + 8i Z^{ M \alpha NP} {L_N}^{ik} L_{Pjk} \left({\bar\epsilon}^j \gamma_{[\mu|} \psi_{i|\nu]} +{\bar\epsilon}_i \gamma_{[\mu} \psi^j_{\nu]}\right)       \\
        & +  \left( \xi^M_\beta + \theta^M_\beta \right) M^{\alpha \beta} \left({\bar\epsilon}^i \gamma_{[\mu|} \psi_{i|\nu]} +{\bar\epsilon}_i \gamma_{[\mu} \psi^i_{\nu]}\right) \nonumber \\[2mm] 
        & + 2 {Z^{ M \alpha}}_{NP}  
       \epsilon_{\beta \gamma} A^{N \beta}_{[\mu} \delta_{\epsilon} A^{P \gamma}_{\nu]} - \left(\xi^M_\beta + \theta^M_\beta \right) \eta_{NP} A^{N(\alpha|}_{[\mu}\delta_{\epsilon} A^{P|\beta)}_{\nu]} \nonumber . 
   \end{align}
   
Before giving the local supersymmetry transformations of the fermions, we introduce the symplectic vector ${\cal G}^{M \alpha}_{\mu \nu} = (H^{\Lambda}_{\mu \nu}, {\cal G}_{\Lambda \mu \nu})$, where $H^{\Lambda}_{\mu \nu}=\Pi^{\Lambda}{}_{M \alpha} H^{M \alpha}_{\mu \nu}$ are the field strengths of the electric vector fields $A^{\Lambda}_{\mu}$ and ${\cal G}_{\Lambda \mu \nu}$ are their magnetic duals, defined by 
\begin{equation}
    \label{GLg}
    {\cal G}_{\Lambda \mu \nu} \equiv {\cal R}_{\Lambda \Sigma} H^{\Sigma}_{\mu \nu} - \frac{1}{2} \epsilon_{\mu \nu \rho \sigma} {\cal I}_{\Lambda \Sigma} H^{\Sigma \rho \sigma} + \text{fermions} \, ,
\end{equation}
where ${\cal R}_{\Lambda \Sigma}$ and ${\cal I}_{\Lambda \Sigma}$ are the kinetic matrices of the electric vector fields in the Lagrangian for the standard gauged $D=4$, ${\cal N}=4$ supergravity. The expressions for these matrices depend on the choice of the symplectic frame and can be deduced from the following decomposition of the $2(n+6) \times 2(n+6)$ matrix ${\cal M}_{\cal M \cal N} = M_{MN}   M_{\alpha \beta}  $ in a given symplectic frame:
\begin{equation}
	\label{Mmatrix}
    {\cal M}_{\mathcal{M} \mathcal{N}} = 
    \begin{pmatrix}
        {\cal M}_{\Lambda \Sigma} & {{\cal M}_{\Lambda}}^{\Sigma} \\[2mm]
        {{\cal M}^{\Lambda}}_{\Sigma} &{\cal M}^{\Lambda \Sigma}
    \end{pmatrix}=
    \begin{pmatrix}
    -({\cal I}+{\cal R}{\cal I}^{-1}{\cal R})_{\Lambda \Sigma} & {({\cal R} {\cal I}^{-1})_\Lambda}^{\Sigma} \\[2mm]
    {({\cal I}^{-1}{\cal R})^\Lambda}_\Sigma & - ({\cal I}^{-1})^{\Lambda \Sigma} 
    \end{pmatrix}.
\end{equation}
The fermionic part of \eqref{GLg} is explicitly given in \cite{DallAgata:2023ahj}. We also point out that ${\cal G}^{M \alpha}_{\mu \nu}$ satisfies the twisted self-duality condition \cite{DallAgata:2023ahj}
\begin{equation}
    \label{twist}
    \epsilon_{\mu \nu \rho \sigma} {\cal G}^{M \alpha \rho \sigma} =  2 \eta^{M N} \epsilon^{\alpha \beta} M_{NP} M_{\beta \gamma} {\cal G}^{P \gamma}_{\mu \nu} +  \text{two-fermion terms} \, .
\end{equation}

The local supersymmetry transformation rules for the fermionic fields $\psi_{i \mu}$, $\chi_i$ and $\lambda_{\underline{a}i}$ in the trombone gauged 
four-dimensional half-maximal supergravity  can be written in a 
manifestly $\text{SL}(2,\mathbb{R}) \times \text{SO}(6,n)$-covariant form by means of the symplectic vector ${\cal G}^{M \alpha}_{\mu \nu}$ and, up to terms quadratic in the fermions, which can be found in \cite{DallAgata:2023ahj}, are given by
   \begin{align}
       \label{dpsig}
     \delta_{\epsilon} \psi_{i \mu} = \, & {\hat D}_{\mu} \epsilon_i - \frac{i}{8} {\cal V}_\alpha L_{M ij} {\cal G}^{M \alpha}_{\nu \rho} \gamma^{\nu \rho} \gamma_\mu  \epsilon^j - \frac{1}{3} g {\bar A}_{1ij} \gamma_\mu \epsilon^j + \frac{g}{2} \epsilon_{ijkl} B^{kl} \gamma_\mu \epsilon^j ,  \\
      \label{dchig}
      \delta_{\epsilon} \chi_i =  & - \frac{i}{4} {\cal V}_{\alpha}^* L_{M ij} {\cal G}^{M \alpha}_{\mu \nu} \gamma^{\mu \nu} \epsilon^j  +{\hat P}_{\mu}^* \gamma^\mu \epsilon_i    + \frac{2}{3} g {\bar A}_{2ij} \epsilon^j - g {\bar B}_{ij} \epsilon^j , \\
      \label{dlambdag}
      \delta_{\epsilon} \lambda_{\underline{a}i} = & \, \frac{i}{8} {\cal V}_{\alpha}^* L_{M \underline{a}} {\cal G}^{M \alpha}_{\mu \nu} \gamma^{\mu \nu} \epsilon_i 
     - {\hat P}_{\underline{a}ij\mu} \gamma^\mu \epsilon^j   + g {\bar A}_{2 \underline{a}}{}^j{}_i \epsilon_j - \frac{1}{4} g {\bar B}_{\underline{a}} \epsilon_i \, ,
    \end{align}
where the fermion shift tensors read
\begin{align}     
     \label{dilshift}
   A_2^{ij} & =  f_{\alpha MNP} {\cal V}^\alpha {L^M}_{kl} L^{Nik} L^{P jl} + \frac{3}{2} \xi_{\alpha M} {\cal V}^\alpha L^{M ij}, \\
    \label{gaushift}
    A_{2 \underline{a} i}{}^j & =  f_{\alpha MNP} {\cal V}^\alpha L^M{}_{\underline{a}} {L^N}_{ik} L^{Pjk} - \frac{1}{4} \delta^j_i \xi_{\alpha M}  {\cal V}^\alpha L^M{}_{\underline{a}}\, , \\ 
    \label{gravshift}
   A_1^{ij} & =   f_{\alpha MNP} ({\cal V}^\alpha)^* L^M{}_{kl} L^{Nik} L^{Pjl}  , \\
   \label{Bijmain}
   B^{ij} & = \theta_{\alpha M} {\cal V}^\alpha L^{Mij} , \\
   \label{Bamain}
   B^{\underline{a}} & = \theta_{\alpha M} {\cal V}^{\alpha} L^{M \underline{a}} .
      \end{align}
In particular, the $A$ tensors \eqref{dilshift}-\eqref{gravshift} were first introduced in \cite{Schon:2006kz}. Furthermore, the covariant derivatives of the supersymmetry transformation parameters $\epsilon_i$ in \eqref{dpsig} are explicitly given by
\begin{equation}
    \label{Dhatmepsilon}
    {\hat D}_\mu \epsilon_i \equiv \partial_\mu \epsilon_i + \frac{1}{4} \omega_{\mu a b} (e,A,\psi) \gamma^{ab} \epsilon_i - \frac{i}{2} {\hat{\cal A}}_\mu \epsilon_i - {\hat \omega}_i{}^j{}_{\mu} \epsilon_j - \frac{g}{2} \theta_{\alpha M} A^{M \alpha}_\mu \epsilon_i ,
\end{equation}
where 
\begin{align}
    \label{spinconn}
    \omega_{\mu}{}^{a b} (e,A,\psi) = & \, 2 e^{\nu [a} \partial_{[\mu} e^{b]}_{\nu]} - e^{\nu [a} e^{b]\rho} e_{c \mu} \partial_\nu e^c_\rho + {\bar\psi}^i_\mu \gamma^{[a} \psi^{b]}_i +  {\bar\psi}^{i [a} \gamma^{b]} \psi_{i \mu} + {\bar\psi}^{i [a} \gamma_\mu \psi^{b]}_i \nonumber \\ 
    & - 2 g e^{[a}_\mu e^{b]\nu} \theta_{\alpha M} A^{M \alpha }_\nu
\end{align}
is the solution for the spin connection ${\omega_\mu}^{ab}$ of the restriction of the supertorsion constraint \eqref{Thata=0}, ${\hat T}^a=0$, to spacetime.

The commutator of two consecutive local supersymmetry transformations, $\delta_Q (\epsilon_1)$ and $\delta_Q (\epsilon_2)$, parametrized by left-handed Weyl spinors $\epsilon_1^i$ and $\epsilon_2^i$ respectively and their charge conjugates, reads
\begin{align}
  \label{QQcomg}
  [\delta_Q (\epsilon_1), \delta_Q (\epsilon_2)] = & \, \delta_{\text{cgct}}(\xi^\mu) \,  + \delta_{\text{Lorentz}} (\lambda_{ab}) \, + \delta_Q (\epsilon_3) + \delta_{\text{SO}(2)} (\Lambda) \nonumber \\ 
  &   + \delta_{\text{SU}(4)} ({\Lambda_i}^j) + \delta_{\text{SO}(n)} ({\Lambda_{\underline{a}}}^{\underline{b}}) + \delta_{\text{gauge}}(\zeta^{M \alpha}) + \delta_{\text{tensor}}(\Xi^{MN}_\mu, \Xi^{\alpha \beta}_\mu) \, , 
  \end{align}
where the first term denotes a covariant general coordinate transformation \cite{deWit:1975veh, Jackiw:1978ar,Freedman:2012zz}  with diffeomorphism parameter 
\begin{equation}
     \label{xim}
     \xi^\mu = \bar{\epsilon}_{2 i} \gamma^\mu \epsilon^i_1 + \bar{\epsilon}^i_2 \gamma^\mu \epsilon_{1i} \, . 
 \end{equation}
We refer the reader to \cite{DallAgata:2023ahj} for the explicit form of this transformation. The expressions for the parameters of the remaining transformations that appear on the right-hand side of \eqref{QQcomg} can be found in the same reference. Here, we only give 
the parameters of the vector and tensor gauge transformations:
\begin{align}
\label{zMa}
    \zeta^{M \alpha} = & -2 ({\cal V}^\alpha)^* L^{Mij} \bar{\epsilon}_{1i} \epsilon_{2j} + \text{c.c.}\,, \\
    \label{XMN}
        \Xi_{MN \mu} = & \, 4 i {L_{[M}}^{ik} L_{N] jk} \left( {\bar\epsilon}_{1 i} \gamma_\mu \epsilon_2^j - {\bar\epsilon}_{2 i} \gamma_\mu \epsilon_1^j\right), \\
        \label{Xab}
    \Xi_{\alpha \beta \mu} = & \, M_{\alpha \beta} \left( 
{\bar\epsilon}_{1 i} \gamma_\mu \epsilon_2^i - {\bar\epsilon}_{2 i} \gamma_\mu \epsilon_1^i \right)  .
\end{align}

The supersymmetry algebra \eqref{QQcomg} has been verified in \cite{DallAgata:2023ahj} for the standard gaugings of four-dimensional ${\cal N}=4$ supergravity, for which $\theta_{\alpha M}=0$. In the presence of a gauging of the scaling symmetry, the algebra \eqref{QQcomg} still closes on the vielbein $e^a_{\mu}$ and on the representatives ${\cal V}_{\alpha}$, $L_{Mij}$ and $L_{M \underline{a}}$ of the coset manifold parametrized by the scalar fields of the theory. Of course, in the latter case, the action of the commutator $ [\delta_Q (\epsilon_1), \delta_Q (\epsilon_2)]$ on these fields yields additional terms that involve the embedding tensor components $\theta_{\alpha M}$. For example, for the vielbein we find 
\begin{equation}
    \label{scaleviel}
    [\delta_Q (\epsilon_1), \delta_Q (\epsilon_2)] e^a_{\mu} \supset - 2 g \theta_{\alpha M} \left( {\cal V}^{\alpha} L^M{}_{ij} {\bar\epsilon}^i_1 \epsilon^j_2 + ({\cal V}^{\alpha})^* L^{Mij} {\bar\epsilon}_{1i} \epsilon_{2j}  \right) e^a_{\mu} \, , 
\end{equation}
which is precisely a scaling gauge transformation of the vielbein with parameters given by \eqref{zMa}. 

Furthermore, the action of the commutator of two local supersymmetry transformations on the vector gauge fields gives 
\begin{align}
\label{QQAMa}
  [\delta_Q(\epsilon_1), \delta_Q (\epsilon_2)] A^{M \alpha}_\mu = & \, \delta_{\text{cgct}}(\xi^\nu) A^{M \alpha}_{\mu}  + \delta_Q (\epsilon_3) A^{M \alpha}_\mu + \delta_{\text{gauge}} (\zeta^{N \beta})  A^{M \alpha}_\mu \nonumber \\
  &+ \delta_{\text{tensor}}(\Xi^{NP}_\nu, \Xi^{\beta \gamma}_\nu) A^{M \alpha}_\mu      - \xi^\nu \left({\cal G}^{M \alpha}_{\mu \nu} - H^{M \alpha}_{\mu \nu} \right) ,
\end{align}
which has exactly the same form as in the standard gauged theory \cite{DallAgata:2023ahj}.
Since ${\cal G}^{\Lambda}_{\mu \nu} \equiv H^{\Lambda}_{\mu \nu}$, \eqref{QQAMa} implies that the ${\cal N}=4$ supersymmetry algebra \eqref{QQcomg} closes on the electric vector fields $A^{\Lambda}_{\mu}$. It also closes on the linear combinations $\Pi^{\Lambda}{}_{M \alpha} \theta^{\alpha M} A_{\Lambda \mu}$ and $\Pi^{\Lambda}{}_{M \alpha} {\hat\Theta}^{\alpha M A} A_{\Lambda \mu}$ of the magnetic gauge fields provided 
\begin{equation}
\label{H=G1}
    \theta_{\cal M} \left( {\cal G}^{\cal M}_{\mu \nu} - H^{\cal M}_{\mu \nu}  \right) = 0 
\end{equation}
and 
\begin{equation}
    \label{H=G2}
     {\hat\Theta}_{\cal M}{}^A   \left( {\cal G}^{\cal M}_{\mu \nu} - H^{\cal M}_{\mu \nu}  \right) = 0 
\end{equation}
respectively. Equations \eqref{H=G1} and \eqref{H=G2} are duality equations between the electric and the magnetic vector fields projected with the components $\theta_{\cal M}$ and  ${\hat\Theta}_{\cal M}{}^A$ of the embedding tensor respectively.

In addition, for the two-form gauge fields $B^{M \alpha}_{\mu \nu}$ we find 
  \begin{align}
      [\delta_Q(\epsilon_1), \delta_Q (\epsilon_2)] B^{M \alpha}_{\mu \nu} =& \,  \delta_Q (\epsilon_3) B^{M \alpha}_{\mu \nu} + \delta_{\text{gauge}} (\zeta^{N \beta}) B^{M \alpha}_{\mu \nu} +  \delta_{\text{tensor}}(\Xi^{NP}_\rho, \Xi^{\beta \gamma}_\rho) B^{M \alpha}_{\mu \nu} \nonumber \\
      & - \delta_Q (\xi^\rho \psi^i_\rho) B^{M \alpha}_{\mu \nu} + \epsilon_{\mu \nu \rho \sigma} \xi^{\rho} J^{M \alpha \sigma}\\
& -2  {Z^{ M \alpha}}_{NP} \epsilon_{\beta \gamma} \xi^\rho A^{N \beta}_{[\mu} {\cal G}^{P \gamma}_{\nu] \rho} + \left( \xi^M_\beta + \theta^M_\beta \right) \eta_{NP} \xi^\rho A^{N(\alpha|}_{[\mu} {\cal G}^{P |\beta)}_{\nu] \rho} \, ,   \nonumber
\end{align}
where the vector gauge transformations \eqref{dzBMN} and \eqref{dzBab} of the two-form gauge fields  have been modified as
\begin{equation}
    \label{dzBMNmod}
    \delta_{\zeta} B^{MN}_{\mu \nu} = - 2 \epsilon_{\alpha \beta} \left( \zeta^{[M|\alpha} {\cal G}^{|N] \beta}_{\mu \nu} - A^{[M|\alpha}_{[\mu} \delta_{\zeta} A^{|N]\beta}_{\nu]} \right)
\end{equation}
and 
\begin{equation}
    \label{dzBabmod}
    \delta_{\zeta} B^{\alpha \beta}_{\mu \nu} = 2 \eta_{MN} \left( \zeta^{M(\alpha|} {\cal G}^{N|\beta)}_{\mu \nu} - A^{M(\alpha|}_{[\mu} \delta_{\zeta} A^{N |\beta)}_{\nu]} \right)
\end{equation}
respectively, and 
\begin{align}
      \label{JMa}
     J^{M \alpha \mu} \equiv & \, - 2 Z^{M \alpha}{}_{NP} \Big{[} {L^N}_{\underline{a}} {L^P}_{ij} {\hat P}^{\underline{a} i j \mu} + {L^N}_{ik} L^{P jk} \big{(} {\bar\chi}_j \gamma^\mu \chi^i + 2  {\bar\lambda}^{\underline{a}}_j \gamma^\mu \lambda^i_{\underline{a}} \nonumber \\
      & + 2i \epsilon^{\mu \nu \rho \sigma}   {\bar\psi}^i_\nu
      \gamma_\rho  \psi_{j \sigma}  \big{)}   + 2 L^{N \underline{a}} L^{P \underline{b}} {\bar\lambda}_{\underline{a}i} \gamma^\mu \lambda^i_{\underline{b}} + 2 {L^N}_{\underline{a}} {L^P}_{ij} \left({\bar\lambda}^{\underline{a}i} \psi^{j \mu} - {\bar\lambda}^{\underline{a}i} \gamma^{\mu \nu} \psi^j_\nu \right) \nonumber \\ 
      & + 2 {L^N}_{\underline{a}} L^{Pij} \left( {\bar\lambda}^{\underline{a}}_i \psi^{\mu}_j - {\bar\lambda}^{\underline{a}}_i \gamma^{\mu \nu} \psi_{j \nu}  \right) \Big{]} + \left( \xi^M_{\beta} + \theta^M_{\beta} \right) \bigg{[} \frac{i}{2} {\cal V}^\alpha {\cal V}^\beta ({\hat P}^\mu)^*  \\ & - \frac{i}{2} ( {\cal V}^\alpha)^* ({\cal V}^\beta)^* {\hat P}^\mu  + M^{\alpha \beta} \left( \frac{3i}{4} {\bar\chi}_i \gamma^\mu \chi^i + \frac{i}{2} {\bar\lambda}^{\underline{a}}_i \gamma^\mu {\lambda}_{\underline{a}}^i      + \frac{1}{2} \epsilon^{\mu \nu \rho \sigma} {\bar\psi}^i_\nu
      \gamma_\rho  \psi_{i \sigma} \right) \nonumber \\ & - \frac{i}{2} {\cal V}^{\alpha} {\cal V}^{\beta} \left( {\bar\chi}_i \psi^{i \mu} - {\bar\chi}_i \gamma^{\mu \nu} \psi^i_\nu \right) + \frac{i}{2} ( {\cal V}^\alpha)^* ({\cal V}^\beta)^* \left( {\bar\chi}^i \psi_i^\mu - {\bar\chi}^i \gamma^{\mu \nu} \psi_{i \nu} \right) \bigg{]} \nonumber \, .
\end{align}
Using the Bianchi identity \eqref{2Bianchi}, the fact that ${\cal G}^{\Lambda}_{\mu \nu} = H^{\Lambda}_{\mu \nu}$ and the quadratic constraints \eqref{q2}-\eqref{q11} on the embedding tensor, one can show that if the equations of motion 
\begin{equation}
    \label{veceomfull}
    \frac{1}{2} \epsilon^{\mu \nu \rho \sigma} {\hat D}_{\nu} {\cal G}^{M \alpha}_{\rho \sigma} = g J^{M \alpha \mu} , 
\end{equation}
where ${\hat D}_{\mu} {\cal G}^{M \alpha}_{\nu \rho} \equiv \partial_\mu {\cal G}^{M \alpha}_{\nu \rho} + g X_{N \beta P \gamma}{}^{M \alpha} A^{N \beta}_\mu {\cal G}^{P \gamma}_{\nu \rho}$, are satisfied, then
\begin{align}
    \label{2formclos}
    [\delta_Q(\epsilon_1), \delta_Q (\epsilon_2)] \left( \Pi^{\Lambda}{}_{\cal M} B^{\cal M}_{\mu \nu}\right) = & \, \delta_{\text{cgct}}(\xi^{\rho}) \left( \Pi^{\Lambda}{}_{\cal M} B^{\cal M}_{\mu \nu} \right) + \delta_Q (\epsilon_3) \left( \Pi^{\Lambda}{}_{\cal M} B^{\cal M}_{\mu \nu} \right) 
 \nonumber \\ & + \delta_{\text{gauge}} (\zeta^{N \beta}) \left( \Pi^{\Lambda}{}_{\cal M} B^{\cal M}_{\mu \nu} \right) \nonumber \\ & + \delta_{\text{tensor}}(\Xi^{NP}_\rho, \Xi^{\beta \gamma}_\rho) \left( \Pi^{\Lambda}{}_{\cal M} B^{\cal M}_{\mu \nu} \right) \\
 & - \Pi^{\Lambda}{}_{\cal M} X_{({\cal N} {\cal P})}{}^{\cal M} \xi^{\rho} A^{\cal N}_{\rho} \left( {\cal G}^{\cal P}_{\mu \nu} - H^{\cal P}_{\mu \nu}\right) \nonumber \\
 & + 3 \Pi^{\Lambda}{}_{\cal M} X_{[{\cal N} {\cal P}]}{}^{\cal M} \xi^{\rho} A^{\cal N}_{[\mu} \left( {\cal G}^{\cal P}_{\nu \rho]} - H^{\cal P}_{\nu \rho]} \right)  . \nonumber
\end{align}
Therefore, the ${\cal N}=4$ supersymmetry algebra \eqref{QQcomg} closes on the antisymmetric tensor fields $\Pi^{\Lambda}{}_{\cal M} B^{\cal M}_{\mu \nu}$ up to the last two terms on the right-hand side of \eqref{2formclos}, which, 
in the absence of a gauging of the scaling symmetry, correspond to a gauge invariance of the Lagrangian of the standard gauged $D=4$, ${\cal N}=4$ supergravity that acts only on the two-form gauge fields (see for example \cite{deWit:2007kvg,deVroome:2007unr}). Moreover, equations \eqref{veceomfull} are identified with the equations of motion for the vector gauge fields $A^{M \alpha}_{\mu}$ and their bosonic sector will be reproduced in the next section as well.  

Finally, the supersymmetry algebra \eqref{QQcomg} closes on the fermionic fields, provided their equations of motion hold. These equations are given in the next section.

\section{Equations of Motion}
\label{sec:eom}

Having specified the local supersymmetry transformation rules, we are now ready to proceed to the derivation of the equations of motion of the most general four-dimensional ${\cal N}=4$ matter-coupled supergravity with local scaling symmetry. Since this theory does not admit an action, it must be constructed directly on the level of the equations of motion. By explicitly working out these field equations, we will also verify that the quadratic constraints \eqref{q2}-\eqref{q11} ensure compatibility of the theory with ${\cal N}=4$ supersymmetry.    

\subsection{Fermionic Field Equations}
The equations of motion for the fermions can be straightforwardly determined with the use of the rheonomic approach, since they are identified with the restrictions to spacetime of the constraints \eqref{choeomg}, \eqref{lameomg} and \eqref{psieomg}, imposed on the inner components of the fermionic supercurvatures by the Bianchi identities. In particular, the equations of motion for the dilatini are the spacetime projection of the superspace equations \eqref{choeomg} and read 
\begin{align}
\label{dileom}
   ({\cal E}_{\chi})_i \equiv & - \gamma^\mu {\hat D}_\mu \chi_i + \gamma^\mu \gamma^\nu \psi_{i \mu}  {\hat P}_\nu^*  - \frac{i}{4} {\cal V}_\alpha^* L_{M ij} { {\cal G}}^{M \alpha}_{\nu \rho} \gamma^\mu \gamma^{\nu \rho} \psi^j_\mu + \frac{i}{4} {\cal V}_\alpha^* L_{M \underline{a}} {\cal G}^{M \alpha}_{\mu \nu} \gamma^{\mu \nu} \lambda^{\underline{a}}_i \nonumber \\
    & + \frac{2}{3} g {\bar A}_{2 ij} \gamma^\mu \psi^j_\mu - 2 g {\bar A}_2{}^{\underline{a}j}{}_i \lambda_{\underline{a}j} + 2 g {\bar A}_2{}^{\underline{a}j}{}_j \lambda_{\underline{a}i} - g {\bar B}_{ij} \gamma^\mu \psi^j_\mu + \frac{5}{2} g {\bar B}^{\underline{a}} \lambda_{\underline{a} i} = 0 \, , 
    \end{align}
  where 
  \begin{equation}
  \label{Dchi}
      {\hat D}_{\mu} \chi_i \equiv  \, \partial_\mu \chi_i + \frac{1}{4} {\omega_{\mu}}^{ab}(e,A,\psi) \gamma_{ab} \chi_i + \frac{3i}{2} \hat{\cal A}_{\mu} \chi_i -  {\hat\omega}_i{}^j{}_\mu \chi_j 
 + \frac{g}{2} \theta_{\alpha M} A^{M \alpha}_\mu \chi_i  \, .
  \end{equation}
 Furthermore, by restricting \eqref{lameomg} to four-dimensional spacetime, we can specify the equations of motion for the gaugini, which are given by     
\begin{align}
    \label{gaueom}
  ({\cal E}_{\lambda})_{\underline{a} i} \equiv & - \gamma^\mu {\hat D}_\mu \lambda_{\underline{a} i} - \gamma^\mu \gamma^\nu \psi^j_\mu  {\hat P}_{\underline{a}ij \nu} +  \frac{i}{8} {\cal V}_\alpha^* L_{M \underline{a}} {\cal G}^{M \alpha}_{\nu \rho} \gamma^\mu \gamma^{\nu \rho} \psi_{i \mu} + \frac{i}{4} {\cal V}_\alpha^* L_{Mij} {\cal G}^{M \alpha}_{\mu \nu} \gamma^{\mu \nu} \lambda^j_{\underline{a}} \nonumber \\
    & + \frac{i}{8} {\cal V}_\alpha L_{M \underline{a}} {\cal G}^{M \alpha}_{\mu \nu} \gamma^{\mu \nu} \chi_i + g {\bar A}_{2 \underline{a}}{}^j{}_i \gamma^\mu \psi_{j \mu} - g A_{2 \underline{a} i}{}^j \chi_j + g A_{2 \underline{a} j}{}^j \chi_i +  2 g {\bar A}_{\underline{a} \underline{b} ij} \lambda^{\underline{b} j} \\
    &+ \frac{2}{3} g {\bar A}_{2(ij)} \lambda^j_{\underline{a}} - \frac{g}{4} {\bar B}_{\underline{a}} \gamma^\mu \psi_{i \mu} - 2 g {\bar B}_{ij} \lambda^j_{\underline{a}} - \frac{3}{4} g B_{\underline{a}} \chi_i  = 0   \nonumber \, , 
\end{align}  
where
\begin{equation}
    {\hat D}_{\mu} \lambda_{\underline{a}i} \equiv  \, \partial_\mu \lambda_{\underline{a}i} + \frac{1}{4} {\omega_{\mu}}^{ab}(e,A,\psi) \gamma_{ab} \lambda_{\underline{a}i} + \frac{i}{2} \hat{\cal A}_{\mu} \lambda_{\underline{a}i} -  {\hat\omega}_i{}^j{}_\mu \lambda_{\underline{a}j} + {\hat\omega}_{\underline{a}}{}^{\underline{b}}{}_\mu \lambda_{\underline{b}i}  + \frac{g}{2} \theta_{\alpha M} A^{M \alpha}_\mu \lambda_{\underline{a}i}
\end{equation}
and 
\begin{equation}
     \label{Aabij}
    A_{\underline{a} \underline{b}}{}^{ij} \equiv f_{\alpha MNP} {\cal V}^{\alpha} {L^M}{}_{\underline{a}} {L^N}{}_{\underline{b}} L^{Pij} .
\end{equation}
Finally, the projection of the ${\cal N}=4$ superspace equations \eqref{psieomg} on spacetime gives the following equations of motion for the gravitini:
\begin{align}
    \label{graviteom}
    ({\cal E}_{\psi})_{i \nu} \equiv & - \gamma^\mu {\hat\rho}_{i \mu \nu}  + {\hat P}_{\nu} \chi_i  + 2 {\hat P}_{\underline{a} ij \nu} \lambda^{\underline{a} j} - \frac{i}{8} {\cal V}_\alpha L_{Mij} {\cal G}^{M \alpha}_{\rho \sigma} \gamma^\mu \gamma^{\rho \sigma} \gamma_\nu \psi^j_\mu \nonumber \\ & - \frac{i}{8} {\cal V}_\alpha L_{M \underline{a}} {\cal G}^{M \alpha}_{\mu \rho} \gamma^{\mu \rho} \gamma_\nu \lambda^{\underline{a}}_i  + \frac{i}{8} {\cal V}_\alpha^* L_{Mij} {\cal G}^{M \alpha}_{\mu \rho} \gamma^{\mu \rho} \gamma_\nu \chi^j + g {\bar A}_{1ij} \psi^j_\nu  \nonumber \\
    & - \frac{g}{3}  {\bar A}_{1ij} \gamma_{\mu \nu} \psi^{j \mu} + \frac{g}{3} {\bar A}_{2ji} \gamma_\nu \chi^j + g A_{2 \underline{a}i}{}^j \gamma_\nu \lambda^{\underline{a}}_j - \frac{3}{2} g \epsilon_{ijkl} B^{kl} \psi^j_{\nu} \\ &  + \frac{g}{2}  \epsilon_{ijkl} B^{kl} \gamma_{\mu \nu} \psi^{j \mu} - \frac{3}{2} g {\bar B}_{ij} \gamma_{\nu} \chi^j + \frac{7}{4} g B^{\underline{a}} \gamma_\nu \lambda_{\underline{a} i} = 0 \nonumber \, , 
\end{align}
where 
\begin{equation}
\label{rhomn}
    {\hat\rho}_{i \mu \nu} \equiv \, 2 \partial_{[\mu|} \psi_{i|\nu]} + \frac{1}{2} {\omega_{[\mu|}}^{ab}(e,A,\psi) \gamma_{ab} \psi_{i|\nu]} - i \hat{\cal A}_{[\mu|} \psi_{i|\nu]} - 2 {\hat\omega}_i{}^j{}_{[\mu|} \psi_{j|\nu]} - g \theta_{\alpha M} A^{M \alpha}_{[\mu|} \psi_{i|\nu]}.
\end{equation}
The fermionic field equations also contain terms cubic in the fermions, which are ignored in this work. For $\theta_{\alpha M}=0$, the equations of motion \eqref{dileom}, \eqref{gaueom} and \eqref{graviteom} arise from the variation of the action of the standard gauged $D=4$, ${\cal N}=4$ supergravity \cite{DallAgata:2023ahj} with respect to the fermionic fields $\chi^i$, $\lambda^{\underline{a}i}$ and $\psi^i_{\mu}$ respectively. However, in the presence of a gauging of the scaling symmetry, there is no action that reproduces these equations via the variational principle, since the fermion mass matrices that can be read off from \eqref{dileom}, \eqref{gaueom} and \eqref{graviteom} are not symmetric (see \eqref{gravmass}-\eqref{Maibj}). 

\subsection{Bosonic Field Equations}

The equations of motion for the bosonic fields follow from the requirement that the fermionic field equations be invariant under local supersymmetry transformations. We will restrict ourselves to the derivation of the bosonic sectors of the bosonic field equations, so we will suppress any term quadratic or quartic in the fermions in the supersymmetry variations of the fermionic field equations. For this purpose, we only need to consider the local supersymmetry transformations of the fermions in \eqref{dileom}, \eqref{gaueom} and 
\eqref{graviteom}. 

Particularly useful for the computation of the supersymmetry variations of the equations of motion for the fermions are the following gradient flow relations, which give the covariant derivatives of the fermion shift tensors:
\begin{align}
    \label{Drhoij}
    {\hat D} A_1^{ij}  = & \,A_2^{(ij)} {\hat P}^* +3 
    {\bar A}_2{}^{\underline{a}(i}{}_k {\hat P}_{\underline{a}}{}^{j)k}, \\
    \label{DVij}
    {\hat D} A_2^{ij}  = & - 3 A_2{}^{\underline{a}}{}_k{}^{(i} {\hat P}_{\underline{a}}{}^{j)k} - \frac{3}{2} A_2{}^{\underline{a}}{}_k{}^{k}  {\hat P}_{\underline{a}}{}^{ij} + \frac{1}{2} \epsilon^{ijkl} {\bar A}_{2kl} {\hat P} + A_1^{ij}{\hat P}, \\
    \label{DLaij}
    {\hat D} A_2{}^{\underline{a}}{}_j{}^i  = & - {\bar A}_2{}^{\underline{a}i}{}_j \hat{P} + \frac{1}{2} \delta^i_j {\bar A}_2{}^{\underline{a}k}{}_k \hat{P}  + 2 A^{\underline{a} \underline{b} ik} {\hat P}_{\underline{b} jk} - \frac{1}{2}  \delta^i_j A^{\underline{a} \underline{b} kl} {\hat P}_{\underline{b} kl} \nonumber \\[2mm]
    & - \frac{1}{6} \delta^i_j A_2^{kl} {\hat P}^{\underline{a}}{}_{kl} - \frac{2}{3} {\bar A}_{1jk} {\hat P}^{\underline{a} ik} + \frac{2}{3} A_2^{(ik)} {\hat P}^{\underline{a}}{}_{jk}\,, \\
	\label{DMabij}
	{\hat D} A_{\underline{a} \underline{b}}{}^{ij} =& \, \frac{1}{2} \epsilon^{ijkl} {\bar A}_{\underline{a} \underline{b} kl} {\hat P} - 4 A_{2[\underline{a}|k}{}^{[i} {\hat P}_{|\underline{b}]}{}^{j]k}  - A_{2[\underline{a}|k}{}^k {\hat P}_{|\underline{b}]}{}^{ij} + A_{\underline{a} \underline{b} \underline{c}} {\hat P}^{\underline{c} i j} , \\
 \label{DBij}
 {\hat D} B^{ij} = & \, \frac{1}{2} \epsilon^{ijkl} {\bar B}_{kl} {\hat P} + B_{\underline{a}} {\hat P}^{\underline{a} ij} , \\ 
 \label{DBa}
{\hat D} B_{\underline{a}} = & \,  {\bar B}_{\underline{a}} {\hat P} + B^{ij} {\hat P}_{\underline{a} ij} \, ,
\end{align}
where
\begin{equation}
	    \label{Sabc}
	    A_{\underline{a} \underline{b} \underline{c}} \equiv f_{\alpha MNP} {\cal V}^{\alpha} L^M{}_{\underline{a}} L^N{}_{\underline{b}} L^P{}_{\underline{c}} .
\end{equation}
The first four of the above gradient flow equations were derived in \cite{DallAgata:2023ahj}. 

Using the Bianchi identity \eqref{DhatPhat}, the twisted self-duality condition \eqref{twist}, the duality equations \eqref{H=G2},  equations \eqref{DVij} and \eqref{DBij} and the quadratic constraints \eqref{T8}, \eqref{T15}, \eqref{T17}, \eqref{T24} and \eqref{T29} on the $A$ and $B$ tensors, we can write the supersymmetry variation of the equations of motion for the dilatini as 
\begin{align}
    \label{dEchi}
    \delta_{\epsilon} ({\cal E}_{\chi})_i   = & \, {\cal E} \epsilon_i \, + \bigg{[} \frac{1}{2} {\cal V}_{\alpha}^* L_{M i j} \epsilon^{\mu \nu \rho \sigma} ( {\hat D}_{\nu} {\cal G}^{M \alpha}_{\rho \sigma}  ) - g {\bar A}_{2}{}^{\underline{a}k}{}_i {\hat P}_{\underline{a} jk}{}^{\mu} + g {\bar A}_{2}{}^{\underline{a}k}{}_j {\hat P}_{\underline{a} ik}{}^{\mu} \nonumber \\
    & - g {\bar A}_{2}{}^{\underline{a}k}{}_k {\hat P}_{\underline{a} ij}{}^{\mu} - \frac{3}{2} g {\bar B}^{\underline{a}} {\hat P}_{\underline{a} ij}{}^{\mu} - \frac{g}{3} \epsilon_{ijkl} A_2^{kl} ({\hat P}^{\mu})^* - \frac{g}{2} \epsilon_{ijkl} B^{kl}  ({\hat P}^{\mu})^* \bigg{]} \gamma_{\mu} \epsilon^j ,
\end{align}
where 
\begin{align}
   \label{Etau}
   {\cal E}\equiv & - e^{-1} {\hat D}_{\mu} \left( e ({\hat P}^{\mu})^* \right) + \frac{1}{8} {\cal V}_{\alpha}^* {\cal V}_{\beta}^*  M_{MN} {\cal G}^{M \alpha}_{\mu \nu} {\cal G}^{N \beta \mu \nu} + g^2 \bigg{(} - \frac{2}{9} A_1^{ij} {\bar A}_{2 ij} \nonumber \\ 
   & + \frac{1}{9} \epsilon^{ijkl} {\bar A}_{2 ij} {\bar A}_{2 kl} - \frac{1}{2} {\bar A}_{2}{}^{\underline{a}i}{}_j {\bar A}_{2 \underline{a}}{}^j{}_i - \frac{3}{8} \epsilon^{ijkl} {\bar A}_{2 ij} {\bar B}_{kl} + \frac{1}{8} {\bar A}_{2 \underline{a}}{}^i{}_i {\bar B}^{\underline{a}} \\ &  + \frac{3}{16} \epsilon^{ijkl} {\bar B}_{ij} {\bar B}_{kl} \bigg{)} , \nonumber 
\end{align}
where $ {\hat D}_{\mu} \left( e ({\hat P}^{\mu})^* \right) \equiv \partial_{\mu} \left( e ({\hat P}^{\mu})^* \right) + 2 i e {\hat{\cal A}}_{\mu} ({\hat P}^{\mu})^* - 2 g e \theta_{\alpha M} A^{M \alpha}_\mu ({\hat P}^{\mu})^* $. Therefore, the equation of motion for the complex scalar of the ${\cal N}=4$ supergravity multiplet reads
\begin{equation}
    \label{taueom}
    {\cal E} = 0 \, .
\end{equation}

In order to specify the equations of motion for the scalar fields of the vector multiplets in four-dimensional ${\cal N}=4$ matter-coupled supergravity with local scaling symmetry, we need to compute the supersymmetry variation of the corresponding equations for the gaugini. Using the Bianchi identity \eqref{scalarBianchi3}, condition \eqref{twist}, the duality equations \eqref{H=G2}, the gradient flow relations \eqref{DLaij}
 and \eqref{DBa} as well as the $T$-identities of appendix \ref{sec:Tid} in the $\mathbf{(10,n)_0}$ and $\mathbf{(6,n)_0}$ representations of SU(4) $\times$ SO($n$) $\times$ SO(2),  we find 
\begin{align}
    \label{deElam}
    \delta_{\epsilon} ({\cal E}_{\lambda})_{\underline{a} i } = & \, {\cal E}_{\underline{a} ij} \epsilon^j + \bigg{[} \frac{1}{4} {\cal V}_{\alpha}^* L_{M \underline{a}} \epsilon^{\mu \nu \rho \sigma} \left( {\hat D}_{\nu} {\cal G}^{M \alpha}_{\rho \sigma}   \right)+ \frac{g}{2}  {\bar A}_{\underline{a} \underline{b} jk} {\hat P}^{\underline{b} jk \mu} + \frac{g}{6} {\bar A}_{2 jk} {\hat P}_{\underline{a}}{}^{jk \mu} \nonumber \\ &  - \frac{3}{4} g {\bar B}_{jk} {\hat P}_{\underline{a}}{}^{jk \mu} + \frac{g}{2} A_{2 \underline{a}j}{}^j ({\hat P}^{\mu})^* - \frac{g}{2} B_{\underline{a}} ({\hat P}^{\mu})^* \bigg{]} \gamma_{\mu} \epsilon_i \, ,
    \end{align} 
where 
\begin{align}
    \label{Eaij}
    {\cal E}_{\underline{a} ij} \equiv  e^{-1} {\hat D}_{\mu} \left(e {\hat P}_{\underline{a} ij}{}^{\mu} \right)  - \frac{1}{2} M_{\alpha \beta} L_{M \underline{a}} L_{Nij} {\cal G}^{M \alpha}_{\mu \nu} {\cal G}^{N \beta \mu \nu} + g^2 \left( {\cal C}_{\underline{a}ij}  + \frac{1}{2} \epsilon_{ijkl} \bar{\cal C}_{\underline{a}}{}^{kl}  \right) , 
    \end{align}
which is manifestly self-dual, i.e. ${\cal E}_{\underline{a}}{}^{ij} = ({\cal E}_{\underline{a} ij})^* = \frac{1}{2} \epsilon^{ijkl} {\cal E}_{\underline{a} kl}  $, with ${\hat D}_{\mu} \left(e {\hat P}_{\underline{a} ij}{}^{\mu} \right) \equiv \partial_{\mu} \left(e {\hat P}_{\underline{a} ij}{}^{\mu} \right) + e {\hat\omega}_{\underline{a}}{}^{\underline{b}}{}_{\mu} {\hat P}_{\underline{b} ij}{}^{\mu} - e {\hat\omega}_i{}^k{}_{\mu} {\hat P}_{\underline{a} kj}{}^{\mu} - e {\hat\omega}_j{}^k{}_{\mu} {\hat P}_{\underline{a} ik}{}^{\mu} - 2 g e \theta_{\alpha M} A^{M \alpha}_{\mu} {\hat P}_{\underline{a} ij}{}^{\mu}$ and 
\begin{align}
    \label{Caij}
    {\cal C}_{\underline{a}ij} = &  - \frac{2}{3} {\bar A}_{2 \underline{a}}{}^k{}_{[i} {\bar A}_{1 j]k} - \frac{1}{6} A_{2 \underline{a}[i}{}^k {\bar A}_{2 j]k} - \frac{1}{2}  A_{2 \underline{a}[i|}{}^k {\bar A}_{2 k|j]}  + {\bar A}_{\underline{a} \underline{b} [i|k} A_2{}^{\underline{b}}{}_{|j]}{}^k + \frac{1}{3} A_{2 \underline{a}k}{}^k {\bar A}_{2 [ij]} \nonumber \\ 
    & + \frac{5}{2} A_{2 \underline{a}[i}{}^k {\bar B}_{j]k} + \frac{1}{2} A_{2 \underline{a}k}{}^k {\bar B}_{ij} - \frac{1}{4} {\bar A}_{\underline{a} \underline{b} ij} B^{\underline{b}} - \frac{1}{4} {\bar A}_{2 [ij]} B_{\underline{a}} + \frac{1}{8} {\bar B}_{ij} B_{\underline{a}} \, . 
    \end{align}
Therefore, invariance of the equations of motion for the gaugini under local supersymmetry transformations imposes the following equations of motion for the $6n$ scalars of the $n$ vector multiplets:  
\begin{equation}
    \label{eomscalvec}
    {\cal E}_{\underline{a} ij} = 0 \, .
\end{equation}

Moreover, the Einstein equations of four-dimensional ${\cal N}=4$ supergravity with local scaling symmetry can be derived by requiring supersymmetry invariance of the equations of motion for the gravitini, \eqref{graviteom}. We note that the supersymmetry variation of the field strengths of the gravitini, ${\hat\rho}_{i \mu \nu}$, gives rise to the commutator 
\begin{equation}
    \label{comm}
    -\gamma^{\mu} \left[ {\hat D}_{\mu}, {\hat D}_{\nu} \right] \epsilon_i = - \gamma^{\mu} \left( \frac{1}{4} {\hat R}_{\mu \nu}{}^{ab} \gamma_{a b} \epsilon_i - \frac{i}{2} {\hat F}_{\mu \nu} \epsilon_i - {\hat R}_i{}^j{}_{\mu \nu} \epsilon_j  - \frac{g}{2} \theta_{\alpha M} H^{M \alpha}_{\mu \nu} \epsilon_i \right) , 
\end{equation}
where the gauged SO(2) and SU(4) curvatures ${\hat F}_{\mu \nu}$ and ${\hat R}_i{}^j{}_{\mu \nu}$ are given by equations \eqref{gSO(2)curv} and \eqref{SU4curv} respectively, and 
\begin{equation}
    \label{Rhatmnab}
    {\hat R}_{\mu \nu}{}^{ab} \equiv 2 \partial_{[\mu} \omega_{\nu]}{}^{ab}(e,A,\psi) + 2 \omega_{[\mu}{}^{a c} (e,A,\psi) \omega_{\nu] c}{}^b  (e,A,\psi) 
\end{equation}
is the Riemann curvature tensor associated with the spin connection \eqref{spinconn}. The relevant Ricci tensor and scalar are defined as usual by ${\hat R}_{\mu \nu} = e_{a \nu} e^{\rho}_b {\hat R}_{\mu 
\rho}{}^{ab}$ and ${\hat R}=g^{\mu \nu} {\hat R}_{\mu \nu}$ respectively. If ${ R}_{\mu \nu}{}^{ab} (e) \equiv 2 \partial_{[\mu} \omega_{\nu]}{}^{ab}(e) + 2 \omega_{[\mu}{}^{a c} (e) \omega_{\nu] c}{}^b  (e) $ is the Riemann tensor associated with the torsion-free spin connection, $\omega_{\mu}{}^{ab}(e)$, then using \eqref{spinconn} we find that up to two- and four-gravitini terms,
\begin{align}
   \label{Rmnab}
    {\hat R}_{\mu \nu}{}^{ab} = & \, { R}_{\mu \nu}{}^{ab} (e)  + 4 g {\theta}_{\alpha M} e^{[a|}_{[\mu} {\cal D}_{\nu]} A^{M \alpha |b]} \nonumber \\  & + 4 g^2  {\theta}_{\alpha M} {\theta}_{\beta N} \left( e^{[a|}_{[\mu} A^{M \alpha}_{\nu]} A^{N \beta |b]}  - \frac{1}{2} e^a_{[\mu} e^b_{\nu]} A^{M \alpha}_{\rho} A^{N \beta \rho}  \right) , \\
    \label{Rmn}
    {\hat R}_{\mu \nu} = &  \, R_{\mu \nu} (e) + g \theta_{\alpha M} \left(  2 e_{a \nu} {\cal D}_{\mu} A^{M \alpha a} + g_{\mu \nu} e^{\rho}_a {\cal D}_{\rho} A^{M \alpha a} \right) \nonumber \\ 
    & + 2 g^2 {\theta}_{\alpha M} {\theta}_{\beta N} \left( A^{M \alpha}_{\mu} A^{N \beta}_{\nu} - g_{\mu \nu} A^{M \alpha}_{\rho} A^{N \beta \rho} \right),  \\
    \label{Riccisc}
    {\hat R} = & \, R(e)  + 6 g {\theta}_{\alpha M} e^{\mu}_a {\cal D}_{\mu} A^{M \alpha a} - 6 g^2 {\theta}_{\alpha M} {\theta}_{\beta N}  A^{M \alpha}_{\mu} A^{N \beta \mu} ,
\end{align}
where ${\cal D}_{\mu} A^{M \alpha a} \equiv \partial_{\mu} A^{M \alpha a} + \omega_{\mu}{}^a{}_{b} (e) A^{M \alpha b}$. The Riemann tensor $\eqref{Rhatmnab}$ also satisfies the Bianchi identity 
\begin{equation}
    \label{RBianchi}
    {\hat R}_{\mu [\nu \rho \sigma]} = - g \theta_{\alpha M} g_{\mu [\nu} H^{M \alpha}_{\rho \sigma]}  
\end{equation}
up to terms involving the gravitini. From $\eqref{Rmn}$ and $\eqref{RBianchi}$ it follows that the commutator \eqref{comm} equals
\begin{align}
     -\gamma^{\mu} \left[ {\hat D}_{\mu}, {\hat D}_{\nu} \right] \epsilon_i = &  - \gamma^{\mu} \bigg{(} \frac{1}{2} {\hat R}_{(\mu \nu)} \epsilon_i - \frac{i}{2} {\hat F}_{\mu \nu} \epsilon_i - {\hat R}_i{}^j{}_{\mu \nu} \epsilon_j - g {\theta}_{\alpha M} H^{M \alpha}_{\mu \nu} \epsilon_i \nonumber \\ &  + \frac{i}{4} g  {\theta}_{\alpha M} \epsilon_{\mu \nu \rho \sigma} H^{M \alpha \rho \sigma} \epsilon_i \bigg{)}  . 
\end{align}
Using the last equation, identities \eqref{gSO(2)curv} and \eqref{SU4curv}, the twisted self-duality condition \eqref{twist}, the duality equations \eqref{H=G1} and \eqref{H=G2}, the gradient flow equations \eqref{Drhoij} and \eqref{DBij} as well as the quadratic constaints on the $A$ and $B$ tensors of appendix \ref{sec:Tid} in the $\mathbf{(15,1)_0}$ and $\mathbf{(1,1)_0}$ representations of SU(4) $\times$ SO($n$) $\times$ SO(2), we can write the bosonic sector of the supersymmetry variation of the equations of motion for the gravitini as 
\begin{align}
    \label{dEpsi}
    \delta_{\epsilon} \left( {\cal E}_{\psi}\right)_{i \nu} = & - \bigg{[} \frac{1}{4} {\cal V}_{\alpha} L_{Mij} \epsilon^{\mu \rho \sigma \lambda} \left( {\hat D}_{\rho} {\cal G}^{M \alpha}_{\sigma \lambda}\right)  + g A_{2 \underline{a}[i}{}^k {\hat P}^{\underline{a}}{}_{j]k}{}^{\mu} - \frac{3}{4} g B^{\underline{a}} {\hat P}_{\underline{a} ij}{}^{\mu} \nonumber \\ & - \frac{g}{3} {\bar A}_{2 [ij]} {\hat P}^{\mu} - \frac{g}{2} {\bar B}_{ij} {\hat P}^{\mu}  \bigg{]} \gamma_{\mu} \gamma_{\nu} \epsilon^j  - \frac{1}{2} \left( {\cal E}_{\text{Einstein}}\right)_{\mu \nu} \gamma^{\mu} \epsilon_i \, ,
\end{align}
where 
\begin{align}
    \label{Eins}
    \left({\cal E}_{\text{Einstein}}\right)_{\mu \nu} \equiv & \,  {\hat R}_{(\mu \nu)} - 2 {\hat P}_{(\mu} {\hat P}^{*}_{\nu)} - {\hat P}_{\underline{a}ij \mu} {\hat P}^{\underline{a} ij}{}_{\nu} - \frac{1}{2} M_{MN} M_{\alpha \beta} {\cal G}^{M \alpha}_{\mu \rho} {\cal G}^{N \beta}{}_{\nu}{}^{\rho} \nonumber \\
    & + g^2 \bigg{(} \frac{1}{3} A_1^{ij} {\bar A}_{1ij} - \frac{1}{9} A_2^{ij} {\bar A}_{2 ij} - \frac{1}{2} A_{2 \underline{a}i}{}^j {\bar A}_2{}^{\underline{a}i}{}_j  \\
    & + \frac{1}{6} A_2^{ij} {\bar B}_{ij} - \frac{3}{2} B^{ij} {\bar B}_{ij} + \frac{1}{8} B^{\underline{a}} {\bar B}_{\underline{a}} \bigg{)} g_{\mu \nu} \, . \nonumber 
\end{align}
In particular, $A_2^{ij} {\bar B}_{ij}$ is real as a result of the quadratic constraint \eqref{q2} (see appendix \ref{sec:Tid}). Therefore, the bosonic sector of the Einstein equations reads
\begin{equation}
    \left({\cal E}_{\text{Einstein}}\right)_{\mu \nu} = 0 \, . 
\end{equation}
From \eqref{Eins} one can read off the effective cosmological constant, which is given by 
\begin{equation}
     \label{cosm}
    \Lambda =  g^2 \left( - \frac{1}{3} A_1^{ij} {\bar A}_{1ij} + \frac{1}{9} A_2^{ij} {\bar A}_{2 ij} + \frac{1}{2} A_{2 \underline{a}i}{}^j {\bar A}_2{}^{\underline{a}i}{}_j  - \frac{1}{6} A_2^{ij} {\bar B}_{ij} + \frac{3}{2} B^{ij} {\bar B}_{ij} - \frac{1}{8} B^{\underline{a}} {\bar B}_{\underline{a}}  \right)  .
\end{equation}
For $\theta_{\alpha M}=0$, \eqref{cosm} consistently reduces to the scalar potential of the standard gauged $D=4$, ${\cal N}=4$ matter-coupled supergravity \cite{Schon:2006kz}. On the other hand, the gauging of the trombone symmetry induces an additional contribution to the effective cosmological constant consisting of the last three terms on the right-hand side of \eqref{cosm}, which is not always positive, unlike its counterpart in the four-dimensional ${\cal N}=8$ supergravity \cite{LeDiffon:2011wt}.

From the supersymmetry variations of the fermionic field equations we can deduce the equations of motion for the vector gauge fields as well. For this purpose, we note that the terms of $\delta_{\epsilon} \left( {\cal E}_{\chi}\right)_i$, $\delta_{\epsilon} \left( {\cal E}_{\lambda}\right)_{\underline{a}i}$ and $\delta_{\epsilon}\left( {\cal E}_{\psi}\right)_{i \nu}$ that are proportional to $\gamma_{\mu} \epsilon^j$, $\gamma_{\mu} \epsilon_i$ and $\gamma_{\mu} \gamma_{\nu} \epsilon^j $ respectively can be written as 
\begin{align}
\label{chivector}
\delta_{\epsilon} \left( {\cal E}_{\chi}\right)_i & \supset {\cal V}_{\alpha}^* L_{Mij} \left( {\cal E}_{\text{vector}} \right)^{M \alpha \mu} \gamma_{\mu} \epsilon^j ,\\
\label{lambdavector}
 \delta_{\epsilon} ({\cal E}_{\lambda})_{\underline{a} i } & \supset \frac{1}{2} {\cal V}_{\alpha}^* L_{M \underline{a}}  \left( {\cal E}_{\text{vector}} \right)^{M \alpha \mu} \gamma_{\mu} \epsilon_i \, , \\
 \label{psivector}
 \delta_{\epsilon} \left( {\cal E}_{\psi}\right)_{i \nu} & \supset - \frac{1}{2} {\cal V}_{\alpha} L_{M ij}  \left( {\cal E}_{\text{vector}} \right)^{M \alpha \mu} \gamma_{\mu} \gamma_{\nu} \epsilon^j  , 
    \end{align}
where 
\begin{align}
    \label{Evector}
    \left( {\cal E}_{\text{vector}} \right)^{M \alpha \mu} \equiv & \, \frac{1}{2} \epsilon^{\mu \nu \rho \sigma} {\hat D}_{\nu} {\cal G}^{M \alpha}_{\rho \sigma} + 2 g Z^{M \alpha NP} L_{N \underline{a}} L_{P ij} {\hat P}^{\underline{a} ij \mu} \nonumber \\
    & - \frac{i}{2} g \left( \xi^M_{\beta} + \theta^M_{\beta} \right) {\cal V}^{\alpha} {\cal V}^{\beta} ({\hat P}^{\mu})^* + \frac{i}{2} g \left( \xi^M_{\beta} + \theta^M_{\beta} \right) ({\cal V}^{\alpha})^* ({\cal V}^{\beta})^* {\hat P}^{\mu} .
\end{align}
Thus, supersymmetry invariance of the equations of motion for the fermions requires
\begin{equation}
\label{VLijv}
    {\cal V}_{\alpha}^* L_{Mij} \left( {\cal E}_{\text{vector}} \right)^{M \alpha \mu} = 0 
\end{equation}
and 
\begin{equation}
    \label{VLav}
    {\cal V}_{\alpha}^* L_{M \underline{a}}  \left( {\cal E}_{\text{vector}} \right)^{M \alpha \mu} = 0 \, .
\end{equation}
Contracting \eqref{VLijv} with ${\cal V}_{\beta} L_N{}^{ij}$ and \eqref{VLav} with ${\cal V}_{\beta} L_N{}^{\underline{a}}$ and then subtracting the two resulting equations and using \eqref{LSO6SU4} we obtain
\begin{equation}
  \eta_{MN}  {\cal V}_{\alpha}^* {\cal V}_{\beta} \left( {\cal E}_{\text{vector}} \right)^{M \alpha \mu} = 0 \, . 
\end{equation}
Due to the constraint \eqref{VV*-V*V=e}, the imaginary part of the above equation implies that 
\begin{equation}
\label{veceom}
    \left( {\cal E}_{\text{vector}} \right)^{M \alpha \mu} = 0 \, ,
\end{equation}
which is but the bosonic sector of the equations of motion for the vector fields. This completes the derivation of the full set of field equations for the trombone gauged half-maximal supergravity in four spacetime dimensions. 
\section{Maximally Symmetric Solutions and Mass Matrices}
\label{sec:mass}

From the equations of motion for the scalar fields, \eqref{taueom} and \eqref{eomscalvec}, it follows that a solution to the field equations of the previous section with constant scalar and vanishing vector, two-form and fermionic fields satisfies the following two conditions:
\begin{align}
    \label{cond1}
    &- \frac{2}{9} A_1^{ij} {\bar A}_{2 ij}  + \frac{1}{9} \epsilon^{ijkl} {\bar A}_{2 ij} {\bar A}_{2 kl} - \frac{1}{2} {\bar A}_{2}{}^{\underline{a}i}{}_j {\bar A}_{2 \underline{a}}{}^j{}_i \nonumber \\ & - \frac{3}{8} \epsilon^{ijkl} {\bar A}_{2 ij} {\bar B}_{kl} + \frac{1}{8} {\bar A}_{2 \underline{a}}{}^i{}_i {\bar B}^{\underline{a}}   + \frac{3}{16} \epsilon^{ijkl} {\bar B}_{ij} {\bar B}_{kl}  = 0 
\end{align}
and
\begin{equation}
\label{cond2}
    {\cal C}_{\underline{a} ij} + \frac{1}{2} \epsilon_{ijkl} \bar{\cal C}_{\underline{a}}{}^{kl}  = 0 \, , 
\end{equation}
where the tensor ${\cal C}_{\underline{a}ij}$ is given by \eqref{Caij}. For the standard gaugings, for which $\theta_{\alpha M}=0$, these conditions reproduce the extremization conditions of the scalar potential \cite{DallAgata:2023ahj}. Any constant solution to \eqref{cond1} and \eqref{cond2} corresponds to a solution to the field equations with maximally symmetric four-dimensional spacetime and cosmological constant given by \eqref{cosm}. 

The mass spectrum of the theory around such a solution can be obtained by linearizing the field equations. The fluctuations of the coset representatives ${\cal V}_{\alpha}$ and $L_M{}^{\underline{M}}$ around such a solution can be written as \cite{DallAgata:2023ahj} 
\begin{equation}
    \label{cosetrepfluct}
    \delta {\cal V}_{\alpha} = \Sigma {\cal V}_{\alpha}^*, \qquad \delta L_M{}^{ij} = \Sigma_{\underline{a}}{}^{ij} L_M{}^{\underline{a}}, \qquad \delta  L_M{}^{\underline{a}} = \Sigma^{\underline{a}}{}_{ij} L_M{}^{ij} , 
\end{equation}
where $\Sigma$ denotes the complex SL(2,$\mathbb{R}$)/SO(2) scalar fluctuation and $\Sigma_{\underline{a}ij}$ are the self-dual SO(6,$n$)/(SO(6)$\times$SO($n$)) scalar fluctuations. Using the gradient flow relations \eqref{Drhoij}-\eqref{DBa}, we find that in terms of the real scalar fluctuations
\begin{equation}
    \label{realfl}
    \Sigma_1 = \sqrt{2} \text{Re} \Sigma, \qquad \Sigma_2 = \sqrt{2} \text{Im} \Sigma, \qquad \Sigma^{\underline{a} \underline{m}} = - \Gamma^{\underline{m}ij} \Sigma^{\underline{a}}{}_{ij} \, ,  
\end{equation}
the linearized form of the equation of motion for the complex scalar of the ${
\cal N
}=4$ supergravity multiplet, \eqref{taueom}, is given by the following two real equations: 
\begin{align}
    \label{S1}
    e^{-1} \partial_{\mu} \left( e \partial^{\mu} \Sigma_1\right) & = ({\cal M}_0^2)^{1,1} \Sigma_1   + ({\cal M}_0^2)^{1,\underline{a} \underline{m}} \delta_{\underline{a} \underline{b}} \delta_{\underline{m} \underline{n}} \Sigma^{\underline{b} \underline{n}} , \\
    \label{S2}
     e^{-1} \partial_{\mu} \left( e \partial^{\mu} \Sigma_2\right) & = ({\cal M}_0^2)^{2,2} \Sigma_2   + ({\cal M}_0^2)^{2,\underline{a} \underline{m}} \delta_{\underline{a} \underline{b}} \delta_{\underline{m} \underline{n}} \Sigma^{\underline{b} \underline{n}} \, ,
\end{align}
while the linearized equations of motion for the scalars of the vector multiplets read 
\begin{equation}
    \label{Sam}
     e^{-1} \partial_{\mu} \left( e \partial^{\mu} \Sigma^{\underline{a} \underline{m}}\right)  = ({\cal M}_0^2)^{\underline{a} \underline{m},1} \Sigma_1 +  ({\cal M}_0^2)^{\underline{a} \underline{m},2} \Sigma_2 +  ({\cal M}_0^2)^{\underline{a} \underline{m}, \underline{b} \underline{n}} \delta_{\underline{n} \underline{p}} \delta_{\underline{b} \underline{c}} \Sigma^{\underline{c} \underline{p}} \, , 
\end{equation}
where the entries of the squared mass matrix for the scalars, ${\cal M}_0^2$, are given by 
\begin{align}
    \label{M11}
    ({\cal M}_0^2)^{1,1} = ({\cal M}_0^2)^{2,2} = & \,  g^2 \bigg{(} - \frac{2}{9} A_1^{ij} {\bar A}_{1ij} - \frac{2}{9} A_2^{(ij)} {\bar A}_{2ij} + \frac{2}{9} A_2^{[ij]} {\bar A}_{2ij} + A_{2 \underline{a} i}{}^j {\bar A}_2{}^{\underline{a}i}{}_j \nonumber \\
    & - \frac{11}{6} A_2^{ij} {\bar B}_{ij} - \frac{1}{4} A_{2 \underline{a} i}{}^i {\bar B}^{\underline{a}} + \frac{3}{4} B^{ij} {\bar B}_{ij} \bigg{)}, \\
    \label{M1am}
     ({\cal M}_0^2)^{1,\underline{a} \underline{m}} = & \, \sqrt{2} g^2 \bigg{(} \frac{1}{4} {\bar A}_{2ij} {\bar A}_2{}^{\underline{a} k}{}_k - {\bar A}^{\underline{a} \underline{b}}{}_{ik} {\bar A}_{2 \underline{b}}{}^k{}_j + \frac{1}{4}  {\bar A}^{\underline{a} \underline{b}}{}_{ij} {\bar A}_{2 \underline{b}}{}^k{}_k \nonumber \\
     & - \frac{5}{8} {\bar A}_2{}^{\underline{a}k}{}_k {\bar B}_{ij}  + \frac{5}{12} {\bar A}_{2ij} {\bar B}^{\underline{a}} - \frac{3}{8} {\bar B}_{ij} {\bar B}^{\underline{a}} \bigg{)} \Gamma^{\underline{m}ij} + c.c. \, ,  \\
     \label{M2am}
      ({\cal M}_0^2)^{2,\underline{a} \underline{m}} = & \, i\sqrt{2} g^2 \bigg{(} \frac{1}{4} {\bar A}_{2ij} {\bar A}_2{}^{\underline{a} k}{}_k - {\bar A}^{\underline{a} \underline{b}}{}_{ik} {\bar A}_{2 \underline{b}}{}^k{}_j + \frac{1}{4}  {\bar A}^{\underline{a} \underline{b}}{}_{ij} {\bar A}_{2 \underline{b}}{}^k{}_k \nonumber \\
     & - \frac{5}{8} {\bar A}_2{}^{\underline{a}k}{}_k {\bar B}_{ij}  + \frac{5}{12} {\bar A}_{2ij} {\bar B}^{\underline{a}} - \frac{3}{8} {\bar B}_{ij} {\bar B}^{\underline{a}} \bigg{)} \Gamma^{\underline{m}ij} + c.c. \, , \\
     \label{Mam1}
      ({\cal M}_0^2)^{\underline{a} \underline{m},1} = & \, \sqrt{2} g^2 \bigg{(} \frac{1}{4} {\bar A}_{2ij} {\bar A}_2{}^{\underline{a} k}{}_k - {\bar A}^{\underline{a} \underline{b}}{}_{ik} {\bar A}_{2 \underline{b}}{}^k{}_j + \frac{1}{4}  {\bar A}^{\underline{a} \underline{b}}{}_{ij} {\bar A}_{2 \underline{b}}{}^k{}_k \nonumber \\
      & - 2 {\bar A}_2{}^{\underline{a}k}{}_i {\bar B}_{jk} - \frac{5}{8} {\bar A}_2{}^{\underline{a}k}{}_k {\bar B}_{ij} - \frac{1}{4} {\bar A}_{2ij} {\bar B}^{\underline{a}}  + \frac{1}{8} {\bar B}_{ij} {\bar B}^{\underline{a}} \bigg{)} \Gamma^{\underline{m}ij} + c.c. \, , \\
      \label{Mam2}
       ({\cal M}_0^2)^{\underline{a} \underline{m},2} = & \, i\sqrt{2} g^2 \bigg{(} \frac{1}{4} {\bar A}_{2ij} {\bar A}_2{}^{\underline{a} k}{}_k - {\bar A}^{\underline{a} \underline{b}}{}_{ik} {\bar A}_{2 \underline{b}}{}^k{}_j + \frac{1}{4}  {\bar A}^{\underline{a} \underline{b}}{}_{ij} {\bar A}_{2 \underline{b}}{}^k{}_k \nonumber \\
      & - 2 {\bar A}_2{}^{\underline{a}k}{}_i {\bar B}_{jk} - \frac{5}{8} {\bar A}_2{}^{\underline{a}k}{}_k {\bar B}_{ij} - \frac{1}{4} {\bar A}_{2ij} {\bar B}^{\underline{a}}  + \frac{1}{8} {\bar B}_{ij} {\bar B}^{\underline{a}} \bigg{)} \Gamma^{\underline{m}ij} + c.c. \, .  
 \end{align}
Furthermore, the symmetric part of the submatrix $({\cal M}_0^2)^{\underline{a} \underline{m}, \underline{b} \underline{n}}$ reads
\begin{align}
\label{symm}
    \frac{1}{2} ({\cal M}_0^2)^{\underline{a} \underline{m}, \underline{b} \underline{n}}  + \frac{1}{2} ({\cal M}_0^2)^{ \underline{b} \underline{n},\underline{a} \underline{m} } = & \,  g^2 \delta^{\underline{a} \underline{b}} \delta^{\underline{m} \underline{n}} \bigg{(} - \frac{2}{9} A_1^{ij} {\bar A}_{1ij}  + \frac{2}{9} A_2^{(ij)} {\bar A}_{2 ij} \nonumber \\
    & + \frac{1}{2} A_{2 \underline{c} i}{}^j {\bar A}_2{}^{\underline{c}i}{}_j - \frac{1}{4}  A_{2 \underline{c} i}{}^i {\bar A}_2{}^{\underline{c}j}{}_j \bigg{)} \nonumber \\
    &  + g^2 \delta^{\underline{m} \underline{n}} \bigg{(} {\bar A}^{(\underline{a}}{}_{\underline{c}ij} A^{\underline{b}) \underline{c} ij} - \frac{1}{2} A_2{}^{(\underline{a}|}{}_i{}^j {\bar A}_2{}^{|\underline{b})i}{}_j + A_2{}^{(\underline{a}|}{}_i{}^i {\bar A}_2{}^{|\underline{b})j}{}_j \nonumber \\
    & - \frac{1}{2} A_2{}^{(\underline{a}|}{}_i{}^i {\bar B}^{|\underline{b})} - \frac{1}{4} B^{(\underline{a}} {\bar B}^{\underline{b})} \bigg{)} \nonumber \\ 
    & + g^2 \bigg{(} - {\bar A}^{(\underline{a}}{}_{\underline{c}kl} A^{\underline{b})\underline{c}ij} - 2 A_2{}^{(\underline{a}|}{}_k{}^i {\bar A}_2{}^{|\underline{b})j}{}_l + \frac{1}{3} \delta^{\underline{a} \underline{b}} A_2^{ij} {\bar A}_{2 kl} \nonumber \\
    & + 2 \delta^{\underline{a} \underline{b}} A_{2 \underline{c}k}{}^i {\bar A}_2{}^{\underline{c}j}{}_l + \frac{1}{6} \delta^{\underline{a} \underline{b}} A_2^{ij} {\bar B}_{kl} + \frac{1}{6}  \delta^{\underline{a} \underline{b}} {\bar A}_{2 kl} B^{ij} \\
    & - \frac{1}{4}  \delta^{\underline{a} \underline{b}} B^{ij} {\bar B}_{kl} \bigg{)} \Gamma^{(\underline{m}}{}_{ij} \Gamma^{\underline{n})kl} \nonumber \\ 
    & + g^2 \bigg{(} - 6 {\bar A}^{[\underline{a}}{}_{\underline{c}kl} A^{\underline{b}] \underline{c} jl} + \frac{8}{3} A^{\underline{a} \underline{b} jl} {\bar A}_{2 (kl)} - \frac{8}{3} {\bar A}^{\underline{a} \underline{b}}{}_{kl} A_2^{(jl)}  \nonumber \\ 
    & + \frac{2}{3}  A^{\underline{a} \underline{b} jl} {\bar A}_{2 [kl]} - \frac{2}{3} {\bar A}^{\underline{a} \underline{b}}{}_{kl} A_2^{[jl]} - 3 A_2{}^{[\underline{a}|}{}_k{}^l {\bar A}_2{}^{|\underline{b}]j}{}_l \nonumber \\
    & - 5 A_2{}^{[\underline{a}|}{}_l{}^j {\bar A}_2{}^{|\underline{b}]l}{}_k + 3 A_2{}^{[\underline{a}|}{}_k{}^j {\bar A}_2{}^{|\underline{b}]l}{}_l + 3 A_2{}^{[\underline{a}|}{}_l{}^l {\bar A}_2{}^{|\underline{b}]j}{}_k \nonumber \\ 
    & - A^{\underline{a} \underline{b} jl} {\bar B}_{kl} + {\bar A}^{\underline{a} \underline{b}}{}_{kl} B^{jl} - 3 A_2{}^{[\underline{a}|}{}_k{}^j {\bar B}^{|\underline{b}]} + 3 {\bar A}_2{}^{[\underline{a}|j}{}_k B^{|\underline{b}]} \bigg{)} \Gamma^{[\underline{m}}{}_{ij} \Gamma^{\underline{n}]ik}, \nonumber  
\end{align}
while its antisymmetric part is given by 
\begin{align}
    \label{antisym}
    \frac{1}{2} ({\cal M}_0^2)^{\underline{a} \underline{m}, \underline{b} \underline{n}}  - \frac{1}{2} ({\cal M}_0^2)^{ \underline{b} \underline{n},\underline{a} \underline{m} } = & \,  g^2 \delta^{\underline{m} \underline{n}} \bigg{(} A^{\underline{a} \underline{b} ij} {\bar B}_{ij} + {\bar A}^{\underline{a} \underline{b}}{}_{ij} B^{ij} + \frac{1}{2} A_2{}^{[\underline{a}|}{}_i{}^i {\bar B}^{|\underline{b}]}  + \frac{1}{2} {\bar A}_2{}^{[\underline{a}|i}{}_i B^{|\underline{b}]} \bigg{)} \nonumber \\
    & - g^2 \left( A^{\underline{a} \underline{b} ij} {\bar B}_{kl} + {\bar A}^{\underline{a} \underline{b}}{}_{kl} B^{ij} \right) \Gamma^{(\underline{m}}{}_{ij} \Gamma^{\underline{n})kl} \nonumber \\
    & + g^2 \bigg{(} - 2 A_2{}^{(\underline{a}|}{}_k{}^j {\bar B}^{|\underline{b})} + 2 {\bar A}_2{}^{(\underline{a}|j}{}_k B^{|\underline{b})} + 2 \delta^{\underline{a} \underline{b}} A_{2 \underline{c}k}{}^j {\bar B}^{\underline{c}} \\ 
    & - 2 \delta^{\underline{a} \underline{b}} {\bar A}_2{}^{\underline{c}j}{}_k B_{\underline{c}} - \frac{2}{3} \delta^{\underline{a} \underline{b}} A_2^{[jl]} {\bar B}_{kl} + \frac{2}{3}  \delta^{\underline{a} \underline{b}} {\bar A}_{2 [kl]} B^{jl} \bigg{)} \Gamma^{[\underline{m}}{}_{ij} \Gamma^{\underline{n}]ik} \nonumber . 
\end{align}
For the derivation of the formulae \eqref{M11}-\eqref{antisym} we have made use of many of the $T$-identities of appendix \ref{sec:Tid}. Since the scalar mass matrix \eqref{M11}-\eqref{antisym} is not symmetric, it cannot arise from a scalar potential. In the absence of a gauging of the scaling symmetry, the elements of the squared mass matrix of the scalars consistently reduce to the expressions provided in \cite{DallAgata:2023ahj} and its antisymmetric part vanishes. 

Moreover, using the duality relation \eqref{twist} and the fact that ${\cal G}^{M \alpha}_{\mu \nu}$ is on-shell identified with $H^{M \alpha}_{\mu \nu}$ by virtue of the duality equations \eqref{H=G1} and \eqref{H=G2}, we can write the equations of motion for the vector fields, \eqref{veceom}, as 
\begin{equation}
    \label{linvec}
    e^{-1} \partial_{\nu} ( e H^{M \alpha \nu \mu} ) = ({\cal M}_1^2)^{M \alpha}{}_{N \beta} A^{N \beta \mu} + \dots \, , 
\end{equation}
where 
\begin{align}
    \label{M1}
    ({\cal M}_1^2)^{M \alpha}{}_{N \beta}= & \, \frac{i}{4} g^2 \left(\xi_{\gamma P} + \theta_{\gamma P} \right) \left(\xi^{\delta}_N - \theta^{\delta}_N \right)  M^{MP}\left( ({\cal V}^\alpha)^*  ({\cal V}^\gamma)^* {\cal V}_\beta {\cal V}_\delta - {\cal V}^\alpha  {\cal V}^\gamma  {\cal V}_\beta^* {\cal V}_\delta^* \right) \nonumber \\
     & - 2  g^2 Z_{P \gamma QR} {\hat\Theta}_{\beta NST} M^{MP} M^{\alpha \gamma} L^Q{}_{\underline{a}} L^{S \underline{a}} L^R{}_{ij} L^{T ij}
\end{align}
is the squared mass matrix of the vector fields and the ellipses represent terms of higher order in the fields that are not relevant for the present analysis. 

In addition, to specify the mass matrices of the fermions, we first note that the linearized fermionic field equations read
\begin{align}
    \label{dillin}
    \gamma^{\mu} {\cal D}_{\mu} \chi_i = & \, g \left( \frac{2}{3} {\bar A}_{2ij} - {\bar B}_{ij}\right) \gamma^{\mu} \psi^j_{\mu} - 2 g {\bar A}_2{}^{\underline{a}j}{}_i \lambda_{\underline{a}j} +  2 g {\bar A}_2{}^{\underline{a}j}{}_j \lambda_{\underline{a}i} + \frac{5}{2} g {\bar B}^{\underline{a}} \lambda_{\underline{a}i} \, , \\
    \label{gauglin}
    \gamma^{\mu} {\cal D}_{\mu} \lambda^{\underline{a}i} = & \, g \left( A_2{}^{\underline{a}}{}_j{}^i - \frac{1}{4} \delta^i_j B^{\underline{a}} \right) \gamma^{\mu} \psi^j_{\mu} - g {\bar A}_2{}^{\underline{a}i}{}_j \chi^j + g {\bar A}_2{}^{\underline{a}j}{}_j \chi^i - \frac{3}{4} g {\bar B}^{\underline{a}} \chi^i   \nonumber \\
    & + 2 g A^{\underline{a} \underline{b} ij} \lambda_{\underline{b}j} + \frac{2}{3} g A_2^{(ij)} \lambda^{\underline{a}}_j - 2 g B^{ij} \lambda^{\underline{a}}_j \, , \\
    \label{gravlin}
    \gamma^{\mu \nu \rho} {\cal D}_{\nu} \psi_{i \rho} = & - \frac{2}{3} g
 \left( {\bar A}_{1ij} - \frac{3}{2} \epsilon_{ijkl} B^{kl} \right) \gamma^{\mu \nu} \psi^j_{\nu} - \frac{1}{2} g \gamma^{\mu} G_i \, , 
 \end{align}
where 
\begin{equation}
{\cal D}_\mu \psi_{i \nu} \equiv \partial_\mu \psi_{i \nu} + \frac{1}{4} \omega_{\mu a b}(e) \gamma^{ab} \psi_{i \nu}
\end{equation}
and similarly for the spin-1/2 fermions and
\begin{align}
    \label{goldstini}
    G_i \equiv \left( \frac{2}{3} {\bar A}_{2ji} - 3 {\bar B}_{ij} \right) \chi^j + \left( 2 A_{2 \underline{a}i}{}^j + \frac{7}{2} \delta_i^j B_{\underline{a}} \right) \lambda^{\underline{a}}_j 
\end{align}
are the goldstini of the broken supersymmetries. The mass matrix of the gravitini can be read off from \eqref{gravlin} and is given by 
\begin{equation}
    \label{gravmass}
    ({\cal M}_{\frac{3}{2}})_{ij} = - \frac{2}{3} g
 \left( {\bar A}_{1ij} - \frac{3}{2} \epsilon_{ijkl} B^{kl} \right).
\end{equation}
On the other hand, from \eqref{dillin} and \eqref{gauglin} it follows that in the $(\chi_i, \sqrt{2} \lambda^{\underline{a}i})$ basis, the entries of the mass matrix for the spin-1/2 fermions, ${\cal M}_{1/2}$,
read
\begin{align}
\label{Mij}
       ({\cal M}_{\frac{1}{2}})_{ij} = & \, 0 \, ,\\
       \label{Miaj}
     ({\cal M}_{\frac{1}{2}})_i{}^{\underline{a}j} = & - \sqrt{2} g {\bar A}_2{}^{\underline{a}j}{}_i + \sqrt{2} g \delta^j_i  {\bar A}_2{}^{\underline{a}k}{}_k  + \frac{5 \sqrt{2}}{4} g \delta^j_i {\bar B}^{\underline{a}} , \\
     \label{Maij}
          ({\cal M}_{\frac{1}{2}})^{\underline{a}i}{}_j = & - \sqrt{2} g {\bar A}_2{}^{\underline{a}i}{}_j + \sqrt{2} g \delta^i_j  {\bar A}_2{}^{\underline{a}k}{}_k - \frac{3 \sqrt{2}}{4} g \delta^i_j {\bar B}^{\underline{a}} , \\
          \label{Maibj}
           ({\cal M}_{\frac{1}{2}})^{\underline{a}i, \underline{b}j} = & \, 2 g A^{\underline{a} \underline{b} ij} + \frac{2}{3} g \delta^{\underline{a} \underline{b}} A_2^{(ij)} - 2 g B^{ij} \delta^{\underline{a} \underline{b}} .
\end{align}
Of course, for a given solution to \eqref{cond1} and \eqref{cond2}, the matrix ${\cal M}_{1/2}$ must be computed after the elimination of the goldstini that are eaten by the massive gravitini.  
\section{Conclusions and Outlook}

We have constructed the most general four-dimensional ${\cal N}=4$ supergravity coupled to an arbitrary number $n$ of vector multiplets with a gauge symmetry that is the direct product of a subgroup of $\text{SL}(2,\mathbb{R}) \times \text{SO}(6,n) $ and the on-shell scaling symmetry of the corresponding ungauged theory. In the embedding tensor formalism, such gaugings are parametrized by three real constant $\text{SL}(2,\mathbb{R}) \times \text{SO}(6,n)$ tensors $f_{\alpha MNP}$, $\xi_{\alpha M}$ and $\theta_{\alpha M}$, which are subject to a set of quadratic consistency constraints. We have explicitly derived a general solution to these constraints in the presence of at least one vector multiplet, that is for $n \geq 1$, by decomposing them with respect to the subgroup $\text{SO}(1,1)_B \times  \text{SO}(1,1)_A \times \text{SO}(5,n-1)$ of  $\text{SL}(2,\mathbb{R}) \times \text{SO}(6,n)$. We have also specified the local supersymmetry transformation rules of half-maximal supergravity with local scaling symmetry in four spacetime dimensions and the associated equations of motion. The latter are completely fixed by supersymmetry and cannot be obtained from an action. 

It is worth pointing out that in the absence of a gauging of the scaling symmetry, the quadratic constraints \eqref{q2}-\eqref{q11}, which in this case reproduce the quadratic consistency constraints on the standard gaugings provided in  \cite{Schon:2006kz} (equations $(2.20)$ thereof), can still be solved by decomposing them with respect to the subgroup $\text{SO}(1,1)_B \times  \text{SO}(1,1)_A \times \text{SO}(5,n-1)$ of $ \text{SL}(2,\mathbb{R}) \times \text{SO}(6,n)$, assuming of course that $n \geq 1$. Indeed, for $\theta_{\alpha M} = 0$, the embedding tensor still contains parameters in the fundamental representation of $ \text{SL}(2,\mathbb{R}) \times \text{SO}(6,n)$, namely $\xi_{\alpha M}$, whose decomposition contains an $\text{SO}(5,n-1)$ singlet, for example $\xi_{+ \oplus}$, which, if non-vanishing, will allow us to explicitly construct a general solution to the quadratic identities $(2.20)$ of \cite{Schon:2006kz}. Such a solution will describe a class of standard gaugings of four-dimensional ${\cal N}=4$ matter-coupled supergravity with non-vanishing $\xi_{\alpha M}$, which have not been extensively studied in the literature to date.  

Solutions with non-vanishing $\xi_{\alpha M}$ to the quadratic constraints on the embedding tensor that parametrizes the standard gaugings of $D=4$, ${\cal N}=4$ supergravity  have also been found in \cite{Schon:2006kz} for the following two cases:
\begin{enumerate}
    \item[\textbf{i.}] for $f_{\alpha MNP}=0$, in which case $\xi_{\alpha M}$ must be of the form $\xi_{\alpha M} = v_{\alpha} w_{M}$, with $v_{\alpha}$ arbitrary and $w_M$ lightlike, i.e. $w_M w^M = 0$. As pointed out in \cite{Schon:2006kz}, this simple solution describes a gauging that can  be obtained by a Scherk-Schwarz reduction from five dimensions with a non-compact SO(1,1) twist. The relevant gauged $D=4$, ${\cal N}=4$ supergravity model with a single vector multiplet was explicitly constructed in \cite{Villadoro:2004ci} by a generalized dimensional reduction  of the aforementioned type of the ungauged pure $D=5$, ${\cal N}=4$ supergravity.
    \item[\textbf{ii.}] for electric gaugings of the four-dimensional ${\cal N}=4$ supergravity model with $n=6$ vector multiplets that originates from compactification of type IIB supergravity on a $T^6/\mathbb{Z}_2$ orientifold \cite{Frey:2002hf,Kachru:2002he}, which has an off-shell global SL(2,$\mathbb{R}$) $\times$ GL(6,$\mathbb{R}$) symmetry. The higher-dimensional origin of such gaugings with non-vanishing $\xi_{\alpha M}$ is not yet clear, however. 
\end{enumerate}

On the other hand, by decomposing the $ \text{SL}(2,\mathbb{R}) \times \text{SO}(6,n)$ tensors $f_{\alpha M N P
}$ and $\xi_{\alpha M}$ with respect to $\text{SO}(1,1)_B \times  \text{SO}(1,1)_A \times \text{SO}(5,n-1)$, one could find a broader class of consistent standard gaugings of $D=4$, ${\cal N}=4$ supergravity coupled to at least one vector multiplet with non-vanishing $\xi_{\alpha M}$  and a potential five-dimensional origin. Then, it would be very interesting to search for new vacua of four-dimensional ${\cal N}=4$ matter-coupled supergravity that arise from these gaugings and to specify the corresponding mass spectra using the conditions satisfied by the critical points of the scalar potential and the explicit formulae for the mass matrices of the various fields of the theory, given in \cite{DallAgata:2023ahj}. An analogous analysis could be performed for the gaugings described by the general solution to the quadratic constraints \eqref{q2}-\eqref{q11} given in subsection \ref{sec:solmain}, which involve the scaling symmetry, by looking for maximally symmetric solutions to the corresponding equations of motion with constant scalar fields.   

\section*{Acknowledgments}

 The author would like to thank Alex Kehagias and Mario Trigiante for comments on the manuscript. The research work was supported by the Hellenic Foundation for Research and Innovation (HFRI) under the 3rd Call for HFRI PhD Fellowships (Fellowship Number: 6554).
\begin{center}
\includegraphics[scale=0.4]{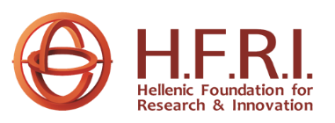}
\end{center}
\appendix 

\section{ Solution of the Bianchi Identities}
\label{sec:Bianchi}
The local supersymmetry transformation rules of four-dimensional ${\cal N}=4$ supergravity with local scaling symmetry can be derived with the use of the geometric or rheonomic approach (for a review see \cite{Castellani:1991eu}). 

The first step in this approach is the extension of  the spacetime fields of the theory to superfields in ${\cal N}=4$ superspace: this means that the spacetime zero-forms $\chi^i$, $\chi_i$, $\lambda^{\underline{a}i}$, $\lambda^{\underline{a}}_i$, ${\cal V}_\alpha$, ${\cal V}_{\alpha}^*$, ${L_M}^{ij}$ and ${L_M}^{\underline{a}}$, the spacetime one-forms $e^a=e^a_{\mu} dx^\mu$, $\psi^i=\psi^i_\mu dx^\mu$, $\psi_i=\psi_{i\mu} dx^\mu$, $A^{M \alpha} = A^{M \alpha}_\mu dx^\mu$ and $\omega_{ab} = \omega_{\mu a b} dx^\mu$, where $\omega_{\mu a b}$ is the spin connection, and the spacetime two-forms $B^{MN} = \frac{1}{2} B^{MN}_{\mu \nu} dx^{\mu} \wedge dx^{\nu} $ and $B^{\alpha \beta} = \frac{1}{2} B^{\alpha \beta}_{\mu \nu} dx^{\mu} \wedge dx^{\nu}$ are promoted to super-zero-forms,  super-one-forms and super-two-forms in  ${\cal N}=4$ superspace respectively. 
These superforms depend on the superspace coordinates $(x^\mu, \theta^i_\alpha, \theta_{i \alpha})$ (where $\theta^i_\alpha$ and $\theta_{i \alpha}$, $i,\alpha=1,2,3,4$, are anticommuting fermionic coordinates and are the components of left-handed Weyl spinors $\theta^i$ and their charge conjugates $\theta_i$ respectively) in such a way that their projections on the spacetime submanifold, i.e. the $\theta^i=d \theta^i = 0 $ hypersurface, are equal to the corresponding spacetime quantities. 

A basis of one-forms in ${\cal N}=4$ superspace is given by the supervielbein $\{ e^a, \psi^i_\alpha, \psi_{i \alpha} \}$, where $e^a$ is the bosonic vielbein, while $\psi^i_\alpha$ and $\psi_{i \alpha}$, which are the spinor components of the left-handed gravitino super-one-forms $\psi^i$ and their charge conjugates $\psi_i$ respectively, constitute the fermionic vielbein.

Given the scaling weights of the various super-$p$-forms of the theory, the appropriate definitions of the corresponding gauged supercurvatures in the presence of a gauging of the trombone symmetry read 
\begin{align}
\label{Rabapp}
R^{ab}= \,  & d \omega^{ab} + \omega^{ac} \wedge {\omega_c}^b,  \\[2mm]
{\hat T}^a = \, &   d e^a  + {\omega^a}_b \wedge e^b - g \theta_{\alpha M} A^{M \alpha} \wedge e^a - {\bar\psi}^i \wedge \gamma^a \psi_i = \hat{D} e^a  - {\bar\psi}^i \wedge \gamma^a \psi_i \, ,  \\[2mm]
\hat{\rho}_i  = \hat{D} \psi_i = \, &  d \psi_i + \frac{1}{4} \omega^{ab} \wedge \gamma_{ab} \psi_i - \frac{i}{2} \hat{\mathcal{A}} \wedge \psi_i - {\hat{\omega}_i}{}^{j} \wedge \psi_j - \frac{g}{2} \theta_{\alpha M} A^{M \alpha} \wedge \psi_i \, , \\[2mm]
{\hat V}_i = \hat{D} \chi_i = \,  & d \chi_i + \frac{1}{4} \omega^{ab} \gamma_{ab} \chi_i + \frac{3 i}{2} \hat{\mathcal{A}} \chi_i -  {\hat{\omega}_i}{}^{j} \chi_j  + \frac{g}{2} \theta_{\alpha M} A^{M \alpha} \chi_i \, , \\[2mm]
 {\hat \Lambda}_{\underline{a}i} = \hat{D} \lambda_{\underline{a}i}= \, & d \lambda_{\underline{a}i} + \frac{1}{4} \omega^{ab}  \gamma_{ab}  \lambda_{\underline{a}i}  + \frac{i}{2} \hat{\mathcal{A}} \lambda_{\underline{a}i} - {\hat{\omega}_i}{}^{j} \lambda_{\underline{a}j} + {\hat{\omega}_{\underline{a}}}{}^{\underline{b}} \lambda_{\underline{b}i} + \frac{g}{2} \theta_{\alpha M} A^{M \alpha} \lambda_{\underline{a}i} \, , \\[2mm]
 \label{Hdef}
{\cal H}^{M \alpha}  = \, & d A^{M \alpha} + \frac{g}{2} X_{N \beta P \gamma}{}^{M \alpha} A^{N \beta} \wedge A^{P \gamma} - \frac{g}{2} {{\Theta}^{\alpha M}}_{NP} B^{NP} + \frac{3}{2} g \theta^{\alpha}_N B^{MN} \nonumber \\ &+ \frac{g}{2} \left(\xi^M_\beta + \theta^M_\beta \right) B^{\alpha \beta} 
- (\mathcal{V}^\alpha)^* L^{Mij} {\bar\psi}_i \wedge \psi_j - \mathcal{V}^\alpha  {L^M}_{ij} {\bar \psi}^i \wedge \psi^j, \\[2mm]
{\cal H}^{(3) MN} = \, & {\hat d}B^{MN}   + \epsilon_{\alpha \beta} A^{[M|\alpha} \wedge \left( dA^{|N]\beta} + \frac{g}{3} {X_{P \gamma Q \delta}}^{|N]\beta} A^{P \gamma} \wedge A^{Q \delta} \right), \\[2mm]
{\cal H}^{(3) \alpha \beta} = \, & {\hat d}B^{\alpha \beta} - \eta_{MN} A^{M(\alpha|} \wedge \left(  dA^{N|\beta)} + \frac{g}{3} {X_{P \gamma Q \delta}}^{N|\beta)} A^{P \gamma} \wedge A^{Q \delta} \right) \,, \\[2mm]
{\hat P} = \, & \frac{i}{2} \epsilon^{\alpha \beta} {\cal V}_{\alpha} {\hat d} {\cal V}_\beta \, , \\[2mm]
\label{Paijapp}
{\hat P}_{\underline{
a}ij} = \, & L^M{}_{\underline{a}} {\hat d} L_{Mij} \, ,
\end{align} 
where $\hat{\mathcal{A}}$, ${\hat{\omega}_i}{}^{j}$ and ${\hat{\omega}_{\underline{a}}}{}^{\underline{b}}$ are the extensions of the gauged SO(2), SU(4) and SO($n$) connections to ${\cal N}=4$ superspace respectively and $\hat{D}$ is the exterior derivative that is covariant with respect to local Lorentz, SO(2), SU(4), SO($n$) and gauge transformations. 

By acting on the gauged supercurvatures with the exterior derivative $d$ and using the fact that $d^2=0$, we obtain the following Bianchi identities:
\begin{align}
\label{DhatRab}
    {\hat D} R^{ab} =& \, 0 \, ,\\[2mm]
    \label{DhatRa}
{\hat D} {\hat T}^a =& \,  {R^a}_b \wedge e^b +  {\bar\psi}_i \wedge \gamma^a {\hat{\rho}}^i + {\bar\psi}^i \wedge \gamma^a {\hat\rho}_i \nonumber \\
& - g \theta_{\alpha M} \left( {\cal H}^{M \alpha} + ({\cal V}^\alpha)^* L^{Mij} {\bar\psi}_i \wedge \psi_j + {\cal V}^{\alpha} L^M{}_{ij} {\bar\psi}^i \wedge \psi^j \right) \wedge e^a, \\[2mm]
\label{Dhatrho}
\hat{D} {\hat\rho}_i   =& \, \frac{1}{4} R^{ab} \wedge \gamma_{ab} \psi_i - \frac{i}{2} \hat{F} \wedge \psi_i - {\hat{R}_i}{}^{ j} \wedge \psi_j \nonumber\\ & - \frac{g}{2} \theta_{\alpha M} \psi_i \wedge \left( {\cal H}^{M \alpha} + ({\cal V}^\alpha)^* L^{Mjk} {\bar\psi}_j \wedge \psi_k + {\cal V}^{\alpha} L^M{}_{jk} {\bar\psi}^j \wedge \psi^k \right)  , \\ 
\label{DhatX}
\hat{D} \hat{V}_i   =& \, \frac{1}{4} R^{ab}  \gamma_{ab} \chi_i + \frac{3i}{2}  \hat{F} \chi_i - {\hat{R}_i}{}^{j}   \chi_j \nonumber \\
& + \frac{g}{2} \theta_{\alpha M} \chi_i \left( {\cal H}^{M \alpha} + ({\cal V}^\alpha)^* L^{Mjk} {\bar\psi}_j \wedge \psi_k + {\cal V}^{\alpha} L^M{}_{jk} {\bar\psi}^j \wedge \psi^k \right)  , \\
\label{DhatL}
\hat{D}  \hat{\Lambda}_{\underline{a}i} = & \, \frac{1}{4} R^{ab}  \gamma_{ab} \lambda_{\underline{a}i} + \frac{i}{2}  \hat{F} \lambda_{\underline{a}i} -  {\hat{R}_i}{}^{j}\lambda_{\underline{a}j} +   {{\hat{R}}_{\underline{a}}}{}^{\underline{b}} \lambda_{\underline{b}i} \nonumber \\
& + \frac{g}{2} \theta_{\alpha M} \lambda_{\underline{a}i}  \left( {\cal H}^{M \alpha} + ({\cal V}^\alpha)^* L^{Mjk} {\bar\psi}_j \wedge \psi_k + {\cal V}^{\alpha} L^M{}_{jk} {\bar\psi}^j \wedge \psi^k \right) ,        \\
\label{DhatH}
\hat{D} {\cal H}^{M \alpha} =& - \mathcal{V}^\alpha L^{Mij}  {\hat{P}}^* \wedge \bar\psi_i \wedge \psi_j - (\mathcal{V}^\alpha )^* L^{M \underline{a}} {{\hat{P}}_{\underline{a}}}^{\hspace{0.15cm} ij} \wedge \bar\psi_i \wedge \psi_j + 2 (\mathcal{V}^\alpha)^* L^{Mij}  \bar\psi_i \wedge {\hat\rho}_j \nonumber  \\[2mm] 
 & - (\mathcal{V}^\alpha)^* {L^M}_{ij}  \hat{P} \wedge \bar\psi^i \wedge \psi^j - \mathcal{V}^\alpha L^{M \underline{a}} \hat{P}_{\underline{a}ij} \wedge \bar\psi^i \wedge \psi^j + 2 \mathcal{V}^\alpha  {L^M}_{ij}  \bar\psi^i \wedge {\hat \rho}^j \\[2mm]
 & - \frac{g}{2} {{\Theta}^{\alpha M}}_{NP} {\cal H}^{(3)NP}  + \frac{3}{2} g \theta^{\alpha}_N {\cal H}^{(3)MN}+ \frac{g}{2} \left(\xi^M_{\beta} + \theta^M_\beta \right) {\cal H}^{(3) \alpha \beta} ,  \nonumber \\[2mm] 
 \label{2formBianchi}
 -\frac{1}{2} {{\Theta}^{\alpha M}}_{NP} & \hat{D} {\cal H}^{(3)NP} + \frac{3}{2} \theta^{\alpha}_N \hat{D} {\cal H}^{(3) MN}+  \frac{1}{2} \left(\xi^M_\beta + \theta^M_\beta\right)\hat{D} {\cal H}^{(3) \alpha \beta} \nonumber = \\  &   {X_{N \beta P \gamma}}^{M \alpha} \left( {\cal H}^{N \beta} + ({\cal V}^\beta)^* L^{Nij} {\bar \psi}_i \wedge \psi_j + {\cal V}^\beta {L^N}_{ij} {\bar \psi}^i \wedge \psi^j \right) \\ 
 & \wedge \left( {\cal H}^{P \gamma}  + ({\cal V}^\gamma)^* L^{Pkl} {\bar \psi}_k \wedge \psi_l + {\cal V}^\gamma {L^P}_{kl} {\bar \psi}^k \wedge \psi^l \right) ,\nonumber \\[2mm]
 \label{DhatP}
 \hat{D} \hat{P} = & \, \frac{i}{2} g \left( \xi_{\alpha M} - \theta_{\alpha M} \right) {\cal V}^\alpha  {\cal V}_\beta {\cal H}^{M \beta} - g \left(\xi_{\alpha M} - \theta_{\alpha M} \right) {\cal V}^\alpha L^{Mij} {\bar \psi}_i \wedge \psi_j \, , \\[2mm]
 \label{DhatPaij}
  \hat{D} {{\hat{P}}_{\underline{a} ij}}  = \, & g {{\hat\Theta}_{\alpha M }}{}^{N P} L_{N \underline{a}} L_{P ij} \left({\cal H}^{M \alpha} + ({\cal V}^\alpha)^* L^{Mkl} {\bar\psi}_k \wedge \psi_l + {\cal V}^\alpha {L^M}_{kl} {\bar\psi}^k \wedge \psi^l\right) ,  
 \end{align}
 where $\hat{F}$, ${\hat{R}_i}{}^{ j}$ and ${{\hat{R}}_{\underline{a}}}{}^{ \underline{b}}$ are the superspace gauged SO(2), SU(4) and SO($n$) curvatures respectively, given by equations \eqref{gSO(2)curv}, \eqref{SU4curv} and \eqref{SOncurv}, which are now to be viewed as superspace equations.

The above Bianchi identities can be solved by suitable expansions of the supercurvatures along the bases of one-, two- and three-forms in ${\cal N}=4$ superspace that are built out of the supervielbein $\{ e^a, \psi^i, \psi_i \}$ by means of the wedge product. These expansions must obey the rheonomy principle, which means that all the components of the supercurvatures along the basis elements that involve at least one of the gravitino super-one-forms, $\psi^i$, $\psi_i$ (outer components) must be expressed in terms of the supercurvature components along the basis elements $e^a$, $e^a \wedge e^b$ and $e^a \wedge e^b \wedge e^c$ (inner components) and the physical superfields. This requirement ensures that no new degrees of freedom are introduced in the theory. Furthermore, the expansions of the supercurvatures along the bases of one-, two- and three-forms in superspace constructed out of the supervielbein are referred to as rheonomic parametrizations of the supercurvatures. 

To solve the Bianchi identities \eqref{DhatRab}-\eqref{DhatPaij}, we first impose the kinematic  constraint 
\begin{equation}
\label{Thata=0}
    {\hat T}^a = 0 \, , 
\end{equation}
which amounts to the vanishing of the supertorsion. The restriction of \eqref{Thata=0} to four-dimensional spacetime allows us to express the spin connection $\omega_{\mu a b}$ in terms of the vielbein $e^a_{\mu}$, the gravitini $\psi^i_{\mu}$ and the linear combination ${\cal B}_{\mu} \equiv g \theta_{\alpha M} A^{M \alpha}_{\mu}$ of the vector gauge fields, which gauges the scaling symmetry, according to \eqref{spinconn}.  

The rheonomic parametrizations of the super-field strengths of the scalars and the vectors, $\hat{P}$, $\hat{P}_{\underline{a}ij}$ and ${\cal H}^{M \alpha}$, in the trombone gauged four-dimensional half-maximal supergravity are the same as those in the standard gauged theory and are given by \cite{DallAgata:2023ahj}
\begin{align}
     \label{hatP}
     \hat{P}  = & \, {\hat{P}}_a e^a + {\bar\psi}_i \chi^i, \\
 \label{hatPaij}    
{\hat{P}}_{\underline{a}ij} = & \, {\hat{P}}_{\underline{a}ij a} e^a + 2 {\bar{\psi}}_{[i|} \lambda_{\underline{a} |j]} + \epsilon_{ijkl} {\bar{\psi}}^k \lambda^l_{\underline{a}}, \\
{\cal H}^{M \alpha}  =& \, \frac{1}{2} {\cal H}^{M \alpha}_{ab} e^a \wedge e^b + \bigg{(}-\frac{1}{4} \mathcal{V}^\alpha L^{Mij} {\bar\lambda}_{\underline{a}i} \gamma_{ab} \lambda^{\underline{a}}_j \, e^a \wedge e^b + \frac{1}{4} \mathcal{V}^\alpha L^{M \underline{a}} {\bar\chi}_i \gamma_{ab} \lambda^i_{\underline{a}} \, e^a \wedge e^b \nonumber  \\
\label{Hrheon}
&+ (\mathcal{V}^\alpha)^* {L^M}_{ij} {\bar \chi}^i \gamma_a \psi^j \wedge e^a +  (\mathcal{V}^\alpha)^* L^{M \underline{a}} {\bar\lambda}^i_{\underline{a}} \gamma_a \psi_i \wedge e^a  + c.c. \bigg{)},
\end{align}
where the inner components ${\cal H}^{M \alpha}_{a b}$ satisfy the constraint 
\begin{equation}
\label{moddualg}
\epsilon_{abcd} {\cal H}^{M \alpha cd}= -2   {M^M}_N {{M}^{\alpha}}_{\beta} { \cal H}^{N \beta}_{ab} .   
\end{equation}

On the other hand, in the presence of a gauging of the scaling symmetry, the rheonomic parametrizations of the fermionic supercurvatures, ${\hat V}_i$, ${\hat\Lambda}_{\underline{a} i}$ and $\hat{\rho}_i$, contain all the terms of their counterparts in the standard gauged theory, which can be found in \cite{DallAgata:2023ahj}, as well as additional ones proportional to the embedding tensor components $\theta_{\alpha M}$, whose form is dictated by the representations of $\text{SU}(4) \times \text{SO}(n) \times \text{SO}(2)$ carried by the fermionic (super)fields. Thus, the correct ansatzes for the rheonomic parametrizations of the super-field strengths of the fermions in $D=4$, ${\cal N}=4$ supergravity with local scaling symmetry read (up to three-fermion terms provided in \cite{DallAgata:2023ahj}) 
\begin{align}
    \label{hatVi}
 {\hat{V}}_i  = & \, {\hat{V}}_{ia} e^a  - \frac{i}{4} L_{Mij} \mathcal{V}_\alpha^* {\cal H}^{M \alpha}_{ab}\gamma^{ab} \psi^j + \gamma^a {\hat{P}}_a^* \psi_i + \frac{2}{3} g {\bar A}_{2ij}\psi^j + \alpha g {\bar B}_{ij} \psi^j ,
\\ \label{hatLai}
 {\hat{\Lambda}}_{\underline{a}i} = & \, {\hat{\Lambda}}_{\underline{a}i a}  e^a - {\hat{P}}_{\underline{a}ij a} \gamma^a \psi^j + \frac{i}{8} L_{M \underline{a}} \mathcal{V}_\alpha^*  {\cal H}^{M \alpha}_{ab}\gamma^{ab} \psi_i  + g {\bar A}_{2 \underline{a}}{}^j{}_i \psi_j + \beta g {\bar B}_{\underline{a}} \psi_i \, ,\\
\label{hatrhoi}
{\hat{\rho}}_i =& \, \frac{1}{2} {\hat{\rho}}_{iab} e^a \wedge e^b - \frac{i}{8} L_{Mij} \mathcal{V}_\alpha {\cal H}^{M \alpha}_{bc} \gamma^{bc} \gamma_a \psi^j \wedge e^a   - \frac{1}{3} g {\bar A}_{1ij}  \gamma_a \psi^j \wedge e^a \nonumber \\
& + \frac{1}{2} \gamma g \epsilon_{ijkl} B^{kl} \gamma_a \psi^j \wedge e^a,  
\end{align}
where we have introduced the fermion shift matrices
\begin{align}     
     \label{dilshiftapp}
   A_2^{ij} & =  f_{\alpha MNP} {\cal V}^\alpha {L^M}_{kl} L^{Nik} L^{P jl} + \frac{3}{2} \xi_{\alpha M} {\cal V}^\alpha L^{M ij}, \\
    \label{gaushiftapp}
    A_{2 \underline{a} i}{}^j & =  f_{\alpha MNP} {\cal V}^\alpha L^M{}_{\underline{a}} {L^N}_{ik} L^{Pjk} - \frac{1}{4} \delta^j_i \xi_{\alpha M}  {\cal V}^\alpha L^M{}_{\underline{a}} \, , \\
    \label{gravshiftapp}
   A_1^{ij} & =   f_{\alpha MNP} ({\cal V}^\alpha)^* L^M{}_{kl} L^{Nik} L^{Pjl} 
   , \\
   \label{Bij}
   B^{ij} & = \theta_{\alpha M} {\cal V}^\alpha L^{Mij} , \\
   \label{Ba}
   B^{\underline{a}} & = \theta_{\alpha M} {\cal V}^{\alpha} L^{M \underline{a}} \, ,
      \end{align}
 the first three of which were first defined in \cite{Schon:2006kz}, while $\alpha$, $\beta$ and $\gamma$ are constant coefficients, whose values are determined by requiring closure of the Bianchi identities and are thus found to be 
 \begin{equation}
     \label{abc}
     \alpha = - 1, \qquad \beta = - \frac{1}{4}, \qquad \gamma
      = 1 \, .
 \end{equation}

Furthermore, from the torsion Bianchi identity \eqref{DhatRa} one obtains the rheonomic parametrization of the Riemann supercurvature $R_{ab}$:
  \begin{align}
\label{Rabg}
R_{ab} = & \,  \frac{1}{2} R_{cdab} e^c \wedge e^d + {\bar{\hat{\theta}}}^i_{abc} \psi_i \wedge e^c + {\bar{\hat{\theta}}}_{iabc} \psi^i \wedge e^c \nonumber \\
& + \frac{i}{4} \mathcal{V}_\alpha L_{Mij} {\cal H}^{M \alpha}_{ab} {\bar\psi}^i \wedge \psi^j + \frac{1}{8} \mathcal{V}_\alpha L_{Mij} \epsilon_{abcd} {\cal H}^{M \alpha cd} {\bar\psi}^i \wedge \psi^j \nonumber \\
& - \frac{i}{4} \mathcal{V}_\alpha^* {L_M}^{ij} {\cal H}^{M \alpha}_{ab} {\bar\psi}_i \wedge \psi_j + \frac{1}{8}  \mathcal{V}_\alpha^* {L_M}^{ij}  \epsilon_{abcd} {\cal H}^{M \alpha cd} {\bar\psi}_i \wedge \psi_j  \\
& + \frac{1}{3} g {\bar A}_{1ij} {\bar\psi}^i \wedge \gamma_{ab} \psi^j + \frac{1}{3} g A_1^{ij} {\bar\psi}_i \wedge \gamma_{ab} \psi_j \nonumber \\
& + \text{four-fermion terms} \, ,  \nonumber
\end{align}
where
\begin{equation}
    {\hat{\theta}}_{iabc}  =  \gamma_{[a|} {\hat{\rho}}_{i|b]c} - \frac{1}{2} \gamma_c {\hat{\rho}}_{iab} - 2 g {\bar B}_{ij} \eta_{c[a} \gamma_{b]} \chi^j + 2 g B^{\underline{a}} \eta_{c [a} \gamma_{b]} \lambda_{\underline{a}i}
\end{equation}
and the omitted four-fermion terms are the same as in the standard gauged ${\cal N}=4$ supergravity in four dimensions (see \cite{DallAgata:2023ahj}). 

Moreover, the rheonomic parametrizations of the super-three-forms ${\cal H}^{(3) M \alpha} \equiv  - \frac{1}{2} {{\Theta}^{\alpha M}}_{NP} {\cal H}^{(3) NP} + \frac{3}{2} \theta^\alpha_N {\cal H}^{(3)MN} + \frac{1}{2} \left( \xi^M_\beta + \theta^M_\beta \right) {\cal H}^{(3) \alpha \beta}$ read
\begin{align}
\label{H3Ma}
{\cal H}^{(3) M \alpha}
=  & \, \frac{1}{6} {\cal H}^{(3) M \alpha}_{abc} e^a \wedge e^b \wedge e^c + i {\Theta}^{\alpha M N P} {L_N}^{\underline{a}} {L_P}^{ij} {\bar{\lambda}}_{\underline{a}i} \gamma_{ab} \psi_j \wedge e^a \wedge e^b  \nonumber \\
&  - 3 i \theta^{\alpha}_N L^{[M}{}_{\underline{a}} L^{N]ij} {\bar\lambda}_i^{\underline{a}} \gamma_{ab} \psi_j \wedge e^a \wedge e^b  \nonumber \\ 
& - \frac{1}{4} \left( \xi^M_\beta + \theta^M_\beta \right) ({\cal V}^\alpha)^*  ({\cal V}^\beta)^* {\bar\chi}^i \gamma_{ab} \psi_i \wedge e^a \wedge e^b \nonumber \\
&- i  {\Theta}^{\alpha M N P} {L_N}^{\underline{a}} L_{Pij} {\bar\lambda}^i_{\underline{a}} \gamma_{ab} \psi^j \wedge e^a \wedge e^b + 3 i \theta^{\alpha}_N L^{[M}{}_{\underline{a}} L^{N]}{}_{ij} {\bar\lambda}^{\underline{a}i} \gamma_{ab} \psi^j \wedge e^a \wedge e^b \\
& - \frac{1}{4} \left( \xi^M_\beta + \theta^M_\beta \right){\cal V}^\alpha {\cal V}^\beta {\bar\chi}_i \gamma_{ab} \psi^i \wedge e^a \wedge e^b \nonumber
\\ & + 2i {\Theta}^{\alpha MNP} {{L_N}}^{ik} L_{Pjk} {\bar\psi}^j \wedge \gamma_a \psi_i \wedge e^a - 6 i \theta^{\alpha}_N L^{[M|ik} L^{|N]}{}_{jk} {\bar\psi}^j \wedge \gamma_a \psi_i \wedge e^a \nonumber \\
& - \frac{1}{2} \left( \xi^M_\beta + \theta^M_\beta \right) M^{\alpha \beta}  {\bar\psi}^i \wedge \gamma_a \psi_i \wedge e^a . \nonumber
 \end{align}

In addition,  the Bianchi identities impose differential constraints on the inner components of the supercurvatures, whose projections on spacetime are identified with the equations of motion of the theory. Indeed, the closure of the Bianchi identities is equivalent to the closure of the ${\cal N}=4$ supersymmetry algebra on the spacetime fields, which happens only when the equations of motion are satisfied. In particular, the ${\bar\psi}^i \wedge \gamma^a \psi_i$ sector of the Bianchi identity \eqref{DhatX} implies the following superspace equations of motion for the dilatini:
\begin{align}
    \label{choeomg}
    \gamma^a {\hat{V}}_{i a} = & \, \frac{i}{4} \mathcal{V}_{\alpha}^* L_{M \underline{a}} {\cal H}^{M \alpha}_{ab} \gamma^{ab} \lambda^{\underline{a}}_i -2 g {\bar A}_2{}^{\underline{a} j}{}_i \lambda_{\underline{a}j} + 2 g {\bar A}_2{}^{\underline{a} j}{}_j  \lambda_{\underline{a}i} + \frac{5}{2} g {\bar B}^{\underline{a}} \lambda_{\underline{a}i} \\
    & + \text{three-fermion terms} \, ,  \nonumber
\end{align}
 while the corresponding sector of the Bianchi identity \eqref{DhatL} gives the following superspace equations of motion for the gaugini:
 \begin{align}
    \label{lameomg}
     \gamma^a {\hat \Lambda}_{\underline{a} i a} = & \, \frac{i}{4} \mathcal{V}_{\alpha}^* L_{Mij}  {\cal H}^{M \alpha}_{ab} \gamma^{ab} \lambda^j_{\underline{a}} + \frac{i}{8} \mathcal{V}_{\alpha} L_{M \underline{a}} {\cal H}^{M \alpha}_{ab} \gamma^{ab} \chi_i  \nonumber \\
       & - g A_{2 \underline{a} i}{}^j \chi_j + g A_{2 \underline{a} j}{}^j \chi_i + 2 g {\bar A}_{\underline{a} \underline{b} ij} \lambda^{\underline{b}j} + \frac{2}{3} g {\bar A}_{2 (ij)} \lambda^j_{\underline{a}} \\
    & - 2 g {\bar B}_{ij} \lambda^j_{\underline{a}} - \frac{3}{4} g B_{\underline{a}} \chi_i + \text{three-fermion terms} \nonumber \, , 
\end{align}
where 
\begin{equation}
         \label{Aabijapp}
    A_{\underline{a} \underline{b}}{}^{ij} \equiv f_{\alpha MNP} {\cal V}^{\alpha} {L^M}{}_{\underline{a}} {L^N}{}_{\underline{b}} L^{Pij}.
\end{equation}
On the other hand, by considering the ${\bar\psi}^i \wedge \gamma^a \psi_i \wedge e^b$ sector of the Bianchi identity \eqref{Dhatrho}, one can specify the superspace equations of motion for the gravitini in the trombone gauged $D=4$, ${\cal N}=4$ supergravity, which read
\begin{align}
 \label{psieomg}
 \gamma^b \hat{\rho}_{iba} = & \, \frac{i}{2}  \mathcal{V}_\alpha L_{M \underline{a}} {\cal H}^{M \alpha}_{ab} \gamma^b \lambda^{\underline{a}}_i - \frac{i}{2}  \mathcal{V}_{\alpha}^*  L_{Mij}  {\cal H}^{M \alpha}_{ab} \gamma^b \chi^j \nonumber \\[2mm] &  + {\hat{P}}_a \chi_i + 2 {\hat{P}}_{\underline{a} ija} \lambda^{\underline{a}j} + \frac{1}{3} g {\bar A}_{2ji} \gamma_a \chi^j + g A_{2 \underline{a} i}{}^j \gamma_a \lambda^{\underline{a}}_j   \\ & - \frac{3}{2} g {\bar B}_{ij} \gamma_a \chi^j + \frac{7}{4} g B^{\underline{a}} \gamma_a \lambda_{\underline{a}i} + \text{three-fermion terms} \, .  \nonumber
 \end{align}
We refer the reader to \cite{DallAgata:2023ahj} for the suppressed three-fermion terms on the right-hand sides of \eqref{choeomg}, \eqref{lameomg} and \eqref{psieomg}. 

From the rheonomic parametrizations of the supercurvatures \eqref{Rabapp}-\eqref{Paijapp} we can determine the local supersymmetry transformations of the spacetime fields in four-dimensional ${\cal N}=4$ supergravity with local scaling symmetry. We recall that from the superspace point of view, a local supersymmetry transformation parametrized by left-handed Weyl spinors $\epsilon^i$ and their charge conjugates $\epsilon_i$ is a Lie derivative $\ell_{\epsilon}$ along a tangent vector $\epsilon$ such that 
\begin{equation}
    \label{contr}
    i_{\epsilon} \psi^i = \epsilon^i \qquad  \text{and} \qquad i_{\epsilon} \psi_i = \epsilon_i \,.
\end{equation}

Using Cartan's magic formula, $\ell_{\epsilon} = d i_{\epsilon} + i_{\epsilon} d$, we find for the super-one-forms $e^a$, $\psi_i$ and $A^{M \alpha}$:
\begin{align}
\ell_{\epsilon} e^a = & \, i_{\epsilon} {\hat T}^a + {\bar\epsilon}^i \gamma^a \psi_i + {\bar\epsilon}_i \gamma^a \psi^i, \\
\ell_{\epsilon} \psi_i =  & \,  {\hat D} \epsilon_i + i_{\epsilon} {\hat\rho}_i, \\
\ell_{\epsilon} A^{M\alpha} = & \, i_{\epsilon} {\cal H}^{M \alpha} + 2 ({\cal V}^\alpha)^* L^{Mij} {\bar\epsilon}_i \psi_j + 2 {\cal V}^\alpha {L^M}_{ij} {\bar\epsilon}^i \psi^j  , 
\end{align}
where we have used the definitions of the superspace curvatures ${\hat T}^a$, ${\hat\rho}_i$ and ${\cal H}^{M \alpha}$ and 
\begin{equation}
    \label{Dhatepsilon}
    {\hat D} \epsilon_i \equiv d \epsilon_i + \frac{1}{4} \omega_{ab} \gamma^{ab} \epsilon_i - \frac{i}{2} \hat{\cal A}  \epsilon_i - {\hat\omega}_i{}^j \epsilon_j - \frac{g}{2} \theta_{\alpha M} A^{M \alpha} \epsilon_i.
\end{equation} 
For the super-zero-forms ${\nu}^I \equiv ({\cal V}_\alpha, {\cal V}_\alpha^*, L_{Mij}, L_{M\underline{a}},\chi^i,\chi_i, \lambda^i_{\underline{a}},\lambda_{\underline{a}i})$ we have the simpler result
\begin{equation}
    \ell_{\epsilon} {\nu}^I = i_{\epsilon} {\hat D} {\nu}^I.
\end{equation}
Furthermore, for the super-two-forms $B^{M \alpha} \equiv  - \frac{1}{2} {{\Theta}^{\alpha M}}_{NP} B^{NP} + \frac{3}{2} \theta^{\alpha}_N B^{MN}+ \frac{1}{2} \left(\xi^M_\beta + \theta^M_\beta \right) B^{\alpha \beta}$ we find
\begin{align}
    \ell_{\epsilon} B^{M \alpha}  = & \, i_{\epsilon} {\cal H}^{(3)M\alpha} - \frac{1}{2} {{\Theta}^{\alpha M}}_{NP} \epsilon_{\beta \gamma} A^{N \beta} \wedge \ell_{\epsilon} A^{P \gamma}  + \frac{3}{2} \theta^{\alpha}_N \epsilon_{\beta \gamma} A^{[M|\beta} \wedge \ell_{\epsilon} A^{|N] \gamma} \nonumber \\
    &  - \frac{1}{2} \left( \xi^M_\beta + \theta^M_\beta \right)\eta_{NP} A^{N(\alpha|} \wedge \ell_{\epsilon} A^{P|\beta)}.
 \end{align}   
Using the parametrizations given for the gauged supercurvatures and identifying the local supersymmetry transformation $\delta_{\epsilon}$ of each spacetime $p$-form with the projection of the Lie derivative $ \ell_{\epsilon}$ of the corresponding super-$p$-form on spacetime, it is straightforward to determine the ${\cal N}=4$ local supersymmetry transformations of all the spacetime fields in the trombone gauged theory. 
The results have been presented in section \ref{sec:susy_rules}.
 \section{Solution of the Quadratic Constraints}
\label{solapp}
Our strategy for the solution of the quadratic constraints obeyed by the embedding tensor that parametrizes the consistent gaugings of four-dimensional ${\cal N}=4$ matter-coupled supergravity that involve the scaling symmetry is analogous to that of \cite{LeDiffon:2008sh,LeDiffon:2011wt} for the solution of the corresponding constraints for the trombone gauged maximal supergravities in various dimensions. Assuming the presence of at least one vector multiplet, that is $n\geq1$, we decompose the irreducible components $f_{\alpha M N P}$, $\xi_{\alpha M}$ and $\theta_{\alpha M}$ of the embedding tensor with respect to the subgroup $\text{SO}(1,1)_B \times  \text{SO}(1,1)_A \times \text{SO}(5,n-1)$ of SL(2,$\mathbb{R}$) $\times$ SO(6,$n$) according to \eqref{xdec}-\eqref{fdec}. We take advantage of the fact that the decomposition of $\theta_{\alpha M}$ contains an $\text{SO}(5,n-1)$ singlet, which, if non-vanishing, enables us to explicitly solve the quadratic identities \eqref{q2}-\eqref{q11}. We then write down the decomposition of each of these constraints with respect to $\text{SO}(1,1)_B \times  \text{SO}(1,1)_A \times \text{SO}(5,n-1)$ writing next to each of the resulting identities its weights $w_B$ and $w_A$ with respect to $\text{SO}(1,1)_B$ and $\text{SO}(1,1)_A$ respectively in the form $(w_B,w_A)$. We have: 

For \eqref{q2}:
\begin{align}
    & \xi_{+ \oplus} \theta_{- \oplus} - \xi_{- \oplus} \theta_{+ \oplus} = 0 \, , \qquad  \label{q2,1} \mathbf{(0,2)} \\
    & \theta_{- \oplus} \xi_{+ \hat{m}} + \xi_{+ \oplus} \theta_{- \hat{m}} - \theta_{+ \oplus} \xi_{- \hat{m}} - \xi_{- \oplus} \theta_{+ \hat{m}} = 0 \, ,\qquad  \label{q2,2}  \mathbf{(0,1)} \\
    & \xi_{+ \oplus} \theta_{- \ominus} + \xi_{+ \ominus} \theta_{- \oplus} - \xi_{- \oplus} \theta_{+ \ominus} - \xi_{- \ominus} \theta_{+ \oplus} = 0 \, ,\qquad   \label{q2,3} \mathbf{(0,0)} \\ 
   & \xi_{+ (\hat{m}} \theta_{- \hat{n})} - \xi_{- (\hat{m}} \theta_{+ \hat{n})} = 0 \, ,\qquad \label{q2,4} \mathbf{(0,0)} \\
    & \theta_{- \ominus} \xi_{+ \hat{m}} + \xi_{+ \ominus} \theta_{- \hat{m}} - \theta_{+ \ominus} \xi_{- \hat{m}} - \xi_{- \ominus} \theta_{+ \hat{m}} = 0 \, ,\qquad \label{q2,5} \mathbf{(0,-1)} \\
    & \xi_{+ \ominus} \theta_{- \ominus} - \xi_{- \ominus} \theta_{+ \ominus} = 0 \, .\qquad \label{q2,6} \mathbf{(0,-2)} 
\end{align}

For \eqref{q3}:
\begin{align}
    & - \theta_{+}^{\hat{n}} f_{- \oplus \hat{m} \hat{n} } + \theta_{+ \oplus} f_{- \oplus \ominus \hat{m}} + \theta_{-}^{\hat{n}} f_{+ \oplus \hat{m} \hat{n}} - \theta_{- \oplus} f_{+ \oplus \ominus \hat{m}} \nonumber \\ & + \frac{1}{2} \theta_{- \oplus} \xi_{+ \hat{m}} - \frac{1}{2} \xi_{+ \oplus} \theta_{- \hat{m}} - \frac{1}{2} \theta_{+ \oplus} \xi_{- \hat{m}} +  \frac{1}{2} \xi_{- \oplus} \theta_{+ \hat{m}}\qquad \label{q3,1} \mathbf{(0,1)}  \\
    & - 3 \theta_{- \oplus} \theta_{+ \hat{m}} + 3 \theta_{+ \oplus} \theta_{- \hat{m}} = 0  \, ,  \nonumber \\
    \, \nonumber \\ 
    &  \theta_{+}^{\hat{m}} f_{- \oplus \ominus \hat{m}} - \theta_{-}^{\hat{m}} f_{+ \oplus \ominus \hat{m}} + \frac{1}{2} \xi_{+ \oplus} \theta_{- \ominus} - \frac{1}{2} \xi_{+ \ominus} \theta_{- \oplus} - \frac{1}{2} \xi_{- \oplus} \theta_{+ \ominus} + \frac{1}{2} \xi_{- \ominus} \theta_{+ \oplus} \nonumber \\
    & - 3 \theta_{+ \oplus} \theta_{ - \ominus} + 3 \theta_{- \oplus} \theta_{+ \ominus} = 0 \, , \qquad  \label{q3,2} \mathbf{(0,0)} \\
    \, \nonumber \\ 
     & \theta_{+}^{\hat{p}} f_{- \hat{m} \hat{n} \hat{p}} + \theta_{+ \oplus} f_{- \ominus \hat{m} \hat{n}} + \theta_{+ \ominus} f_{- \oplus \hat{m} \hat{n}} - \theta_{-}^{\hat{p}} f_{+ \hat{m} \hat{n} \hat{p}} - \theta_{- \oplus} f_{+ \ominus \hat{m} \hat{n}} - \theta_{- \ominus} f_{+ \oplus \hat{m} \hat{n}} \nonumber \\
     & + \xi_{+ [\hat{m}} \theta_{- \hat{n}]} - \xi_{- [\hat{m}} \theta_{+ \hat{n}]} - 6 \theta_{+ [\hat{m}} \theta_{- \hat{n}]} = 0 \, , \qquad 
     \label{q3,3}   \mathbf{(0,0)} \\
     & \, \nonumber \\
     & - \theta_{+}^{\hat{n}} f_{- \ominus \hat{m} \hat{n}} - \theta_{+ \ominus} f_{- \oplus \ominus \hat{m}} + \theta_{-}^{\hat{n}} f_{+ \ominus \hat{m} \hat{n}} + \theta_{- \ominus} f_{+ \oplus \ominus \hat{m}} \nonumber \\ 
     & + \frac{1}{2} \theta_{- \ominus} \xi_{+ \hat{m}} - \frac{1}{2} \xi_{+ \ominus} \theta_{- \hat{m}} - \frac{1}{2} \theta_{+ \ominus} \xi_{- \hat{m}} + \frac{1}{2} \xi_{- \ominus} \theta_{+ \hat{m}} \qquad \mathbf{(0,-1)} \label{q3,4} \\
     & - 3 \theta_{- \ominus} \theta_{+ \hat{m}} + 3 \theta_{+ \ominus} \theta_{- \hat{m}} = 0 \, . \nonumber 
\end{align}

For \eqref{q4}:
\begin{align}
    & - \theta_{+}^{\hat{n}} f_{+ \oplus \hat{m} \hat{n}} + \theta_{+ \oplus} f_{+ \oplus \ominus \hat{m}} + \frac{1}{2} \theta_{+ \oplus} \xi_{+ \hat{m}} - \frac{1}{2} \xi_{+ \oplus} \theta_{+ \hat{m}} = 0 \, ,\qquad \label{q4,1} \mathbf{(2,1)} \\
    & \theta_{+}^{\hat{m}} f_{+ \oplus \ominus \hat{m}} + \frac{1}{2} \xi_{+ \oplus} \theta_{+ \ominus} - \frac{1}{2} \xi_{+ \ominus} \theta_{+ \oplus} = 0 \, , \qquad \mathbf{(2,0)} \label{q4,2} \\ 
    & \theta_{+}^{\hat{p}} f_{+ \hat{m} \hat{n} \hat{p}} + \theta_{+ \oplus}
    f_{+ \ominus \hat{m} \hat{n}} + \theta_{+ \ominus}
    f_{+ \oplus \hat{m} \hat{n}}  + \xi_{+ [\hat{m}} \theta_{+ \hat{n}]} = 0 \, ,\qquad \mathbf{(2,0)} \label{q4,3} \\
    & \theta_{+}^{\hat{n}} f_{+ \ominus \hat{m} \hat{n}} + \theta_{+ \ominus} f_{+ \oplus \ominus \hat{m}} - \frac{1}{2} \theta_{+ \ominus} \xi_{+ \hat{m}}  + \frac{1}{2} \xi_{+ \ominus} \theta_{+ \hat{m}} = 0 \, ,\label{q4,4} \qquad \mathbf{(2,-1)} \\
    & \, \nonumber \\
     & - \theta_{+}^{\hat{n}} f_{- \oplus \hat{m} \hat{n} } + \theta_{+ \oplus} f_{- \oplus \ominus \hat{m}} -\theta_{-}^{\hat{n}} f_{+ \oplus \hat{m} \hat{n}} + \theta_{- \oplus} f_{+ \oplus \ominus \hat{m}} \nonumber \\ & + \frac{1}{2} \theta_{- \oplus} \xi_{+ \hat{m}} - \frac{1}{2} \xi_{+ \oplus} \theta_{- \hat{m}} + \frac{1}{2} \theta_{+ \oplus} \xi_{- \hat{m}} -  \frac{1}{2} \xi_{- \oplus} \theta_{+ \hat{m}} = 0 \, ,\qquad \label{q4,5} \mathbf{(0,1)}  \\
       & \, \nonumber \\ 
    &  \theta_{+}^{\hat{m}} f_{- \oplus \ominus \hat{m}} + \theta_{-}^{\hat{m}} f_{+ \oplus \ominus \hat{m}} \nonumber \\ & + \frac{1}{2} \xi_{+ \oplus} \theta_{- \ominus} - \frac{1}{2} \xi_{+ \ominus} \theta_{- \oplus} + \frac{1}{2} \xi_{- \oplus} \theta_{+ \ominus} - \frac{1}{2} \xi_{- \ominus} \theta_{+ \oplus}  = 0  \, ,\qquad \label{q4,6} \mathbf{(0,0)} \\
    \, \nonumber \\ 
     & \theta_{+}^{\hat{p}} f_{- \hat{m} \hat{n} \hat{p}} + \theta_{+ \oplus} f_{- \ominus \hat{m} \hat{n}} + \theta_{+ \ominus} f_{- \oplus \hat{m} \hat{n}} + \theta_{-}^{\hat{p}} f_{+ \hat{m} \hat{n} \hat{p}} + \theta_{- \oplus} f_{+ \ominus \hat{m} \hat{n}} + \theta_{- \ominus} f_{+ \oplus \hat{m} \hat{n}} \nonumber \\
     & + \xi_{+ [\hat{m}} \theta_{- \hat{n}]} + \xi_{- [\hat{m}} \theta_{+ \hat{n}]} = 0 \, ,\qquad \label{q4,7} \mathbf{(0,0)} \\
     & \, \nonumber \\
     &  \theta_{+}^{\hat{n}} f_{- \ominus \hat{m} \hat{n}} + \theta_{+ \ominus} f_{- \oplus \ominus \hat{m}} + \theta_{-}^{\hat{n}} f_{+ \ominus \hat{m} \hat{n}} + \theta_{- \ominus} f_{+ \oplus \ominus \hat{m}} \nonumber \\ 
     & - \frac{1}{2} \theta_{- \ominus} \xi_{+ \hat{m}} + \frac{1}{2} \xi_{+ \ominus} \theta_{- \hat{m}} - \frac{1}{2} \theta_{+ \ominus} \xi_{- \hat{m}} + \frac{1}{2} \xi_{- \ominus} \theta_{+ \hat{m}} = 0 \, ,\qquad \mathbf{(0,-1)} \label{q4,8} \\
     & \, \nonumber \\
      & - \theta_{-}^{\hat{n}} f_{- \oplus \hat{m} \hat{n}} + \theta_{- \oplus} f_{- \oplus \ominus \hat{m}} + \frac{1}{2} \theta_{- \oplus} \xi_{- \hat{m}} - \frac{1}{2} \xi_{- \oplus} \theta_{- \hat{m}} = 0 \, ,\qquad \mathbf{(-2,1)}  \label{q4,9} \\
    & \theta_{-}^{\hat{m}} f_{- \oplus \ominus \hat{m}} + \frac{1}{2} \xi_{- \oplus} \theta_{- \ominus} - \frac{1}{2} \xi_{- \ominus} \theta_{-  \oplus} = 0 \, ,\qquad \mathbf{(-2,0)} \label{q4,10} \\ 
    & \theta_{-}^{\hat{p}} f_{- \hat{m} \hat{n} \hat{p}} + \theta_{- \oplus}
    f_{- \ominus \hat{m} \hat{n}} + \theta_{- \ominus}
    f_{- \oplus \hat{m} \hat{n}}  + \xi_{- [\hat{m}} \theta_{- \hat{n}]} = 0 \, ,\qquad \mathbf{(-2,0)}  \label{q4,11} \\
    & \theta_{-}^{\hat{n}} f_{- \ominus \hat{m} \hat{n}} + \theta_{- \ominus} f_{- \oplus \ominus \hat{m}} - \frac{1}{2} \theta_{- \ominus} \xi_{- \hat{m}}  + \frac{1}{2} \xi_{- \ominus} \theta_{- \hat{m}} = 0 \, .\label{q4,12} \qquad \mathbf{(-2,-1)}
\end{align}

For \eqref{q5}:
\begin{align}
    & \xi_{+}^{\hat{m}} \theta_{+ \hat{m}} + \xi_{+ \oplus} \theta_{+ \ominus} + \xi_{+ \ominus} \theta_{+ \oplus} + \theta^{\hat{m}}_{+} \theta_{+ \hat{m}} + 2 \theta_{+ \oplus} \theta_{+ \ominus} = 0 \, ,\qquad \label{q5,1} \mathbf{(2,0)} \\
    & \, \nonumber \\ 
    & \xi_{+}^{\hat{m}} \theta_{- \hat{m}} + \xi_{+ \oplus} \theta_{- \ominus} + \xi_{+ \ominus} \theta_{- \oplus} + \xi_{-}^{\hat{m}} \theta_{+ \hat{m}} + \xi_{- \oplus} \theta_{+ \ominus} + \xi_{- \ominus} \theta_{+ \oplus} \nonumber  \\
    & + 2 \theta_{+}^{\hat{m}} \theta_{- \hat{m}} + 2 \theta_{+ \oplus} \theta_{- \ominus} + 2 \theta_{+ \ominus} \theta_{- \oplus} = 0 \, ,\qquad \mathbf{(0,0)} \label{q5,2} \\
    & \, \nonumber \\
    &  \xi_{-}^{\hat{m}} \theta_{- \hat{m}} + \xi_{- \oplus} \theta_{- \ominus} + \xi_{- \ominus} \theta_{- \oplus} + \theta^{\hat{m}}_{-} \theta_{- \hat{m}} + 2 \theta_{- \oplus} \theta_{- \ominus} = 0 \, .\qquad \mathbf{(-2,0)}  \label{q5,3} 
\end{align}

For \eqref{q6}:
\begin{align}
     & - \xi_{+}^{\hat{n}} f_{+ \oplus \hat{m} \hat{n}} + \xi_{+ \oplus} f_{+ \oplus \ominus \hat{m}} - \frac{1}{2} \theta_{+ \oplus} \xi_{+ \hat{m}} + \frac{1}{2} \xi_{+ \oplus} \theta_{+ \hat{m}} = 0 \, ,\qquad \mathbf{(2,1)} \label{q6,1} \\
    & \xi_{+}^{\hat{m}} f_{+ \oplus \ominus \hat{m}} - \frac{1}{2} \xi_{+ \oplus} \theta_{+ \ominus} + \frac{1}{2} \xi_{+ \ominus} \theta_{+ \oplus} = 0 \, ,\qquad \mathbf{(2,0)} \label{q6,2} \\ 
    & \xi_{+}^{\hat{p}} f_{+ \hat{m} \hat{n} \hat{p}} + \xi_{+ \oplus}
    f_{+ \ominus \hat{m} \hat{n}} + \xi_{+ \ominus}
    f_{+ \oplus \hat{m} \hat{n}}  - \xi_{+ [\hat{m}} \theta_{+ \hat{n}]} = 0 \, ,\qquad \mathbf{(2,0)} \label{q6,3} \\
    & \xi_{+}^{\hat{n}} f_{+ \ominus \hat{m} \hat{n}} + \xi_{+ \ominus} f_{+ \oplus \ominus \hat{m}} + \frac{1}{2} \theta_{+ \ominus} \xi_{+ \hat{m}}  - \frac{1}{2} \xi_{+ \ominus} \theta_{+ \hat{m}} = 0 \, ,\qquad \mathbf{(2,-1)} \label{q6,4} \\
    & \, \nonumber \\
     & - \xi_{+}^{\hat{n}} f_{- \oplus \hat{m} \hat{n} } + \xi_{+ \oplus} f_{- \oplus \ominus \hat{m}} -\xi_{-}^{\hat{n}} f_{+ \oplus \hat{m} \hat{n}} + \xi_{- \oplus} f_{+ \oplus \ominus \hat{m}} \nonumber \\ & - \frac{1}{2} \theta_{- \oplus} \xi_{+ \hat{m}} + \frac{1}{2} \xi_{+ \oplus} \theta_{- \hat{m}} - \frac{1}{2} \theta_{+ \oplus} \xi_{- \hat{m}} +  \frac{1}{2} \xi_{- \oplus} \theta_{+ \hat{m}} = 0 \, ,\qquad \label{q6,5} \mathbf{(0,1)}  \\
       & \, \nonumber \\ 
    &  \xi_{+}^{\hat{m}} f_{- \oplus \ominus \hat{m}} + \xi_{-}^{\hat{m}} f_{+ \oplus \ominus \hat{m}} \nonumber \\ & - \frac{1}{2} \xi_{+ \oplus} \theta_{- \ominus} + \frac{1}{2} \xi_{+ \ominus} \theta_{- \oplus} - \frac{1}{2} \xi_{- \oplus} \theta_{+ \ominus} + \frac{1}{2} \xi_{- \ominus} \theta_{+ \oplus}  = 0 \, , \qquad \label{q6,6} \mathbf{(0,0)} \\
    \, \nonumber \\ 
     & \xi_{+}^{\hat{p}} f_{- \hat{m} \hat{n} \hat{p}} + \xi_{+ \oplus} f_{- \ominus \hat{m} \hat{n}} + \xi_{+ \ominus} f_{- \oplus \hat{m} \hat{n}} + \xi_{-}^{\hat{p}} f_{+ \hat{m} \hat{n} \hat{p}} + \xi_{- \oplus} f_{+ \ominus \hat{m} \hat{n}} + \xi_{- \ominus} f_{+ \oplus \hat{m} \hat{n}} \nonumber \\
     & - \xi_{+ [\hat{m}} \theta_{- \hat{n}]} - \xi_{- [\hat{m}} \theta_{+ \hat{n}]} = 0 \, ,\qquad \mathbf{(0,0)} \label{q6,7} \\
     & \, \nonumber \\
     &  \xi_{+}^{\hat{n}} f_{- \ominus \hat{m} \hat{n}} + \xi_{+ \ominus} f_{- \oplus \ominus \hat{m}} + \xi_{-}^{\hat{n}} f_{+ \ominus \hat{m} \hat{n}} + \xi_{- \ominus} f_{+ \oplus \ominus \hat{m}} \nonumber \\ 
     & + \frac{1}{2} \theta_{- \ominus} \xi_{+ \hat{m}} - \frac{1}{2} \xi_{+ \ominus} \theta_{- \hat{m}} + \frac{1}{2} \theta_{+ \ominus} \xi_{- \hat{m}} - \frac{1}{2} \xi_{- \ominus} \theta_{+ \hat{m}} = 0 \, ,\qquad \mathbf{(0,-1)} \label{q6,8} \\
     & \, \nonumber \\
      & - \xi_{-}^{\hat{n}} f_{- \oplus \hat{m} \hat{n}} + \xi_{- \oplus} f_{- \oplus \ominus \hat{m}} - \frac{1}{2} \theta_{- \oplus} \xi_{- \hat{m}} + \frac{1}{2} \xi_{- \oplus} \theta_{- \hat{m}} = 0 \, ,\qquad \mathbf{(-2,1)}   \label{q6,9} \\
    & \xi_{-}^{\hat{m}} f_{- \oplus \ominus \hat{m}} - \frac{1}{2} \xi_{- \oplus} \theta_{- \ominus} + \frac{1}{2} \xi_{- \ominus} \theta_{- \oplus} = 0 \, ,\qquad \mathbf{(-2,0)}  \label{q6,10} \\ 
    & \xi_{-}^{\hat{p}} f_{- \hat{m} \hat{n} \hat{p}} + \xi_{- \oplus}
    f_{- \ominus \hat{m} \hat{n}} + \xi_{- \ominus}
    f_{- \oplus \hat{m} \hat{n}}  - \xi_{- [\hat{m}} \theta_{- \hat{n}]} = 0\, ,\qquad \mathbf{(-2,0)} \label{q6,11} \\
    & \xi_{-}^{\hat{n}} f_{- \ominus \hat{m} \hat{n}} + \xi_{- \ominus} f_{- \oplus \ominus \hat{m}} + \frac{1}{2} \theta_{- \ominus} \xi_{- \hat{m}}  - \frac{1}{2} \xi_{- \ominus} \theta_{- \hat{m}} = 0 \, . \qquad \mathbf{(-2,-1)} \label{q6,12} 
     \end{align}
For \eqref{q7}:
\begin{align}
    & \xi_{+}^{\hat{m}} \theta_{+ \hat{m}} + \xi_{+ \oplus} \theta_{+ \ominus} + \xi_{+ \ominus} \theta_{+ \oplus} + \xi^{\hat{m}}_{+} \xi_{+ \hat{m}} + 2 \xi_{+ \oplus} \xi_{+ \ominus} = 0 \, ,\qquad \mathbf{(2,0)} \label{q7,1}\\
    & \, \nonumber \\ 
    & \xi_{+}^{\hat{m}} \theta_{- \hat{m}} + \xi_{+ \oplus} \theta_{- \ominus} + \xi_{+ \ominus} \theta_{- \oplus} + \xi_{-}^{\hat{m}} \theta_{+ \hat{m}} + \xi_{- \oplus} \theta_{+ \ominus} + \xi_{- \ominus} \theta_{+ \oplus} \nonumber \\
    & + 2 \xi_{+}^{\hat{m}} \xi_{- \hat{m}} + 2 \xi_{+ \oplus} \xi_{- \ominus} + 2 \xi_{+ \ominus} \xi_{- \oplus} = 0 \, ,\qquad \mathbf{(0,0)} \label{q7,2} \\
    & \, \nonumber \\
    &  \xi_{-}^{\hat{m}} \theta_{- \hat{m}} + \xi_{- \oplus} \theta_{- \ominus} + \xi_{- \ominus} \theta_{- \oplus} + \xi^{\hat{m}}_{-} \xi_{- \hat{m}} + 2 \xi_{- \oplus} \xi_{- \ominus} = 0 \, . \qquad \mathbf{(-2,0)} \label{q7,3}
\end{align}
For \eqref{q8}:
\begin{align}
    & - \xi_{+}^{\hat{n}} f_{- \oplus \hat{m} \hat{n}} + \xi_{ + \oplus} f_{- \oplus \ominus \hat{m}} + \xi_{-}^{\hat{n}} f_{+ \oplus \hat{m} \hat{n}} - \xi_{ - \oplus} f_{+ \oplus \ominus \hat{m}} + \xi_{- \oplus} \xi_{+ \hat{m}} - \xi_{+ \oplus} \xi_{- \hat{m}} \nonumber \\
    & - \frac{3}{2} \theta_{- \oplus} \xi_{+ \hat{m}} + \frac{3}{2} \xi_{+ \oplus} \theta_{- \hat{m}} + \frac{3}{2} \theta_{+ \oplus} \xi_{- \hat{m}} - \frac{3}{2} \xi_{- \oplus} \theta_{+ \hat{m}} = 0 \, ,\qquad \label{q8,1} \mathbf{(0,1)} \\ &  \, \nonumber \\
    & \xi_{+}^{\hat{m}} f_{- \oplus \ominus \hat{m}} - \xi_{-}^{\hat{m}} f_{+\oplus \ominus \hat{m}} + \xi_{+ \oplus} \xi_{- \ominus} - \xi_{- \oplus} \xi_{+ \ominus} \nonumber \\
    & - \frac{3}{2} \xi_{+ \oplus} \theta_{- \ominus} + \frac{3}{2} \xi_{+ \ominus} \theta_{- \oplus} +  \frac{3}{2} \xi_{- \oplus} \theta_{+ \ominus} - \frac{3}{2} \xi_{- \ominus} \theta_{+ \oplus}  = 0 \, ,\qquad \mathbf{(0,0)} \label{q8,2} \\
    & \, \nonumber \\
    &  \xi_{+}^{\hat{p}} f_{- \hat{m} \hat{n} \hat{p}} + \xi_{+ \oplus} f_{- \ominus \hat{m} \hat{n}} + \xi_{+ \ominus} f_{- \oplus \hat{m} \hat{n}} - \xi_{-}^{\hat{p}} f_{+ \hat{m} \hat{n} \hat{p}} - \xi_{- \oplus} f_{+ \ominus \hat{m} \hat{n}} - \xi_{- \ominus} f_{+ \oplus \hat{m} \hat{n}} \nonumber \\
    & + 2 \xi_{+ [\hat{m}} \xi_{- \hat{n}]} - 3 \xi_{+ [\hat{m}} \theta_{- \hat{n}]} + 3 \xi_{- [\hat{m}} \theta_{+ \hat{n}]}  = 0 \, ,\qquad \mathbf{(0,0)} \label{q8,3} \\
    & \, \nonumber \\
    & -\xi_{+}^{\hat{n}} f_{- \ominus \hat{m} \hat{n}} -  \xi_{ + \ominus} f_{- \oplus \ominus \hat{m}} + \xi_{-}^{\hat{n}} f_{+ \ominus \hat{m} \hat{n}} + \xi_{ - \ominus} f_{+ \oplus \ominus \hat{m}} + \xi_{- \ominus} \xi_{+ \hat{m}} - \xi_{+ \ominus} \xi_{- \hat{m}} \nonumber \\
    & - \frac{3}{2} \theta_{- \ominus} \xi_{+ \hat{m}} + \frac{3}{2} \xi_{+ \ominus} \theta_{- \hat{m}} + \frac{3}{2} \theta_{+ \ominus} \xi_{- \hat{m}} - \frac{3}{2}  \xi_{- \ominus} \theta_{+ \hat{m}} = 0 \, .\qquad \mathbf{(0,-1)} \label{q8,4}
\end{align}
For \eqref{q9}:
\begin{align}
    & 6 f_{+ \hat{r} [\hat{m} \hat{n}} f_{+ \hat{p}] \oplus}{}^{\hat{r}} + 6 f_{+ \oplus [\hat{m} \hat{n}} f_{+ \hat{p}] \oplus \ominus} \nonumber \\ 
    & + 3 \left( 
    \xi_{+[\hat{m}} + \theta_{+ [\hat{m}}    \right) f_{+ \hat{n} \hat{p}] \oplus} - \left( \xi_{+ \oplus} + \theta_{+ \oplus} \right) f_{+ \hat{m} \hat{n} \hat{p}} = 0 \, ,\qquad \mathbf{(2,1)} \label{q9,1} \\
    & \, \nonumber \\
    & 2 f_{+ \hat{m} \hat{n} \hat{p}} f_{+ \oplus \ominus}{}^{\hat{p}} + 4 f_{+ \hat{p} \oplus [\hat{m}} f_{+ \hat{n}] \ominus}{}^{\hat{p}}  + 2 \left( \xi_{+ [\hat{m}} + \theta_{+ [\hat{m}} \right) f_{+ \hat{n}] \oplus \ominus} \nonumber \\ 
    &  + \left( \xi_{+ \oplus} + \theta_{+ \oplus} \right) f_{+ \ominus \hat{m} \hat{n}} - \left( \xi_{+ \ominus} + \theta_{+ \ominus} \right) f_{+ \oplus \hat{m} \hat{n}} = 0 \, ,\qquad \mathbf{(2,0)} \label{q9,2} \\
    & \, \nonumber \\
    & 3 f_{+ \hat{r} [\hat{m} \hat{n}} f_{+ \hat{p} \hat{q}]}{}^{\hat{r}} + 6 f_{+ \oplus [\hat{m} \hat{n}} f_{+ \hat{p} \hat{q}] \ominus} + 2 \left( 
     \xi_{+[\hat{m}} + \theta_{+ [\hat{m}}  \right) f_{+ \hat{n} \hat{p} \hat{q}]} = 0 \, , \qquad \mathbf{(2,0)} \label{q9,3} \\
     & \, \nonumber \\
     & 6 f_{+ \hat{r} [\hat{m} \hat{n}} f_{+ \hat{p}] \ominus}{}^{\hat{r}} - 6 f_{+ \ominus [\hat{m} \hat{n}} f_{+ \hat{p}] \oplus \ominus} \nonumber \\
     & +  3 \left( 
    \xi_{+[\hat{m}} + \theta_{+ [\hat{m}}    \right) f_{+ \hat{n} \hat{p}] \ominus} - \left( \xi_{+ \ominus} + \theta_{+ \ominus} \right) f_{+ \hat{m} \hat{n} \hat{p}} = 0 \, ,\qquad \mathbf{(2,-1)} \label{q9,4} \\
    & \, \nonumber \\
    & 6 f_{+ \hat{r}[\hat{m} \hat{n}} f_{- \hat{p}] \oplus}{}^{\hat{r}} + 6 f_{+ \oplus [\hat{m} \hat{n}} f_{- \hat{p}] \oplus \ominus} + 6 f_{- \hat{r}[\hat{m} \hat{n}} f_{+ \hat{p}] \oplus}{}^{\hat{r}}  + 6 f_{- \oplus [\hat{m} \hat{n}} f_{+ \hat{p}] \oplus \ominus} \nonumber \\
    & + 3 \left( \xi_{+ [\hat{m}}  +\theta_{+ [\hat{m}} \right) f_{- \hat{n} \hat{p}] \oplus} - \left( \xi_{+ \oplus} + \theta_{+ \oplus} \right) f_{- \hat{m} \hat{n} \hat{p}} \qquad \mathbf{(0,1)} \label{q9,5} \\
    & + 3 \left( \xi_{- [\hat{m}}  +\theta_{- [\hat{m}} \right) f_{+ \hat{n} \hat{p}] \oplus} - \left( \xi_{- \oplus} + \theta_{- \oplus} \right) f_{+ \hat{m} \hat{n} \hat{p}} = 0 \, ,\nonumber \\
    & \, \nonumber \\
    & f_{+ \hat{m} \hat{n} \hat{p}} f_{- \oplus \ominus}{}^{\hat{p}} + f_{- \hat{m} \hat{n} \hat{p}} f_{+ \oplus \ominus}{}^{\hat{p}} + 2 f_{+ \hat{p} \oplus [\hat{m}} f_{- \hat{n}] \ominus}{}^{\hat{p}} + 2 f_{- \hat{p} \oplus [\hat{m}} f_{+ \hat{n}] \ominus}{}^{\hat{p}} \nonumber \\
    & + \left( \xi_{+ [\hat{m}} + \theta_{+ [\hat{m}} \right) f_{- \hat{n}] \oplus \ominus} +  \left( \xi_{- [\hat{m}} + \theta_{- [\hat{m}} \right) f_{+ \hat{n}] \oplus \ominus} + \frac{1}{2} \left( \xi_{+ \oplus} + \theta_{+ \oplus} \right) f_{- \ominus \hat{m} \hat{n}}  \qquad \mathbf{(0,0)} \label{q9,6} \\
    &  + \frac{1}{2} \left( \xi_{- \oplus} + \theta_{- \oplus} \right) f_{+ \ominus \hat{m} \hat{n}} - \frac{1}{2} \left( \xi_{+ \ominus} + \theta_{+ \ominus} \right) f_{- \oplus \hat{m} \hat{n}} - \frac{1}{2} \left( \xi_{- \ominus} + \theta_{- \ominus} \right) f_{+ \oplus \hat{m} \hat{n}} = 0 \, ,\nonumber \\
    & \, \nonumber \\
    & 3 f_{+ \hat{r} [\hat{m} \hat{n}} f_{- \hat{p} \hat{q}]}{}^{\hat{r}} + 3 f_{+ \oplus [\hat{m} \hat{n}} f_{- \hat{p} \hat{q}] \ominus} + 3 f_{+ \ominus [\hat{m} \hat{n}} f_{- \hat{p} \hat{q}] \oplus} \nonumber \\
    & + \left( \xi_{+ [\hat{m}} + \theta_{+ [\hat{m}} \right) f_{- \hat{n} \hat{p} \hat{q}]} + \left( \xi_{- [\hat{m}} + \theta_{- [\hat{m}} \right) f_{+ \hat{n} \hat{p} \hat{q}]} = 0 \, ,\qquad \mathbf{(0,0)} \label{q9,7} \\
    & \, \nonumber \\
    &  6 f_{+ \hat{r}[\hat{m} \hat{n}} f_{- \hat{p}] \ominus}{}^{\hat{r}} - 6 f_{+ \ominus [\hat{m} \hat{n}} f_{- \hat{p}] \oplus \ominus} + 6 f_{- \hat{r}[\hat{m} \hat{n}} f_{+ \hat{p}] \ominus}{}^{\hat{r}}  - 6 f_{- \ominus [\hat{m} \hat{n}} f_{+ \hat{p}] \oplus \ominus} \nonumber \\
    & + 3 \left( \xi_{+ [\hat{m}}  +\theta_{+ [\hat{m}} \right) f_{- \hat{n} \hat{p}] \ominus} - \left( \xi_{+ \ominus} + \theta_{+ \ominus} \right) f_{- \hat{m} \hat{n} \hat{p}} \qquad \mathbf{(0,-1)} \label{q9,8} \\
    & + 3 \left( \xi_{- [\hat{m}}  +\theta_{- [\hat{m}} \right) f_{+ \hat{n} \hat{p}] \ominus} - \left( \xi_{- \ominus} + \theta_{- \ominus} \right) f_{+ \hat{m} \hat{n} \hat{p}} = 0 \, ,\nonumber \\
    & \, \nonumber \\
    & 6 f_{- \hat{r} [\hat{m} \hat{n}} f_{- \hat{p}] \oplus}{}^{\hat{r}} + 6 f_{- \oplus [\hat{m} \hat{n}} f_{- \hat{p}] \oplus \ominus} \nonumber \\ 
    & + 3 \left( 
    \xi_{-[\hat{m}} + \theta_{- [\hat{m}}    \right) f_{- \hat{n} \hat{p}] \oplus} - \left( \xi_{- \oplus} + \theta_{- \oplus} \right) f_{- \hat{m} \hat{n} \hat{p}} = 0 \, ,\qquad \mathbf{(-2,1)} \label{q9,9} \\
    & \, \nonumber \\
    & 2 f_{- \hat{m} \hat{n} \hat{p}} f_{- \oplus \ominus}{}^{\hat{p}} + 4 f_{- \hat{p} \oplus [\hat{m}} f_{- \hat{n}] \ominus}{}^{\hat{p}}  + 2 \left( \xi_{- [\hat{m}} + \theta_{- [\hat{m}} \right) f_{- \hat{n}] \oplus \ominus} \nonumber \\ 
    &  + \left( \xi_{- \oplus} + \theta_{- \oplus} \right) f_{- \ominus \hat{m} \hat{n}} - \left( \xi_{- \ominus} + \theta_{- \ominus} \right) f_{- \oplus \hat{m} \hat{n}} = 0 \, ,\qquad \mathbf{(-2,0)} \label{q9,10} \\
    & \, \nonumber \\
    & 3 f_{- \hat{r} [\hat{m} \hat{n}} f_{- \hat{p} \hat{q}]}{}^{\hat{r}} + 6 f_{- \oplus [\hat{m} \hat{n}} f_{- \hat{p} \hat{q}] \ominus} + 2 \left( 
     \xi_{-[\hat{m}} + \theta_{- [\hat{m}}  \right) f_{- \hat{n} \hat{p} \hat{q}]} = 0 \, , \qquad \mathbf{(-2,0)} \label{q9,11} \\
     & \, \nonumber \\
     & 6 f_{- \hat{r} [\hat{m} \hat{n}} f_{- \hat{p}] \ominus}{}^{\hat{r}} - 6 f_{- \ominus [\hat{m} \hat{n}} f_{- \hat{p}] \oplus \ominus} \nonumber \\
     & +  3 \left( 
    \xi_{-[\hat{m}} + \theta_{- [\hat{m}}    \right) f_{- \hat{n} \hat{p}] \ominus} - \left( \xi_{- \ominus} + \theta_{- \ominus} \right) f_{- \hat{m} \hat{n} \hat{p}} = 0 \, . \qquad \label{q9,12} \mathbf{(-2,-1)}
     \end{align}
For \eqref{q10}:     
\begin{align}
    & 3 \theta_{+ [\hat{m}} f_{- \hat{n} \hat{p}] \oplus} - 3 \theta_{- [\hat{m}} f_{+ \hat{n} \hat{p}] \oplus} - \theta_{+ \oplus} f_{- \hat{m} \hat{n} \hat{p}} + \theta_{- \oplus} f_{+ \hat{m} \hat{n} \hat{p}} = 0 \, ,\qquad \mathbf{(0,1)} \label{q10,1} \\
    & 2 \theta_{+ [\hat{m}} f_{- \hat{n}] \oplus \ominus} - 2 \theta_{- [\hat{m}} f_{+\hat{n}] \oplus \ominus} + \theta_{+ 
    \oplus} f_{- \ominus \hat{m} \hat{n}  } - \theta_{- 
    \oplus} f_{+ \ominus \hat{m} \hat{n}  } \nonumber \\ &  - \theta_{+ 
    \ominus} f_{- \oplus \hat{m} \hat{n}  } + \theta_{- 
    \ominus} f_{+ \oplus \hat{m} \hat{n}  } = 0 \, , \qquad \mathbf{(0,0)} \label{q10,2}  \\
    & \theta_{+ [\hat{m}} f_{- \hat{n} \hat{p} \hat{q}]} - \theta_{- [\hat{m}} f_{+ \hat{n} \hat{p} \hat{q}]} = 0 \, ,\qquad \mathbf{(0,0)} \label{q10,3} \\
    & 3 \theta_{+ [\hat{m}} f_{- \hat{n} \hat{p}] \ominus} - 3 \theta_{- [\hat{m}} f_{+ \hat{n} \hat{p}] \ominus} - \theta_{+ \ominus} f_{- \hat{m} \hat{n} \hat{p}} + \theta_{- \ominus} f_{+ \hat{m} \hat{n} \hat{p}} = 0 \, .\qquad \mathbf{(0,-1)} \label{q10,4} 
\end{align}
For \eqref{q11}:
\begin{align}
    & 2 f_{+ \oplus \hat{p} [\hat{m}} f_{- \hat{n}] \oplus}{}^{\hat{p}} + \left( \xi_{+ \oplus} - \theta_{+ \oplus} \right) f_{- \oplus \hat{m} \hat{n}} - \left( \xi_{- \oplus} - \theta_{- \oplus} \right) f_{+ \oplus \hat{m} \hat{n}} = 0 \, , \qquad \mathbf{(0,2)} \label{q11,1} \\ 
       & \, 
       \nonumber \\
        & - f_{+ \oplus \hat{m} \hat{n}} f_{- \oplus \ominus}{}^{\hat{n}} +  f_{- \oplus \hat{m} \hat{n}} f_{+ \oplus \ominus}{}^{\hat{n}} + \left( 
          \xi_{+ \oplus} - \theta_{+ \oplus} \right) f_{- \oplus \ominus \hat{m}} - \left( 
          \xi_{- \oplus} - \theta_{- \oplus} \right) f_{+ \oplus \ominus \hat{m}} \nonumber \\
          & - \frac{1}{4} \xi_{+ \oplus} \xi_{- \hat{m}} + \frac{1}{4} \xi_{- \oplus} \xi_{+ \hat{m}} + \frac{1}{4} \theta_{+ \oplus} \xi_{- \hat{m}} - \frac{1}{4} \theta_{- \oplus} \xi_{+ \hat{m}} \qquad \mathbf{(0,1)} \label{q11,2} \\
          & + \frac{1}{4} \xi_{+ \oplus} \theta_{- \hat{m}} - \frac{1}{4} \xi_{- \oplus} \theta_{+ \hat{m}} + \frac{3}{4} \theta_{+ \oplus} \theta_{- \hat{m}} - \frac{3}{4} \theta_{- \oplus} \theta_{+ \hat{m}} = 0 \, ,\nonumber  \\
          & \, \nonumber \\
          & f_{+ \hat{m} \hat{n} \hat{r}} f_{- \hat{p} \oplus}{}^{\hat{r}} -  f_{- \hat{m} \hat{n} \hat{r}} f_{+ \hat{p} \oplus}{}^{\hat{r}} + f_{+ \oplus \hat{m} \hat{n}} f_{- \oplus \ominus \hat{p}} -  f_{- \oplus \hat{m} \hat{n}} f_{+ \oplus \ominus \hat{p}} \nonumber \\ 
          & - \left( \xi_{+[\hat{m}} - \theta_{+ [\hat{m}} \right) f_{- \hat{n}] \hat{p} \oplus} +  \left( \xi_{-[\hat{m}} - \theta_{- [\hat{m}} \right) f_{+ \hat{n}] \hat{p} \oplus} + \frac{1}{2} \left( 
          \xi_{+ \hat{p}} - \theta_{+ \hat{p}}  \right) f_{- \oplus \hat{m} \hat{n}} \nonumber \\
          & - \frac{1}{2} \left( 
          \xi_{- \hat{p}} - \theta_{- \hat{p}}  \right) f_{+ \oplus \hat{m} \hat{n}} - \frac{1}{2} \left( \xi_{+ \oplus} - \theta_{+ \oplus} \right) f_{- \hat{m} \hat{n} \hat{p}} + \frac{1}{2} \left( \xi_{- \oplus} - \theta_{- \oplus} \right) f_{+ \hat{m} \hat{n} \hat{p}} \qquad \mathbf{(0,1)} \label{q11,3} \\
          & + \frac{1}{2} \xi_{+ \oplus} \xi_{- [\hat{m}} \eta_{\hat{n}] \hat{p}} - \frac{1}{2} \xi_{- \oplus} \xi_{+ [\hat{m}} \eta_{\hat{n}] \hat{p}} - \frac{1}{2} \theta_{+ \oplus} \xi_{- [\hat{m}} \eta_{\hat{n}] \hat{p}} + \frac{1}{2} \theta_{- \oplus} \xi_{+ [\hat{m}} \eta_{\hat{n}] \hat{p}} \nonumber \\
          & - \frac{1}{2} \xi_{+ \oplus} \theta_{- [\hat{m}} \eta_{\hat{n}]\hat{p}} + \frac{1}{2} \xi_{- \oplus} \theta_{+ [\hat{m}} \eta_{\hat{n}]\hat{p}} - \frac{3}{2} \theta_{+ \oplus} \theta_{-[\hat{m}} \eta_{\hat{n}] \hat{p}} + \frac{3}{2} \theta_{- \oplus} \theta_{+[\hat{m}} \eta_{\hat{n}] \hat{p}} = 0 \, ,\nonumber \\
          & \, \nonumber \\
          & f_{+ \hat{m} \hat{n} \hat{p}} f_{- \oplus \ominus}{}^{\hat{p}} - f_{- \hat{m} \hat{n} \hat{p}} f_{+ \oplus \ominus}{}^{\hat{p}} - \left( \xi_{+ [\hat{m}} - \theta_{+[\hat{m}} \right) f_{- \hat{n}] \oplus \ominus} +  \left( \xi_{- [\hat{m}} - \theta_{-[\hat{m}} \right) f_{+ \hat{n}] \oplus \ominus} \nonumber \\
          & + \frac{1}{2} \left(   \xi_{+ \oplus} - \theta_{+ \oplus} \right) f_{- \ominus \hat{m} \hat{n}} - \frac{1}{2} \left(   \xi_{- \oplus} - \theta_{- \oplus} \right) f_{+ \ominus \hat{m} \hat{n}} \qquad \mathbf{(0,0)} \label{q11,4} \\ 
          & - \frac{1}{2} \left( \xi_{+ \ominus} - \theta_{+ \ominus} \right) f_{- \oplus \hat{m} \hat{n}}  + \frac{1}{2} \left( \xi_{- \ominus} - \theta_{- \ominus} \right) f_{+ \oplus \hat{m} \hat{n}} = 0 \, , \nonumber \\ 
          & \, \nonumber \\
          & f_{+ \hat{m} \hat{n} \hat{r}} f_{- \hat{p} \hat{q}}{}^{\hat{r}} - f_{- \hat{m} \hat{n} \hat{r}} f_{+ \hat{p} \hat{q}}{}^{\hat{r}} + f_{+ \oplus \hat{m} \hat{n}} f_{- \ominus \hat{p} \hat{q}} - f_{- \oplus \hat{m} \hat{n}} f_{+ \ominus \hat{p} \hat{q}} + f_{+ \ominus \hat{m} \hat{n}} f_{- \oplus \hat{p} \hat{q}} - f_{- \ominus \hat{m} \hat{n}} f_{+ \oplus \hat{p} \hat{q}} \nonumber \\
          & - \left( \xi_{+ [\hat{m}} - \theta_{+[\hat{m}} \right) f_{- \hat{n}] \hat{p} \hat{q}} + \left( \xi_{- [\hat{m}} - \theta_{-[\hat{m}} \right) f_{+ \hat{n}] \hat{p} \hat{q}} \nonumber \\ & + \left( \xi_{+ [\hat{p}} - \theta_{+[\hat{p}} \right) f_{- \hat{q}] \hat{m} \hat{n}} - \left( \xi_{- [\hat{p}} - \theta_{-[\hat{p}} \right) f_{+ \hat{q}] \hat{m} \hat{n}} \qquad \label{q11,5} \mathbf{(0,0)} \\ 
          & + \xi_{+ [\hat{m}} \xi_{- [\hat{p}} \eta_{\hat{q}] \hat{n}]} -  \xi_{- [\hat{m}} \xi_{+ [\hat{p}} \eta_{\hat{q}] \hat{n}]} -  \xi_{+ [\hat{m}} \theta_{- [\hat{p}} \eta_{\hat{q}] \hat{n}]} + \xi_{- [\hat{m}} \theta_{+ [\hat{p}} \eta_{\hat{q}] \hat{n}]} \nonumber \\
          & + \xi_{+ [\hat{p}} \theta_{- [\hat{m}} \eta_{\hat{n}] \hat{q}]} -  \xi_{- [\hat{p}} \theta_{+ [\hat{m}} \eta_{\hat{n}] \hat{q}]}  - 3 \theta_{+ [\hat{m}} \theta_{- [\hat{p}} \eta_{\hat{q}] \hat{n}]}  + 3 \theta_{- [\hat{m}} \theta_{+ [\hat{p}} \eta_{\hat{q}] \hat{n}]} = 0 \, ,\nonumber \\
          & \, \nonumber \\
          & - f_{+ \oplus \hat{m} \hat{p}} f_{- \hat{n} \ominus}{}^{\hat{p}} + f_{- \oplus \hat{m} \hat{p}} f_{+ \hat{n} \ominus}{}^{\hat{p}} - 2 f_{+ \oplus \ominus [\hat{m}} f_{- \hat{n}] \oplus \ominus} + \left( 
          \xi_{+(\hat{m}} - \theta_{+ (\hat{m}}   \right) f_{- \hat{n}) \oplus \ominus} \nonumber \\
          & - \left( 
          \xi_{-(\hat{m}} - \theta_{- (\hat{m}}   \right) f_{+ \hat{n}) \oplus \ominus} + \frac{1}{2} \left( \xi_{+ \oplus} - \theta_{+ \oplus} \right) f_{- \ominus \hat{m} \hat{n}} -  \frac{1}{2} \left( \xi_{- \oplus} - \theta_{- \oplus} \right) f_{+ \ominus \hat{m} \hat{n}} \nonumber \\
          & +  \frac{1}{2} \left( \xi_{+ \ominus} - \theta_{+ \ominus} \right) f_{- \oplus \hat{m} \hat{n}} - \frac{1}{2} \left( \xi_{- \ominus} - \theta_{- \ominus} \right) f_{+ \oplus \hat{m} \hat{n}} + \frac{1}{2} \xi_{+ [\hat{m}} \xi_{- \hat{n}]} - \frac{1}{2} \xi_{+ [
          \hat{m}} \theta_{- \hat{n}]} \qquad \label{q11,6} \mathbf{(0,0)} \\
           & +\frac{1}{2} \xi_{- [
          \hat{m}} \theta_{+ \hat{n}]} - \frac{3}{2} \theta_{+ [\hat{m}} \theta_{- \hat{n}]}  + \frac{1}{4} \eta_{\hat{m} \hat{n}} \left( 
           \xi_{+ \oplus} \xi_{- \ominus} - \xi_{- \oplus} \xi_{+ \ominus} - \xi_{+ \oplus} \theta_{- \ominus} + \xi_{- \oplus} \theta_{+ \ominus}   \right) 
         \nonumber \\
         & + \frac{1}{4} \eta_{\hat{m} \hat{n}} \left( \xi_{+ \ominus} \theta_{- \oplus} - \xi_{- \ominus} \theta_{+ \oplus} - 3 \theta_{+ \oplus} \theta_{- \ominus} + 3 \theta_{- \oplus} \theta_{+ \ominus} \right) = 0 \, ,\nonumber \\
         & \, \nonumber \\
         & - f_{+ \ominus \hat{m} \hat{n}} f_{- \oplus \ominus}{}^{\hat{n}} +  f_{- \ominus \hat{m} \hat{n}} f_{+ \oplus \ominus}{}^{\hat{n}} + \left( 
          \xi_{+ \ominus} - \theta_{+ \ominus} \right) f_{- \oplus \ominus \hat{m}} - \left( 
          \xi_{- \ominus} - \theta_{- \ominus} \right) f_{+ \oplus \ominus \hat{m}} \nonumber \\
          & + \frac{1}{4} \xi_{+ \ominus} \xi_{- \hat{m}} - \frac{1}{4} \xi_{- \ominus} \xi_{+ \hat{m}} - \frac{1}{4} \theta_{+ \ominus} \xi_{- \hat{m}} + \frac{1}{4} \theta_{- \ominus} \xi_{+ \hat{m}} \qquad \mathbf{(0,-1)} \label{q11,7} \\
          & - \frac{1}{4} \xi_{+ \ominus} \theta_{- \hat{m}} + \frac{1}{4} \xi_{- \ominus} \theta_{+ \hat{m}} - \frac{3}{4} \theta_{+ \ominus} \theta_{- \hat{m}} + \frac{3}{4} \theta_{- \ominus} \theta_{+ \hat{m}} = 0 \, ,\nonumber \\
          & \, \nonumber \\
           &  f_{+ \hat{m} \hat{n} \hat{r}} f_{- \hat{p} \ominus}{}^{\hat{r}} -  f_{- \hat{m} \hat{n} \hat{r}} f_{+ \hat{p} \ominus}{}^{\hat{r}} - f_{+ \ominus \hat{m} \hat{n}} f_{- \oplus \ominus \hat{p}} +  f_{- \ominus \hat{m} \hat{n}} f_{+ \oplus \ominus \hat{p}} \nonumber \\ 
          & - \left( \xi_{+[\hat{m}} - \theta_{+ [\hat{m}} \right) f_{- \hat{n}] \hat{p} \ominus} +  \left( \xi_{-[\hat{m}} - \theta_{- [\hat{m}} \right) f_{+ \hat{n}] \hat{p} \ominus} + \frac{1}{2} \left( 
          \xi_{+ \hat{p}} - \theta_{+ \hat{p}}  \right) f_{- \ominus \hat{m} \hat{n}} \nonumber \\
          & - \frac{1}{2} \left( 
          \xi_{- \hat{p}} - \theta_{- \hat{p}}  \right) f_{+ \ominus \hat{m} \hat{n}} - \frac{1}{2} \left( \xi_{+ \ominus} - \theta_{+ \ominus} \right) f_{- \hat{m} \hat{n} \hat{p}} + \frac{1}{2} \left( \xi_{- \ominus} - \theta_{- \ominus} \right) f_{+ \hat{m} \hat{n} \hat{p}} \qquad \mathbf{(0,-1)} \label{q11,8} \\
          & + \frac{1}{2} \xi_{+ \ominus} \xi_{- [\hat{m}} \eta_{\hat{n}] \hat{p}} - \frac{1}{2} \xi_{- \ominus} \xi_{+ [\hat{m}} \eta_{\hat{n}] \hat{p}} - \frac{1}{2} \theta_{+ \ominus} \xi_{- [\hat{m}} \eta_{\hat{n}] \hat{p}} + \frac{1}{2} \theta_{- \ominus} \xi_{+ [\hat{m}} \eta_{\hat{n}] \hat{p}} \nonumber \\
          & - \frac{1}{2} \xi_{+ \ominus} \theta_{- [\hat{m}} \eta_{\hat{n}]\hat{p}} + \frac{1}{2} \xi_{- \ominus} \theta_{+ [\hat{m}} \eta_{\hat{n}]\hat{p}} - \frac{3}{2} \theta_{+ \ominus} \theta_{-[\hat{m}} \eta_{\hat{n}] \hat{p}} + \frac{3}{2} \theta_{- \ominus} \theta_{+[\hat{m}} \eta_{\hat{n}] \hat{p}} = 0 \, , \nonumber \\
          & \, \nonumber \\
           & 2 f_{+ \ominus \hat{p} [\hat{m}} f_{- \hat{n}] \ominus}{}^{\hat{p}} + \left( \xi_{+ \ominus} - \theta_{+ \ominus} \right) f_{- \ominus \hat{m} \hat{n}} - \left( \xi_{- \ominus} - \theta_{- \ominus} \right) f_{+ \ominus \hat{m} \hat{n}} = 0 \, . 
           \label{q11,9} \qquad \mathbf{(0,-2)}
        \end{align}

 To solve the quadratic constraints \eqref{q2,1}-\eqref{q11,9} in a systematic way, we first consider those with $\text{SO}(1,1)_B \times \text{SO}(1,1)_A$ weights $(w_B,w_A)=(2,1)$, i.e. the constraints \eqref{q4,1}, \eqref{q6,1} and \eqref{q9,1}. If $\theta_{\alpha M}$ is not identically zero, we may assume without loss of generality that $\theta_{+ \oplus} \neq 0 $. Then, from \eqref{q4,1} it follows that
\begin{equation}
    \label{f++-m}
    f_{+ \oplus \ominus \hat{m}} = \frac{1}{\theta_{+ \oplus}} \theta_{+}^{\hat{n}} f_{+ \oplus \hat{m} \hat{n}} - \frac{1}{2} \xi_{+ \hat{m}} + \frac{1}{2} \frac{\xi_{+ \oplus}}{\theta_{+ \oplus}} \theta_{+ \hat{m}} \, . 
    \end{equation}
Substituting the above expression for $f_{+ \oplus \ominus \hat{m} }$ into equation \eqref{q6,1}, the latter becomes 
\begin{equation}
    \label{calO1}
    {\cal O}_{1/2,\,1/2} \cdot \left( \xi_{+ \hat{m}} - \frac{\xi_{+ \oplus}}{\theta_{+ \oplus}} \theta_{+ \hat{m}}\right) = 0 \, , 
\end{equation}
where we have introduced the operator
\begin{equation}
    \label{O1/2}
    {\cal O}_{1/2,\,1/2} \equiv \delta_{f_{+ \oplus}} - \frac{1}{2} \left( \xi_{+ \oplus} + \theta_{+ \oplus} \right) \equiv f_{+ \oplus}{}^{\hat{m} \hat{n}} t_{\hat{m} \hat{n}} - \frac{1}{2} \left( \xi_{+ \oplus} + \theta_{+ \oplus} \right) ,
\end{equation}
where $t_{\hat{m} \hat{n}}$ are the generators of SO(5,$n-1$), whose elements in the fundamental representation have been chosen to be $(t_{\hat{m} \hat{n}})_{\hat{p}}{}^{\hat{q}} = \delta^{\hat{q}}_{[\hat{m}} \eta_{\hat{n}] \hat{p}}$\,. The general solution of \eqref{calO1} for the SO(5,$n-1$) vector $\xi_{+ \hat{m}}$ reads
\begin{equation}
    \label{xi+m}
    \xi_{+ \hat{m}} = \frac{\xi_{+ \oplus}}{\theta_{+ \oplus}} \theta_{+ \hat{m}} + \zeta_{+ \hat{m}} \, ,
\end{equation} 
where $\zeta_{+ \hat{m}}$ is a real zero mode of the operator ${\cal O}_{1/2,\,1/2}$. Using the above equation we can simplify \eqref{f++-m} to 
\begin{equation}
    \label{f++-msim}
    f_{+ \oplus \ominus \hat{m}} = \frac{1}{\theta_{+ \oplus}} \theta_{+}^{\hat{n}} f_{+ \oplus \hat{m} \hat{n}} - \frac{1}{2} \zeta_{+ \hat{m}} \, .
\end{equation}
Furthermore, by plugging \eqref{xi+m} and \eqref{f++-msim} into the constraint \eqref{q9,1}, we can determine the embedding tensor components $f_{+ \hat{m} \hat{n} \hat{p}}$ up to a zero mode $\zeta_{+ \hat{m} \hat{n} \hat{p}} = \zeta_{+ [\hat{m} \hat{n} \hat{p}]}$ of ${\cal O}_{1/2,1/2}$. The relevant result reads
\begin{equation}
    \label{f+mnp}
    f_{+ \hat{m} \hat{n} \hat{p}} = \frac{3}{\theta_{+ \oplus}} \theta_{+ [\hat{m}} f_{+ \oplus \hat{n} \hat{p}]} + \zeta_{+ \hat{m} \hat{n} \hat{p}} \, .  
    \end{equation}
    
We continue with the quadratic constraints with $(w_B,w_A)=(2,0)$. Given \eqref{f++-msim}, equation \eqref{q4,2} determines the component $\xi_{+ \ominus}$ according to    
\begin{equation}
    \label{xi+-}
    \xi_{+ \ominus} = \frac{\xi_{+ \oplus}}{\theta_{+ \oplus}} \theta_{+ \ominus} - \frac{1}{\theta_{+ \oplus}} \zeta_{+}^{\hat{m}} \theta_{+ \hat{m}} \, .  
\end{equation}
Also, using \eqref{xi+m}, \eqref{f++-msim} and \eqref{xi+-} we find that \eqref{q6,2} implies that $\zeta_{+ \hat{m}}$ is lightlike, i.e.
\begin{equation}
\label{nullz}
\zeta^{\hat{m}}_{+} \zeta_{+ \hat{m}} = 0 \, .
\end{equation}
Moreover, from \eqref{q5,1} it follows that 
\begin{equation}
    \label{xi++or}
    \left( 1 + \frac{\xi_{+ \oplus}}{\theta_{+ \oplus}} \right) \left( \theta_{+}^{\hat{m}} \theta_{+ \hat{m}} + 2 \theta_{+ \oplus} \theta_{+ \ominus}  \right) = 0 \, ,
\end{equation}
which guarantees that \eqref{q7,1} holds as well. Furthermore, equation \eqref{q4,3} yields the following expression for the antisymmetric $\text{SO}(5,n-1)$ tensor $f_{+ \ominus \hat{m} \hat{n}}$:
\begin{align}
    \label{f+-mn}
    f_{+ \ominus \hat{m} \hat{n}} = & - \frac{2}{\theta_{+ \oplus}^2} \theta_{+}^{\hat{p}} \theta_{+[\hat{m}} f_{+ \oplus \hat{n}] \hat{p}} - \frac{1}{\theta_{+ \oplus}^2} \left( \theta_{+}^{\hat{p}} \theta_{+ \hat{p}} + \theta_{+ \oplus} \theta_{+ \ominus} \right) f_{+ \oplus \hat{m} \hat{n}} \nonumber \\ & - \frac{1}{\theta_{+ \oplus}} \left( \theta_{+}^{\hat{p}} \zeta_{+ \hat{m} \hat{n} \hat{p}} + \zeta_{+[\hat{m}} \theta_{+ \hat{n}]} \right) ,  
\end{align}
where we have used \eqref{xi+m} and \eqref{f+mnp}. Next, substituting \eqref{f+-mn} into the identity \eqref{q6,3} we obtain the constraint
\begin{equation}
    \label{zmnpzp}
    \zeta_{+ \hat{m} \hat{n} \hat{p}} \zeta_{+}^{\hat{p}} = 0 \, . 
\end{equation}
Finally, the constraint \eqref{q9,2} is satisfied as a result of \eqref{xi+m}-\eqref{xi+-} and \eqref{xi++or}-\eqref{zmnpzp}, while \eqref{q9,3} holds provided 
\begin{equation}
    \label{3zz+2zz}
    3 \zeta_{+ \hat{r}[\hat{m} \hat{n}} \zeta_{+ \hat{p} \hat{q}]}{}^{\hat r} + 2 \zeta_{+ [\hat{m}} \zeta_{+ \hat{n} \hat{p} \hat{q}]} - \frac{3}{\theta_{+ \oplus}^2} \left( \theta_{+}^{\hat{r}} \theta_{+ \hat{r}} + 2 \theta_{+ \oplus} \theta_{+ \ominus} \right) f_{+ \oplus[\hat{m} \hat{n}} f_{+ \oplus \hat{p} \hat{q}]}= 0  \, .
\end{equation}

We proceed to examine the quadratic constraints with $\text{SO}(1,1)_B \times \text{SO}(1,1)_A$ weights $(w_B,w_A)=(2,-1)$. Using \eqref{xi+m}, \eqref{f++-msim}, \eqref{xi+-} and \eqref{f+-mn} we find that equation \eqref{q4,4} implies that
\begin{equation}
    \left( \theta_{+ \ominus} + \frac{1}{2 \theta_{+ \oplus}} \theta_{+}^{\hat{n}} \theta_{+ \hat{n}} \right) \zeta_{+ \hat{m}} = 0 \, , 
\end{equation}
which, combined with previous results, ensures that \eqref{q6,4} and \eqref{q9,4} hold as well. 

Then, we analyze the constraints with $(w_B,w_A)=(0,2)$. It is straightforward to solve \eqref{q2,1} for $\xi_{- \oplus}$, which is given by
\begin{equation}
    \label{xi-+}
    \xi_{- \oplus} = \frac{\xi_{+ \oplus}}{\theta_{+ \oplus}} \theta_{- \oplus} \, .
\end{equation}
By virtue of the last equation, the constraint \eqref{q11,1} can be written as 
\begin{equation}
    \label{O1=0}
    {\cal O}_{-1,\,1} \cdot \left( f_{- \oplus \hat{m} \hat{n}} - \frac{\theta_{- \oplus}}{\theta_{+ \oplus}} f_{+ \oplus \hat{m} \hat{n}} \right) = 0 \, , 
\end{equation}
where we have defined the operator 
\begin{equation}
    \label{O1}
     {\cal O}_{-1,\,1} \equiv \delta_{f_{+ \oplus}} + \xi_{+ \oplus} - \theta_{+ \oplus} \, . 
\end{equation}
The general solution of \eqref{O1=0} reads
\begin{equation}
\label{f-+mn}
    f_{- \oplus \hat{m} \hat{n}} =   \frac{\theta_{- \oplus}}{\theta_{+ \oplus}} f_{+ \oplus \hat{m} \hat{n}} + \zeta_{- \oplus \hat{m} \hat{n}} \, , 
\end{equation}
where $\zeta_{- \oplus \hat{m} \hat{n}} = \zeta_{- \oplus [\hat{m} \hat{n}]}$ denotes a real zero mode of the operator ${\cal O}_{-1,\,1}$. 

Our next task is the solution of the quadratic constraints with  $(w_B,w_A)=(0,1)$. From \eqref{q2,2} it follows that 
\begin{equation}
    \label{xi-m}
    \xi_{- \hat{m}} =  \frac{\theta_{- \oplus}}{\theta_{+ \oplus}} \zeta_{+ \hat{m}} + \frac{\xi_{+ \oplus}}{\theta_{+ \oplus}} \theta_{- \hat{m}} \, ,
    \end{equation}
where we have made use of equations \eqref{xi+m} and \eqref{xi-+}. Also, subtracting the constraint \eqref{q4,5} from \eqref{q3,1} and using \eqref{f++-msim}, \eqref{xi-+} and \eqref{xi-m} we find 
\begin{equation}
    \label{th-m}
    \theta_{- \hat{m}} = \frac{\theta_{- \oplus}}{\theta_{+ \oplus}} \theta_{+ \hat{m}} + \zeta_{- \hat{m}} \, , 
\end{equation}
where $\zeta_{- \hat{m}}$ is a real zero mode of the operator
\begin{equation}
    \label{O3/2}
    {\cal O}_{-1/2,\,3/2} \equiv \delta_{f_{+ \oplus}} + \frac{1}{2} \left( \xi_{+ \oplus} - 3 \theta_{+ \oplus} \right) .
\end{equation}
Therefore, \eqref{xi-m} can be written as
\begin{equation}
    \label{xi-msol}
    \xi_{- \hat{m}} = \frac{\xi_{+ \oplus} \theta_{- \oplus}}{\theta_{+ \oplus}^2} \theta_{+ \hat{m}} + \frac{\theta_{- \oplus}}{\theta_{+ \oplus}} \zeta_{+ \hat{m}} + \frac{\xi_{+ \oplus}}{\theta_{+ \oplus}} \zeta_{- \hat{m}} \, . 
\end{equation}
On the other hand, adding the constraints \eqref{q3,1} and \eqref{q4,5} and using \eqref{xi+m}, \eqref{f-+mn} and \eqref{th-m} we determine the components $f_{- \oplus \ominus \hat{m}}$ of the embedding tensor according to 
\begin{equation}
    \label{f-+-m}
    f_{- \oplus \ominus \hat{m}} = \frac{\theta_{- \oplus}}{\theta_{+ \oplus}^2} f_{+ \oplus \hat{m} \hat{n}} \theta_{+}^{\hat{n}} + \frac{1}{\theta_{+ \oplus}} \zeta_{- \oplus \hat{m} \hat{n}} \theta_{+}^{\hat{n}} - \frac{1}{2} \frac{\theta_{- \oplus}}{\theta_{+ \oplus}} \zeta_{+ \hat{m}} + \frac{1}{2} \left( \frac{\xi_{+ \oplus}}{\theta_{+ \oplus}} - 3  \right) \zeta_{-\hat{m}} \, . 
\end{equation}
Furthermore, given equations \eqref{xi+m}, \eqref{f++-msim}, \eqref{xi-+}, \eqref{f-+mn}, \eqref{th-m}, \eqref{xi-msol} and \eqref{f-+-m}, the constraints \eqref{q6,5} and \eqref{q8,1} are satisfied provided 
\begin{equation}
    \label{z-mnz+n}
    \zeta_{- \oplus \hat{m} \hat{n}} \zeta_{+}^{\hat{n}} = 0 \, . 
\end{equation}

Also, substituting equations \eqref{f+mnp}, \eqref{f-+mn} and \eqref{th-m} into \eqref{q10,1} we derive the following expression for the embedding tensor components $f_{- \hat{m} \hat{n} \hat{p}}$:
\begin{equation}
    \label{f-mnp}
    f_{- \hat{m} \hat{n} \hat{p}} = 3 \frac{\theta_{- \oplus}}{\theta_{+ \oplus}^2} \theta_{+ [\hat{m}} f_{+ \oplus \hat{n} \hat{p}]}  + \frac{\theta_{- \oplus}}{\theta_{+ \oplus}} \zeta_{+ \hat{m} \hat{n} \hat{p}} - \frac{3}{\theta_{+ \oplus}} \zeta_{- [\hat{m}} f_{+ \oplus \hat{n} \hat{p}]} + \frac{3}{\theta_{+ \oplus}} \theta_{+ [\hat{m}} \zeta_{- \oplus \hat{n} \hat{p}]} \, , 
\end{equation}
which implies that the constraint \eqref{q9,5} holds provided 
\begin{equation}
    \label{z-+mqz+npq}
    \zeta_{- \oplus [\hat{m}}{}^{\hat{q}} \zeta_{+ \hat{n} \hat{p}]\hat{q}} - 2 \left( \frac{\xi_{+ \oplus}}{\theta_{+ \oplus}}  - 1 \right) \zeta_{-[\hat{m}} f_{+ \oplus \hat{n} \hat{p}]} = 0 \, . 
\end{equation}

Moreover, equation \eqref{q11,2} is satisfied by virtue of \eqref{xi+m}, \eqref{f++-msim}, \eqref{xi-+}, \eqref{f-+mn}, \eqref{th-m}, \eqref{xi-msol}, \eqref{f-+-m} and \eqref{z-mnz+n}. Finally, the quadratic constraint \eqref{q11,3} implies that
\begin{align}
    \label{zzfeta}
    & \zeta_{- \oplus \hat{p}}{}^{\hat{q}} \zeta_{+ \hat{m} \hat{n} \hat{q}} + \zeta_{+ [\hat{m}} \zeta_{- \oplus \hat{n}] \hat{p}} - \zeta_{+ \hat{p}} \zeta_{- \oplus \hat{m} \hat{n}} - 2 \left( \frac{\xi_{+ \oplus}}{\theta_{+\oplus}} - 1  \right) \zeta_{- [\hat{m}} f_{+ \oplus \hat{n}] \hat{p}} \nonumber \\
    & + \frac{2}{\theta_{+ \oplus}} \zeta_{-[\hat{m}} f_{+ \oplus \hat{n}]}{}^{\hat{q}} f_{+ \oplus \hat{p} \hat{q}} - \frac{1}{2} \left( \frac{\xi_{+ \oplus}^2}{\theta_{+ \oplus}} - 2 \xi_{+ \oplus} - 3 \theta_{+ \oplus} \right) \zeta_{-[\hat{m}} \eta_{\hat{n}] \hat{p}} = 0 \,.
\end{align}
We note that the constraint \eqref{z-+mqz+npq} follows from \eqref{zzfeta} by antisymmetrizing the latter in $\hat{m},\hat{n},\hat{p}$. 

We now focus on the quadratic constraints with $\text{SO}(1,1)_B\times\text{SO}(1,1)_A$ weights $(w_B,w_A)=(0,0)$. Solving \eqref{q2,3} for $\xi_{- \ominus}$ we obtain
\begin{equation}
    \label{xi--}
    \xi_{- \ominus} = - \frac{\theta_{- \oplus}}{\theta_{+ \oplus}^2} \zeta_{+}^{\hat{m}} \theta_{+ \hat{m}} + \frac{\xi_{+ \oplus}}{\theta_{+ \oplus}} \theta_{- \ominus} \, ,
\end{equation}
where we have used equations \eqref{xi+-} and \eqref{xi-+}. Furthermore, given \eqref{f++-msim}, \eqref{xi+-}, \eqref{xi-+}, \eqref{th-m}, \eqref{f-+-m} and \eqref{xi--}, the constraints \eqref{q3,2} and \eqref{q4,6} are satisfied provided 
\begin{equation}
    \label{z+z-}
    \zeta_{+}^{\hat{m}} \zeta_{- \hat{m}} = 0 
\end{equation}
and 
\begin{equation}
    \label{th+z-or}
    \left( \frac{\xi_{+ \oplus}}{\theta_{+ \oplus}} - 3 \right) \left( \theta_{+}^{\hat{m}} \zeta_{- \hat{m}} + \theta_{+ \oplus} \theta_{- \ominus} - \theta_{+ \ominus} \theta_{- \oplus} \right) = 0 \, .
\end{equation}
Also, from the constraint \eqref{q5,2} it follows that  
\begin{equation}
    \label{th+z-}
    \theta_{+}^{\hat{m}} \zeta_{- \hat{m}} + \theta_{+ \oplus} \theta_{- \ominus} - \theta_{+ \ominus} \theta_{- \oplus} = 0 \, ,
\end{equation}
where we have used, among others, equations \eqref{xi++or} and \eqref{th+z-or}. Clearly, \eqref{th+z-} implies \eqref{th+z-or} and can be 
 solved for $\theta_{- \ominus}$. Moreover, the expressions for the embedding tensor components and the constraints on the $\zeta$ tensors that have been derived so far guarantee the validity of equations \eqref{q6,6}, \eqref{q7,2} and \eqref{q8,2}. 

Next, using \eqref{xi+m}, \eqref{th-m} and \eqref{xi-msol} we find that the quadratic constraint \eqref{q2,4} holds provided 
\begin{equation}
    \label{z+mz-n}
    \zeta_{+(\hat{m}} \zeta_{- \hat{n})} = 0 \, ,
\end{equation}
which implies \eqref{z+z-}. 

Furthermore, it is straightforward to solve the quadratic constraint \eqref{q10,2} for the components $f_{- \ominus \hat{m} \hat{n}}$ of the embedding tensor. Using equations \eqref{f++-msim}, \eqref{f-+mn}, \eqref{th-m} and \eqref{f-+-m} we find 
\begin{align}
    \label{f--mn}
    f_{- \ominus \hat{m} \hat{n}} = & - \frac{\theta_{- \ominus}}{\theta_{+ \oplus}} f_{+ \oplus \hat{m} \hat{n}}  - \frac{\theta_{-\oplus}}{\theta_{+ \oplus}^3} \theta_{+}^{\hat{p}} \theta_{+ \hat{p}} f_{+ \oplus \hat{m} \hat{n}} - 2 \frac{\theta_{-\oplus}}{\theta_{+ \oplus}^3} \theta_{+}^{\hat{p}} \theta_{+ [\hat{m}} f_{+ \oplus \hat{n}] \hat{p}} \nonumber \\ & + \frac{2}{\theta_{+ \oplus}^2} \zeta_{-[\hat{m}} f_{+ \oplus \hat{n}] \hat{p}} \theta_{+}^{\hat{p}} + \frac{\theta_{+ \ominus}}{\theta_{+ \oplus}} \zeta_{- \oplus \hat{m} \hat{n}} - \frac{2}{\theta_{+ \oplus}^2} \theta_{+[\hat{m}} \zeta_{- \oplus \hat{n}]\hat{p}} \theta_{+}^{\hat{p}} \nonumber \\
    & - \frac{\theta_{- \oplus}}{\theta_{+ \oplus}^2} \zeta_{+ \hat{m} \hat{n} \hat{p}} \theta_{+}^{\hat{p}} - \frac{\theta_{- \oplus}}{\theta_{+ \oplus}^2} \zeta_{+ [\hat{m}} \theta_{+ \hat{n}]} + \frac{\xi_{+\oplus}}{\theta_{+ \oplus}^2} \zeta_{- [\hat{m}} \theta_{+ \hat{n}]} - \frac{3}{\theta_{+ \oplus}} \zeta_{- [\hat{m}} \theta_{+ \hat{n}]} \\
    & + \frac{1}{\theta_{+ \oplus}} \zeta_{+[\hat{m}} \zeta_{- \hat{n}]} \, . \nonumber
 \end{align}
 Then, equations \eqref{q3,3}, \eqref{q4,7}, \eqref{q6,7} and \eqref{q8,3} imply three additional constraints:
 \begin{align}
     \label{z+mnpz-p} 
     & \zeta_{+ \hat{m} \hat{n} \hat{p}} \zeta_{-}^{\hat{p}} = 0 \, ,  \\
     \label{z+mz-nA}
     & \zeta_{+ [\hat{m}} \zeta_{- \hat{n}]} = 0 \, ,\\
     \label{z-+mnor}
     &      \left( \theta_{+ \ominus} + \frac{1}{2 \theta_{+ \oplus}} \theta_{+}^{\hat{p}} \theta_{+ \hat{p}}  \right) \zeta_{- \oplus \hat{m} \hat{n}} = 0 \, , 
 \end{align}
 while \eqref{q9,6} and \eqref{q11,4} are satisfied by virtue of all the above formulae for the components of the embedding tensor and constraints on the $\zeta$ tensors. 
 
 Also, the quadratic constraint \eqref{q10,3} holds provided
 \begin{equation}
     \label{z-mz+npqA}
     \zeta_{-[\hat{m}} \zeta_{+ \hat{n} \hat{p} \hat{q}]} = 0 \, , 
 \end{equation}
 while \eqref{q9,7}, \eqref{q11,5} and \eqref{q11,6} are satisfied as a result of the derived expressions for the embedding tensor components and constraints on the $\zeta$ tensors. 

 Let us now consider the quadratic constraints with $(w_B,w_A)=(0,-1)$. Our results for the components of the embedding tensor guarantee the validity of \eqref{q2,5} and imply that equations \eqref{q3,4} and \eqref{q4,8} are satisfied provided
 \begin{equation}
     \label{z-mor}
     \left( \theta_{+ \ominus} + \frac{1}{2 \theta_{+ \oplus}} \theta^{\hat{n}}_{+} \theta_{+ \hat{n}} \right) \zeta_{- \hat{m}} = 0 \, .
 \end{equation}
Furthermore, the quadratic constraints \eqref{q6,8}, \eqref{q8,4}, \eqref{q9,8}, \eqref{q10,4}, \eqref{q11,7} and \eqref{q11,8}  hold as a consequence of all the above expressions for the embedding tensor components and constraints on the $\zeta$ tensors. 

Next, we analyze the quadratic constraints with $\text{SO}(1,1)_B \times \text{SO}(1,1)_A$ weights $(w_B,w_A)=(-2,1)$. Using equations \eqref{xi-+}, \eqref{f-+mn}, \eqref{th-m}, \eqref{xi-msol} and \eqref{f-+-m} we find that \eqref{q4,9} is satisfied on condition that 
\begin{equation}
    \label{z-+mnz-n}
    \zeta_{- \oplus \hat{m} \hat{n}} \zeta_{-}^{\hat{n}} = 0 \, , 
\end{equation}
which guarantees that \eqref{q6,9} holds as well. On the other hand, the constraint \eqref{q9,9} is satisfied by virtue of the derived formulae for the components of the embedding tensor and constraints on the $\zeta$ tensors. 

We then examine the quadratic constraints with $(w_B,w_A)=(-2,0)$. Given equations \eqref{xi++or}, \eqref{xi-+}, \eqref{th-m}, \eqref{xi-msol}, \eqref{f-+-m} and \eqref{xi--}, the constraints \eqref{q4,10} and \eqref{q5,3} hold provided 
\begin{equation}
    \label{z-mz-m}
    \zeta^{\hat{m}}_{-} \zeta_{- \hat{m}} = 0 \, , 
\end{equation}
which ensures the validity of \eqref{q6,10} and \eqref{q7,3} as well. Furthermore, equations \eqref{q4,11}, \eqref{q6,11}, \eqref{q9,10} and \eqref{q9,11} are satisfied by virtue of all the above results for the components of the embedding tensor. This is also true for the quadratic constraints \eqref{q4,12}, \eqref{q6,12} and \eqref{q9,12}, which have $\text{SO}(1,1)_B\times\text{SO}(1,1)_A$ weights $(w_B,w_A)=(-2,-1)$, as well as \eqref{q2,6} and \eqref{q11,9}, with $(w_B,w_A)=(0,-2)$. 

We have completed the analysis of the quadratic identities \eqref{q2,1}-\eqref{q11,9}. The results of this appendix give rise to the general solution to the quadratic consistency constraints on the embedding tensor provided in subsection \ref{sec:solmain} and to an additional one with non-vanishing $\theta_{+}^{\hat{m}} \theta_{+ \hat{m}} + 2 \theta_{+ \oplus} \theta_{+ \ominus} = \theta_{+}^M \theta_{+ M}$ and $\xi_{\alpha M} = - \theta_{\alpha M}$. 

 \section{$\mathbf{\it T}$-identities}
\label{sec:Tid}
By appropriately dressing the quadratic constraints \eqref{q2}-\eqref{q11} on the embedding tensor with the representatives of the coset spaces SL(2,$\mathbb{R}$)/SO(2) and SO(6,$n$)/(SO(6)$\times$SO($n$)) parametrized by the scalar fields of the theory, one obtains numerous quadratic constraints on the $A$ and $B$ tensors that appear in the local supersymmetry transformations of the fermions and in the fermionic field equations. Each of these constraints, which are also known as $T$-identities, carries a specific representation of  the isotropy group $H=\text{SO}(2) \times \text{SU}(4) \times \text{SO}(n)$ of the coset space \eqref{Mscalar}. In this appendix, we provide an exhaustive list of the $T$-identities of four-dimensional ${\cal N}=4$ supergravity with local scaling symmetry, classifying them according to their origin (one of \eqref{q2}-\eqref{q11}) and their $H$ representation, using the notation $({\cal R}_{\text{SU}(4)},{\cal R}_{\text{SO}(n)})_{q_{\text{SO}(2)}}$, where ${\cal R}_{\text{SU}(4)}$ and ${\cal R}_{\text{SO}(n)}$ denote the SU(4) and SO($n$) representations respectively and $q_{\text{SO}(2)}$ the SO(2) charge.
\subsection{From \eqref{q2}}
Rep $(\mathbf{(6 \times 6)}_S,\mathbf{1})_{\mathbf{0}}$:
\begin{align}
\label{T2}
    \qquad & A_2^{[kl]} {\bar B}_{ij} - {\bar A}_{2[ij]} B^{kl} - \delta^k_{[i} A_2^{[lm]} {\bar B}_{j]m} + \delta^l_{[i} A_2^{[km]} {\bar B}_{j]m} \nonumber \\ & + \delta^{[k}_i {\bar A}_{2[jm]} B^{l]m} - \delta^{[k}_j {\bar A}_{2[im]} B^{l]m}  
   + \frac{1}{2} \delta^k_{[i} \delta^l_{j]} \left( A_2^{mn} {\bar B}_{mn} - {\bar A}_{2 mn} B^{mn}  \right) = 0  \, . 
\end{align}
The two-fold symmetric tensor product of the $\mathbf{6}$ representation of SU(4) decomposes as 
\begin{equation}
    \mathbf{(6 \times 6)}_S = \mathbf{1} + \mathbf{20'} . 
\end{equation}
In order to specify the component of the quadratic constraint \eqref{T2} that is an SU(4) singlet, we contract \eqref{T2} with $\delta^i_k \delta^j_l$. We find  
\begin{equation}
\label{T2_1}
    \mathbf{(1,1)_0}: \qquad A_2^{ij} {\bar B}_{ij} - {\bar A}_{2ij} B^{ij} = 0 
\end{equation}
Rep $\mathbf{(6,n)_0}$:
\begin{equation}
\label{T3}
     \frac{1}{3} \epsilon_{ijkl} A_2^{kl} {\bar B}_{\underline{a}} - \frac{2}{3} {\bar A}_{2[ij]} B_{\underline{a}} -A_{2 \underline{a} k }{}^k {\bar B}_{ij} + \frac{1}{2} \epsilon_{ijlm} {\bar A}_{2 \underline{a}}{}^k{}_k B^{lm} = 0 
\end{equation}
Rep $\mathbf{(1,n(n+1)/2)_0}$:
\begin{equation}
    \label{T4}
     A_{2(\underline{a}|i}{}^i {\bar B}_{|\underline{b})} - {\bar A}_{2 (\underline{a}|}{}^i{}_i B_{|\underline{b})} =0 
\end{equation}

\subsection{From \eqref{q3}}
Rep $\mathbf{(15,1)_0}$:
\begin{align}
\label{T5}
     & \frac{2}{3} A_2^{(jk)} {\bar B}_{ik} + \frac{2}{3} {\bar A}_{2 (ik)} B^{jk} - \frac{1}{3} \epsilon_{iklm} A_1^{jk} B^{lm} - \frac{1}{3} \epsilon^{jklm} {\bar A}_{1ik} {\bar B}_{lm} \nonumber \\
    & - B_{\underline{a}} \left( 
     {\bar A}_2{}^{\underline{a}j}{}_i - \frac{1}{4} \delta^j_i {\bar A}_2{}^{\underline{a}k}{}_k  \right) - {\bar B}^{\underline{a}} \left( 
      A_{2 \underline{a}i}{}^j - \frac{1}{4} \delta^j_i    A_{2 \underline{a}k}{}^k \right) \nonumber  \\
      & - \frac{2}{3} A_2^{[jk]} {\bar B}_{ik} - \frac{2}{3} {\bar A}_{2 [ik]} B^{jk} + \frac{1}{6} \delta^j_i \left( 
      A_2^{kl} {\bar B}_{kl} + {\bar A}_{2kl} B^{kl}   \right)  \\ 
      & + 6 {\bar B}_{ik} B^{jk} - \frac{3}{2} \delta^j_i B^{kl} {\bar B}_{kl} = 0 \nonumber
   \end{align}
Rep $\mathbf{(6,n)_0}$:   
\begin{align}
\label{T6}
       & 2 A_{2 \underline{a} [i}{}^k {\bar B}_{j]k}  + {\bar A}_{2 \underline{a}}{}^k{}_{[i} \epsilon_{j]klm} B^{lm} + A_{2 \underline{a} k}{}^k {\bar B}_{ij}  + {\bar A}_{\underline{a} \underline{b} ij} B^{\underline{b}} - \frac{1}{2} \epsilon_{ijkl} A_{\underline{a} \underline{b}}{}^{kl} {\bar B}^{\underline{b}} \nonumber \\
    &  - \frac{1}{3} {\bar A}_{2 [ij]} B_{\underline{a}} + \frac{1}{6} \epsilon_{ijkl} A_2^{kl} {\bar B}_{\underline{a}}  + 3 {\bar B}_{ij} B_{\underline{a}} - \frac{3}{2} \epsilon_{ijkl} B^{kl} {\bar B}_{\underline{a}} = 0  
\end{align}
Rep $\mathbf{(1,n(n-1)/2)_0}$:
\begin{align}
    \label{T7}
      &  A_{\underline{a} \underline{b}}{}^{ij} {\bar B}_{ij} - {\bar A}_{\underline{a} \underline{b} ij} B^{ij} - A_{\underline{a} \underline{b} \underline{c}} {\bar B}^{\underline{c}} + {\bar A}_{\underline{a} \underline{b} \underline{c}} B^{\underline{c}}  - A_{2[\underline{a}|i}{}^i {\bar B}_{|\underline{b}]} + {\bar A}_{2[\underline{a}|}{}^i{}_i B_{|\underline{b}]} - 6 B_{[\underline{a}} {\bar B}_{\underline{b}]} = 0 
\end{align}

\subsection{From \eqref{q4}}
Rep $\mathbf{(15,1)_{+2}}$:
\begin{align}
\label{T8}
      & \frac{2}{3} {\bar A}_{1ik} B^{jk} - \frac{1}{3} \epsilon_{iklm} A_2^{(jk)} B^{lm}  + B^{\underline{a}} \left(A_{2 \underline{a}i}{}^j - \frac{1}{4} \delta^j_i A_{2 \underline{a}k}{}^k  \right) \nonumber \\
    & + \frac{1}{6} \epsilon_{iklm} \left( A_2^{lm} B^{jk} - A_2^{[jk]} B^{lm} \right) = 0 
    \end{align}
Rep $\mathbf{(15,1)_0}$:
\begin{align}
\label{T9}
  & \frac{2}{3} {\bar A}_{2 ik} B^{jk} - \frac{2}{3} A_2^{jk} {\bar B}_{ik} + \frac{1}{3} \epsilon^{jklm} {\bar A}_{1ik} {\bar B}_{lm} - \frac{1}{3} \epsilon_{iklm} A_1^{jk} B^{lm} \nonumber \\
& - B_{\underline{a}} \left( {\bar A}_2{}^{\underline{a}j}{}_i - \frac{1}{4} \delta^j_i {\bar A}_2{}^{\underline{a}k}{}_k  \right) + {\bar B}^{\underline{a}} \left(A_{2 \underline{a}i}{}^j - \frac{1}{4} \delta^j_i A_{2 \underline{a}k}{}^k  \right) \\
&  + \frac{1}{6} \delta^j_i \left( A_2^{kl} {\bar B}_{kl} - {\bar A}_{2kl} B^{kl} \right) = 0 \nonumber
    \end{align}    
Rep $\mathbf{(6,n)_{+2}}$:
  \begin{equation}
  \label{T10}
      A_{2 \underline{a} [i}{}^k \epsilon_{j]klm} B^{lm} - \frac{1}{2} \epsilon_{ijkl} A_{\underline{a} \underline{b}}{}^{kl} B^{\underline{b}} - \frac{1}{6} \epsilon_{ijkl} A_2^{kl} B_{\underline{a}} = 0 
\end{equation}
Rep $\mathbf{(6,n)_{-2}}$:
\begin{equation}
    \label{T11}
     2 {\bar A}_{2 \underline{a}}{}^k{}_{[i} {\bar B}_{j]k}  + {\bar A}_{2 \underline{a}}{}^k{}_k {\bar B}_{ij} + {\bar A}_{\underline{a} \underline{b} ij} {\bar B}^{\underline{b}}  + \frac{1}{3} {\bar A}_{2[ij]} {\bar B}_{\underline{a}} = 0 
\end{equation}  
Rep $\mathbf{(6,n)_0}$:
\begin{align}
    \label{T12}
     &- A_{2 \underline{a}[i}{}^k {\bar B}_{j]k}  + \frac{1}{2} {\bar A}_{2 \underline{a}}{}^k{}_{[i} \epsilon_{j]klm} B^{lm}  + \frac{1}{4} \epsilon_{ijlm} {\bar A}_{2 \underline{a}}{}^k{}_k B^{lm} \nonumber \\
    &  + \frac{1}{2} {\bar A}_{\underline{a} \underline{b} ij} B^{\underline{b}} + \frac{1}{4} \epsilon_{ijkl} A_{\underline{a} \underline{b}}{}^{kl} {\bar B}^{\underline{b}} + \frac{1}{6} {\bar A}_{2[ij]} B_{\underline{a}} + \frac{1}{12} \epsilon_{ijkl} A_2^{kl} {\bar B}_{\underline{a}} = 0    
\end{align}
Rep $\mathbf{(1,n(n-1)/2)_{+2}}$:
\begin{equation}
    \label{T13}
       - \frac{1}{2} \epsilon_{ijkl} A_{\underline{a} \underline{b}}{}^{ij} B^{kl} + A_{\underline{a} \underline{b} \underline{c}} B^{\underline{c}} - A_{2[\underline{a}|i}{}^i B_{|\underline{b}]} = 0 
\end{equation}
Rep $\mathbf{(1,n(n-1)/2)_0}$:
\begin{align}
    \label{T14}
      & A_{\underline{a} \underline{b}}{}^{ij} {\bar B}_{ij} + {\bar A}_{\underline{a} \underline{b} ij} B^{ij} - A_{\underline{a} \underline{b} \underline{c}} {\bar B}^{\underline{c}} - {\bar A}_{\underline{a} \underline{b} \underline{c}} B^{\underline{c}}  + A_{2[\underline{a}|i}{}^i {\bar B}_{|\underline{b}]} + {\bar A}_{2 [\underline{a}|}{}^i{}_i  B_{|\underline{b}]} = 0 
\end{align}

\subsection{From \eqref{q5}}
Rep $\mathbf{(1,1)_{+2}}$:
\begin{equation}
\label{T15}
     \frac{1}{3} \epsilon_{ijkl} A_2^{ij} B^{kl} + \frac{1}{2} \epsilon_{ijkl} B^{ij} B^{kl} + A_{2 \underline{a}i}{}^i B^{\underline{a}} - B^{\underline{a}} B_{\underline{a}} = 0 
\end{equation}
Rep $\mathbf{(1,1)_0}$:
\begin{equation}
\label{T16}
     \frac{1}{3} A_2^{ij} {\bar B}_{ij} + \frac{1}{3} {\bar A}_{2 ij} B^{ij} + \frac{1}{2} A_{2 \underline{a} i}{}^i {\bar B}^{\underline{a}} + \frac{1}{2} {\bar A}_2{}^{\underline{a}i}{}_i B_{\underline{a}} + B^{ij} {\bar B}_{ij} - B^{\underline{a}} {\bar B}_{\underline{a}} = 0 
\end{equation}

\subsection{From \eqref{q6}}
Rep $\mathbf{(15,1)_{+2}}$:
\begin{align}
    \label{T17}
 & \frac{4}{9} A_2^{[jk]} {\bar A}_{1ik} - \frac{2}{9} \epsilon_{iklm} A_2^{(jk)} A_2^{lm} - A_2{}^{\underline{a}}{}_k{}^k \left( A_{2 \underline{a} i}{}^j - \frac{1}{4} \delta^j_i A_{2 \underline{a} l}{}^l  \right)    \nonumber \\
& + \frac{1}{6} \epsilon_{iklm} \left( A_2^{[jk]} B^{lm} - A_2^{lm} B^{jk} \right) = 0 
\end{align}
Rep $ \mathbf{(15,1)_0}$:
\begin{align}
    \label{T18}
     & \frac{4}{9} A_2^{[jk]} {\bar A}_{2(ik)} - \frac{4}{9} A_2^{(jk)} {\bar A}_{2[ik]} - \frac{2}{9} \epsilon_{iklm} A_1^{jk} A_2^{lm}  + \frac{2}{9} \epsilon^{jklm} {\bar A}_{1ik} {\bar A}_{2lm} \nonumber \\ 
    & + A_{2 \underline{a} k}{}^k {\bar A}_2{}^{\underline{a}j}{}_i - A_{2 \underline{a}i}{}^j {\bar A}_2{}^{\underline{a} k}{}_k + \frac{2}{3} A_2^{[jk]} {\bar B}_{ik} - \frac{2}{3} {\bar A}_{2[ik]} B^{jk} \\
    & - \frac{1}{6} \delta^j_i \left( A_2^{kl} {\bar B}_{kl} - {\bar A}_{2kl} B^{kl} \right) = 0 \nonumber 
\end{align}
Rep $\mathbf{(6,n)_{+2}}$:
\begin{align}
    \label{T19}
     & \frac{4}{3} A_{2 \underline{a} [i}{}^k \epsilon_{j]klm} A_2^{lm} + \frac{1}{3} \epsilon_{ijkl} A_2^{kl} A_{2 \underline{a} m}{}^m + \epsilon_{ijkl} A_{\underline{a} \underline{b}}{}^{kl} A_2{}^{\underline{b}}{}_m{}^m \nonumber \\
    & + \frac{1}{3} \epsilon_{ijkl} A_2^{kl} B_{\underline{a}} + \frac{1}{2} \epsilon_{ijlm} A_{2 \underline{a} k}{}^k B^{lm} = 0 
\end{align}
Rep $\mathbf{(6,n)_{-2}}$:
\begin{align}
    \label{T20}
     & \frac{2}{3} {\bar A}_{2 [ik]} {\bar A}_{2 \underline{a}}{}^k{}_j -  \frac{2}{3} {\bar A}_{2 [jk]} {\bar A}_{2 \underline{a}}{}^k{}_i - \frac{1}{3} {\bar A}_{2 [ij]} {\bar A}_{2 \underline{a}}{}^k{}_k + {\bar A}_{\underline{a} \underline{b} ij} {\bar A}_2{}^{\underline{b} k}{}_k \nonumber \\
    &  + \frac{1}{3} {\bar A}_{2 [ij]} {\bar B}_{\underline{a}}  + \frac{1}{2} {\bar A}_{2 \underline{a}}{}^k{}_k {\bar B}_{ij} = 0  
\end{align}
Rep $\mathbf{(6,n)_0}$:
\begin{align}
    \label{T21}
    & \frac{2}{3} {\bar A}_{2 \underline{a}}{}^k{}_{[i} \epsilon_{j]klm} A_2^{lm}  + \frac{1}{6} \epsilon_{ijkl} A_2^{kl} {\bar A}_{2 \underline{a}}{}^m{}_m  + \frac{2}{3} {\bar A}_{2[ik]} A_{2 \underline{a}j}{}^k - \frac{2}{3} {\bar A}_{2[jk]} A_{2 \underline{a}i}{}^k 
    \nonumber \\ &  - \frac{1}{3} {\bar A}_{2 [ij]} A_{2 \underline{a} k}{}^k 
    - {\bar A}_{\underline{a} \underline{b} ij} A_2{}^{\underline{b}}{}_k{}^k - \frac{1}{2} \epsilon_{ijkl} A_{\underline{a} \underline{b}}{}^{kl} {\bar A}_2{}^{\underline{b}m}{}_m  \\
    & - \frac{1}{3} {\bar A}_{2[ij]} B_{\underline{a}} - \frac{1}{6} \epsilon_{ijkl} A_2^{kl} {\bar B}_{\underline{a}} - \frac{1}{2} A_{2 \underline{a}k}{}^k {\bar B}_{ij} - \frac{1}{4} \epsilon_{ijlm} {\bar A}_{2 \underline{a}}{}^k{}_k B^{lm} = 0 \nonumber
\end{align}
Rep $\mathbf{(1,n(n-1)/2)_{+2}}$:
\begin{equation}
    \label{T22}
      \frac{1}{3} \epsilon_{ijkl} A_{\underline{a} \underline{b}}{}^{ij} A_2^{kl} + A_{\underline{a} \underline{b} \underline{c}} A_2{}^{\underline{c}}{}_i{}^i - A_{2[\underline{a}|i}{}^i B_{|\underline{b}]} = 0 
\end{equation}
Rep $\mathbf{(1,n(n-1)/2)_0}$:
\begin{align}
    \label{T23}
       &  \frac{2}{3} A_{\underline{a} \underline{b}}{}^{ij} {\bar A}_{2ij} + \frac{2}{3} {\bar A}_{\underline{a} \underline{b} ij } A_2^{ij} + A_{\underline{a} \underline{b} \underline{c}} {\bar A}_2{}^{\underline{c}i}{}_i + {\bar A}_{\underline{a} \underline{b} \underline{c}} A_2{}^{\underline{c}}{}_i{}^i  - A_{2[\underline{a}|i}{}^i {\bar B}_{|\underline{b}]} - {\bar A}_{2[\underline{a}|}{}^i{}_i B_{|\underline{b}]} = 0 
\end{align}

\subsection{From \eqref{q7}}
Rep $\mathbf{(1,1)_{+2}}$:
\begin{equation}
\label{T24}
      \frac{2}{9} \epsilon_{ijkl} A_2^{ij} A_2^{kl} - A_{2 \underline{a} i}{}^i A_2{}^{\underline{a}}{}_j{}^j  + \frac{1}{3} \epsilon_{ijkl} A_2^{ij} B^{kl} + A_{2 \underline{a} i}{}^i B^{\underline{a}} = 0 
\end{equation}
Rep $\mathbf{(1,1)_0}$:
\begin{align}
    \label{T25}
     & \frac{4}{9} A_2^{[ij]} {\bar A}_{2ij} - A_{2 \underline{a}i}{}^i {\bar A}_2{}^{\underline{a}j}{}_j  + \frac{1}{2} A_{2 \underline{a}i}{}^i {\bar B}^{\underline{a}} + \frac{1}{2} {\bar A}_{2 \underline{a}}{}^i{}_i B^{\underline{a}} + \frac{1}{3} A_2^{ij} {\bar B}_{ij} + \frac{1}{3} {\bar A}_{2ij} B^{ij} = 0 
\end{align}
\subsection{From \eqref{q8}}
Rep $\mathbf{(15,1)_0}$:
\begin{align}
    \label{T26}
      & \frac{4}{9} A_2^{[jk]} {\bar A}_{2 (ik)} + \frac{4}{9} A_2^{(jk)} {\bar A}_{2 [ik]} - \frac{2}{9} \epsilon^{jklm} {\bar A}_{1ik} {\bar A}_{2 lm} - \frac{2}{9} \epsilon_{iklm} A_1^{jk} A_2^{lm} \nonumber \\
    & + A_{2 \underline{a} k}{}^k {\bar A}_2{}^{\underline{a} j}{}_i + A_{2 \underline{a} i}{}^j {\bar A}_2{}^{\underline{a} k}{}_k - \frac{1}{2} \delta^j_i A_{2 \underline{a} k}{}^k {\bar A}_2{}^{\underline{a} l}{}_l \nonumber \\ 
    & - \frac{8}{9} A_2^{[jk]} {\bar A}_{2[ik]} + \frac{2}{9} \delta^j_i A_2^{[kl]} {\bar A}_{2kl} + 2 A_2^{[jk]} {\bar B}_{ik} + 2 {\bar A}_{2 [ik]} B^{jk} \\
    & - \frac{1}{2} \delta^j_i \left( A_2^{kl} {\bar B}_{kl} +{\bar A}_{2 kl} B^{kl} \right) =  0 \nonumber
\end{align}
Rep $\mathbf{(6,n)_0}$:
\begin{align}
    \label{T27}
     & - \frac{2}{3} {\bar A}_{2 [ik]} A_{2 \underline{a} j}{}^k +  \frac{2}{3} {\bar A}_{2 [jk]} A_{2 \underline{a} i}{}^k + {\bar A}_{2 [ij]} A_{2 \underline{a} k}{}^k  + \frac{2}{3} {\bar A}_{2 \underline{a}}{}^k{}_{[i} \epsilon_{j]klm} A_2^{lm} \nonumber \\ 
    & - \frac{1}{6} \epsilon_{ijkl} A_2^{kl} {\bar A}_{2 \underline{a}}{}^m{}_m - {\bar A}_{\underline{a} \underline{b} ij} A_2{}^{\underline{b}}{}_k{}^k + \frac{1}{2} \epsilon_{ijkl} A_{\underline{a} \underline{b}}{}^{kl} {\bar A}_2{}^{\underline{b}m}{}_m \\ 
    & + {\bar A}_{2[ij]} B_{\underline{a}} - \frac{1}{2} \epsilon_{ijkl} A_2^{kl} {\bar B}_{\underline{a}} - \frac{3}{2} A_{2 \underline{a} k}{}^k {\bar B}_{ij} + \frac{3}{4} \epsilon_{ijlm} {\bar A}_{2 \underline{a}}{}^k{}_k B^{lm} = 0       \nonumber
\end{align}
Rep $\mathbf{(1,n(n-1)/2)_0}$:
\begin{align}
    \label{T28}
    & \frac{2}{3} A_{\underline{a} \underline{b}}{}^{ij} {\bar A}_{2ij} - \frac{2}{3} {\bar A}_{\underline{a} \underline{b} ij} A_2^{ij} + A_{\underline{a} \underline{b} \underline{c}} {\bar A}_2{}^{\underline{c} i}{}_i  - {\bar A}_{\underline{a} \underline{b} \underline{c} } A_2{}^{\underline{c}}{}_i{}^i \nonumber \\ 
    & + 2 A_{2 [\underline{a}|i}{}^i {\bar A}_{2|\underline{b}]}{}^j{}_j + 3 A_{2 [\underline{a}|i}{}^i {\bar B}_{|\underline{b}]} - 3 {\bar A}_{2 [\underline{a}|}{}^i{}_i B_{|\underline{b}]} = 0 
\end{align}

\subsection{From \eqref{q9}}
Rep $\mathbf{(15,1)_{-2}}$:
\begin{align}
    \label{T29} 
      & \frac{4}{3} A_1^{jk} {\bar A}_{2 (ik)} + 3 {\bar A}_2{}^{\underline{a}j}{}_k {\bar A}_{2 \underline{a}}{}^k{}_i - \frac{3}{2} {\bar A}_2{}^{\underline{a}j}{}_i {\bar A}_{2 \underline{a}}{}^k{}_k \nonumber \\ 
    & + \frac{2}{3}  A_1^{jk} {\bar A}_{2[ik]} + A_1^{jk} {\bar B}_{ik} + \frac{1}{3} \epsilon^{jklm} {\bar A}_{2 (ik)} {\bar A}_{2lm} + \frac{1}{2} \epsilon^{jklm} {\bar A}_{2 (ik)} {\bar B}_{lm} = \\
    & = \frac{1}{3} \delta^j_i A_1^{kl} {\bar A}_{2 kl} + \frac{3}{4} \delta^j_i \left( {\bar A}_2{}^{\underline{a}k}{}_l {\bar A}_{2 \underline{a}}{}^l{}_k - \frac{1}{2} {\bar A}_2{}^{\underline{a}k}{}_k {\bar A}_{2 \underline{a}}{}^l{}_l \right) \nonumber 
        \end{align}
 Rep $\mathbf{(15,1)_0}$:       
  \begin{align}
      \label{T30}
       & \frac{2}{3} A_1^{jk} {\bar A}_{1ik} + \frac{2}{3} A_2^{(jk)} {\bar A}_{2 (ik)} + \frac{1}{3} A_2^{[jk]} {\bar A}_{2 (ik)} + \frac{1}{3} A_2^{(jk)} {\bar A}_{2 [ik]} \nonumber \\
      & + \frac{1}{6} \epsilon^{jklm} {\bar A}_{1ik} {\bar A}_{2 lm} + \frac{1}{6} \epsilon_{iklm} A_1^{jk} A_2^{lm} - \frac{3}{2} A_{2 \underline{a} k}{}^j {\bar A}_2{}^{\underline{a}k}{}_i - \frac{3}{2} A_{2 \underline{a} i}{}^k {\bar A}_2{}^{\underline{a}j}{}_k \nonumber \\
      & + \frac{3}{4}  A_{2 \underline{a} i}{}^j {\bar A}_2{}^{\underline{a}k}{}_k + \frac{3}{4}  A_{2 \underline{a} k}{}^k {\bar A}_2{}^{\underline{a}j}{}_i + \frac{1}{2} B^{jk} {\bar A}_{2 (ik)} + \frac{1}{2} A_2^{(jk)} {\bar B}_{ik}   \\
      &  + \frac{1}{4} \epsilon^{jklm} {\bar A}_{1ik} {\bar B}_{lm} + \frac{1}{4} \epsilon_{iklm} A_1^{jk} B^{lm} = \nonumber \\
      &  = \frac{1}{6} \delta^j_i \left( A_1^{kl} {\bar A}_{1 kl} + A_2^{(kl)} {\bar A}_{2 kl}  \right) - \frac{3}{4} \delta^j_i A_{2 \underline{a}k}{}^l {\bar A}_2{}^{\underline{a} k}{}_l + \frac{3}{8} \delta^j_i A_{2 \underline{a}k}{}^k {\bar A}_2{}^{\underline{a} l}{}_l \nonumber
  \end{align}   
  Rep $\mathbf{(10,n)_{+2}}$:
\begin{align}
    \label{T31}
     & \frac{4}{3} A_{2 \underline{a} (i}{}^k {\bar A}_{1 j) k} + \frac{1}{3} A_{2 \underline{a} (i}{}^k \epsilon_{j)klm} A_2^{lm} + A_2{}^{\underline{b}}{}_{(i}{}^k \epsilon_{j) klm} A_{\underline{a} \underline{b}}{}^{lm} \nonumber \\
    & - \frac{1}{3} {\bar A}_{1ij} B_{\underline{a}} + \frac{1}{2} A_{2 \underline{a} (i}{}^k \epsilon_{j)klm} B^{lm} = 0
\end{align}  
Rep $\mathbf{(10,n)_{-2}}$:
\begin{align}
    \label{T32}
     & 2 {\bar A}_{2 (ik)} {\bar A}_{2 \underline{a} }{}^k{}_j + 2 {\bar A}_{2 (jk)} {\bar A}_{2 \underline{a} }{}^k{}_i - 2 {\bar A}_{2 (ij)} {\bar A}_{2 \underline{a} }{}^k{}_k + 6 {\bar A}_{\underline{a} \underline{b} (i|k } {\bar A}_2{}^{\underline{b} k}{}_{|j)} \nonumber \\
    & + {\bar A}_{2 [ik]} {\bar A}_{2 \underline{a} }{}^k{}_j + {\bar A}_{2 [jk]} {\bar A}_{2 \underline{a} }{}^k{}_i + 3 {\bar A}_{2 \underline{a}}{}^k{}_{(i} {\bar B}_{j)k} + {\bar A}_{2 (ij)} {\bar B}_{\underline{a}} = 0 
\end{align}
Rep $\mathbf{(10,n)_0}$:
\begin{align}
    \label{T33}
     & 2 {\bar A}_{2 (ik)} A_{2 \underline{a} j}{}^k + 2 {\bar A}_{2 (jk)} A_{2 \underline{a} i}{}^k - 4 {\bar A}_{2 \underline{a}}{}^k{}_{(i} {\bar A}_{1j)k} + 2 {\bar A}_{1ij} {\bar A}_{2 \underline{a}}{}^k{}_k \nonumber \\ 
    & + 6 {\bar A}_{\underline{a} \underline{b} (i|k} A_2{}^{\underline{b}}{}_{|j)}{}^k - 3 {\bar A}_2{}^{\underline{b} k}{}_{(i} \epsilon_{j)klm} A_{\underline{a} \underline{b}}{}^{lm} + {\bar A}_{2 [ik]} A_{2 \underline{a} j}{}^k + {\bar A}_{2 [jk]} A_{2 \underline{a} i }{}^k  \\
    & - {\bar A}_{2 \underline{a}}{}^k{}_{(i} \epsilon_{j)klm} A_2^{lm}  + 3 A_{2 \underline{a} (i}{}^k {\bar B}_{j)k} - \frac{3}{2} {\bar A}_{2 \underline{a}}{}^k{}_{(i} \epsilon_{j)klm} B^{lm} - {\bar A}_{1ij} {\bar B}_{\underline{a}}  - {\bar A}_{2 (ij)} B_{\underline{a}} = 0 \nonumber
\end{align}
Rep $\mathbf{(15,n(n-1)/2)_{-2}}$:
\begin{align}
    \label{T34}
     & - 2 \epsilon^{jklm} {\bar A}_{[\underline{a}|\underline{c}ik} {\bar A}_{|\underline{b}]}{}^{\underline{c}}{}_{lm} - 2 {\bar A}_{\underline{a} \underline{b} \underline{c}} \left( {\bar A}_2{}^{\underline{c}j}{}_i - \frac{1}{4} \delta^j_i {\bar A}_2{}^{\underline{c} k}{}_k \right)  - 4 {\bar A}_{2 [\underline{a}|}{}^k{}_i {\bar A}_{2|\underline{b}]}{}^j{}_k \nonumber \\ 
    &  - \frac{4}{3} A_1^{jk} {\bar A}_{\underline{a} \underline{b} ik}  + \frac{2}{3} \epsilon^{jklm} {\bar A}_{2 (ik)} {\bar A}_{\underline{a} \underline{b} lm} - 2 {\bar A}_{2 [\underline{a}|}{}^j{}_i {\bar A}_{2|\underline{b}]}{}^k{}_k \nonumber \\
    & + \frac{1}{3} \epsilon^{jklm} \left( {\bar A}_{2 [ik]} {\bar A}_{\underline{a} \underline{b} lm} - {\bar A}_{\underline{a} \underline{b} ik} {\bar A}_{2 lm} \right)  + 2 {\bar A}_{2 [\underline{a}|}{}^j{}_i {\bar B}_{|\underline{b}]} - \frac{1}{2} \delta^j_i {\bar A}_{2 [\underline{a}|}{}^k{}_k {\bar B}_{|\underline{b}]}  \\ & + \frac{1}{2} \epsilon^{jklm} \left(  {\bar A}_{\underline{a} \underline{b} lm} {\bar B}_{ik} - {\bar A}_{\underline{a} \underline{b} ik} {\bar B}_{ lm} \right) = 0 \nonumber
\end{align}
Rep $\mathbf{(15,n(n-1)/2)_0}$:
\begin{align}
    \label{T35}
     & \frac{2}{3} {\bar A}_{2 ik} A_{\underline{a} \underline{b}}{}^{jk} - \frac{2}{3} A_2^{jk} {\bar A}_{\underline{a} \underline{b} ik} - \frac{1}{6} \delta^j_i \left( {\bar A}_{2 kl} A_{\underline{a} \underline{b}}{}^{kl} - A_2^{kl} {\bar A}_{\underline{a} \underline{b} kl} \right) \nonumber \\ 
    & - \frac{1}{3} \epsilon_{iklm} A_1^{jk} A_{\underline{a} \underline{b}}{}^{lm} + \frac{1}{3} \epsilon^{jklm} {\bar A}_{1 ik} {\bar A}_{\underline{a} \underline{b} lm} + 2 A_{2 [\underline{a}|i}{}^k {\bar A}_{2 |\underline{b}]}{}^j{}_k \nonumber \\ & 
    - 2 A_{2 [\underline{a}|k}{}^j {\bar A}_{2 |\underline{b}]}{}^k{}_i - A_{\underline{a} \underline{b} \underline{c}} \left( {\bar A}_2{}^{\underline{c}j}{}_i - \frac{1}{4} \delta^j_i {\bar A}_2{}^{\underline{c}k}{}_k \right)  \nonumber \\ & + {\bar A}_{\underline{a} \underline{b} \underline{c}} \left( A_2{}^{\underline{c}}{}_i{}^j - \frac{1}{4} \delta^j_i A_2{}^{\underline{c}}{}_k{}^k \right) + 4 A_{[\underline{a}}{}^{\underline{c} jk} {\bar A}_{\underline{b}] \underline{c} ik} - \delta^j_i A_{[\underline{a}}{}^{\underline{c} kl} {\bar A}_{\underline{b}] \underline{c} kl}\\ 
    & + A_{2 [\underline{a}|k}{}^k {\bar A}_{2 |\underline{b}]}{}^j{}_i + A_{2 [\underline{a}|i}{}^j {\bar A}_{2 |\underline{b}]}{}^k{}_k  
     - \frac{1}{2} \delta^j_i   A_{2 [\underline{a}|k}{}^k {\bar A}_{2 |\underline{b}]}{}^l{}_l   \nonumber \\ 
     & + A_{\underline{a} \underline{b}}{}^{jk} {\bar B}_{ik} - {\bar A}_{\underline{a} \underline{b} ik} B^{jk} - \frac{1}{4} \delta^j_i \left(  A_{\underline{a} \underline{b}}{}^{kl} {\bar B}_{kl} - {\bar A}_{\underline{a} \underline{b} kl} B^{kl}  \right) \nonumber \\ 
     & - A_{2 [\underline{a}|i}{}^j {\bar B}_{|\underline{b}]} + {\bar A}_{2 [\underline{a}|}{}^j{}_i B_{|\underline{b}]} + \frac{1}{4} \delta^j_i \left(   A_{2 [\underline{a}|k}{}^k {\bar B}_{|\underline{b}]}  -  {\bar A}_{2 [\underline{a}|}{}^k{}_k B_{|\underline{b}]}  \right) = 0 \nonumber
\end{align}
Rep $\mathbf{(6,n(n-1)(n-2)/6)_{+2}}$:
\begin{align}
    \label{T36}
       & 6 A_{[\underline{a} \underline{b}}{}^{lm} A_{2 \underline{c}][j}{}^k \epsilon_{i]klm} +  \epsilon_{ijkl} \left( 3 A_{[\underline{a} \underline{b}| \underline{d}} A_{|\underline{c}]}{}^{\underline{d} kl} + \frac{1}{3}  A_2^{kl} A_{\underline{a} \underline{b} \underline{c}} \right) \nonumber \\
    &  + \frac{1}{2} \epsilon_{ijkl} \left( A_{\underline{a} \underline{b} \underline{c}} B^{kl} - 3 A_{[\underline{a} \underline{b}}{}^{kl} B_{\underline{c}]} \right) = 0 
\end{align}
Rep $\mathbf{(6,n(n-1)(n-2)/6)_{-2}}$:
\begin{align}
    \label{T37}
         & 6 {\bar A}_{2 [\underline{a}|}{}^k{}_{[i|} {\bar A}_{|\underline{b} \underline{c}]j]k} + 3 {\bar A}_{[\underline{a} \underline{b}| \underline{d}} {\bar A}_{|\underline{c}]}{}^{\underline{d}}{}_{ij} + 3 {\bar A}_{2 [\underline{a}|}{}^k{}_k {\bar A}_{|\underline{b} \underline{c}] ij}   + \frac{1}{3} {\bar A}_{2 [ij]} {\bar A}_{\underline{a} \underline{b} \underline{c}} \nonumber \\ 
    &- \frac{3}{2} {\bar B}_{[\underline{a}} {\bar A}_{\underline{b} \underline{c}]ij} + \frac{1}{2} {\bar A}_{\underline{a} \underline{b} \underline{c}} {\bar B}_{ij} = 0 
\end{align}
Rep $\mathbf{(6,n(n-1)(n-2)/6)_0}$:
\begin{align}
   \label{T38}
     & -12 A_{2[\underline{a}[i|}{}^k {\bar A}_{|\underline{b} \underline{c}]j]k} - 6 A_{[\underline{b} \underline{c}}{}^{lm} {\bar A}_{2 \underline{a}]}{}^k{}_{[j} \epsilon_{i]klm} + 6 A_{[\underline{a} \underline{b}|\underline{d}} {\bar A}_{|\underline{c}]}{}^{\underline{d}}{}_{ij}  + 3 \epsilon_{ijkl} {\bar A}_{[\underline{a} \underline{b}|\underline{d}} A_{|\underline{c}]}{}^{\underline{d} kl} \nonumber \\ 
    & + 3 \epsilon_{ijkl} A_{[\underline{a} \underline{b}}{}^{kl} {\bar A}_{2 \underline{c}]}{}^m{}_m + \frac{2}{3} {\bar A}_{2 [ij]} A_{\underline{a} \underline{b} \underline{c}} + \frac{1}{3} \epsilon_{ijkl} A_2^{kl} {\bar A}_{\underline{a} \underline{b} \underline{c}}  \\
    & + A_{\underline{a} \underline{b} \underline{c}} {\bar B}_{ij} + \frac{1}{2} \epsilon_{ijkl} {\bar A}_{\underline{a} \underline{b} \underline{c}} B^{kl} -3 B_{[\underline{a}} {\bar A}_{\underline{b} \underline{c}]ij} - \frac{3}{2} \epsilon_{ijkl} A_{[\underline{a} \underline{b}}{}^{kl} {\bar B}_{\underline{c}]} = 0 \nonumber
\end{align}
Rep $\mathbf{(1,n(n-1)(n-2)(n-3)/24)_{+2}}$:
\begin{align}
    \label{T39} 
      & - \frac{3}{2} \epsilon_{ijkl} A_{[\underline{a} \underline{b}}{}^{ij} A_{\underline{c} \underline{d}]}{}^{kl} + 3 A_{\underline{e}[\underline{a} \underline{b}} A_{\underline{c} \underline{d}]}{}^{\underline{e}}   + 2 A_{[\underline{a} \underline{b} \underline{c}} A_{2 \underline{d}]i}{}^i - 2  A_{[\underline{a} \underline{b} \underline{c}} B_{\underline{d}]} = 0 
\end{align}
Rep $\mathbf{(1,n(n-1)(n-2)(n-3)/24)_0}$:
\begin{align}
    \label{T40}
      & - 3 A_{[\underline{a} \underline{b}}{}^{ij} {\bar A}_{\underline{c} \underline{d}] ij } + 3 A_{\underline{e} [\underline{a} \underline{b}} {\bar A}_{\underline{c} \underline{d}]}{}^{\underline{e}} + A_{[\underline{a} \underline{b} \underline{c}} {\bar A}_{2 \underline{d}]}{}^i{}_i  + {\bar A}_{[\underline{a} \underline{b} \underline{c}} A_{2 \underline{d}]i}{}^i - A_{[\underline{a} \underline{b} \underline{c}} {\bar B}_{\underline{d}]} - {\bar A}_{[\underline{a} \underline{b} \underline{c}} B_{\underline{d}]} = 0
\end{align}
\subsection{From \eqref{q10}}
Rep $\mathbf{(15,1)_0}$:
\begin{equation}
\label{T41}
     A_2^{(jk)} {\bar B}_{ik} - {\bar A}_{2 (ik)} B^{jk} - \frac{1}{2} \epsilon_{iklm} A_1^{jk} B^{lm} + \frac{1}{2} \epsilon^{jklm} {\bar A}_{1ik} {\bar B}_{lm} = 0 
\end{equation}
Rep $\mathbf{(10,n)_0}$:
\begin{equation}
\label{T42}
      {\bar A}_{2 (ij)} B_{\underline{a}} - {\bar A}_{1ij} {\bar B}_{\underline{a}} + 3 A_{2 \underline{a} (i}{}^k {\bar B}_{j)k} + \frac{3}{2} {\bar A}_{2 \underline{a}}{}^k{}_{(i} \epsilon_{j)klm} B^{lm} = 0 
 \end{equation}
Rep $\mathbf{(15,n(n-1)/2)_0}$:
\begin{align}
\label{T43}
     & 2 A_{\underline{a} \underline{b}}{}^{jk} {\bar B}_{ik} + 2 {\bar A}_{\underline{a} \underline{b} ik} B^{jk} - \frac{1}{2} \delta^j_i \left( A_{\underline{a} \underline{b}}{}^{kl} {\bar B}_{kl} + {\bar A}_{\underline{a} \underline{b} kl} B^{kl}\right)  \nonumber \\ 
    & - 2 A_{2[\underline{a}|i}{}^j {\bar B}_{|\underline{b}]} - 2 {\bar A}_{2 [\underline{a}|}{}^j{}_i B_{|\underline{b}]} + \frac{1}{2} \delta^j_i \left(  A_{2 [\underline{a}|k}{}^k {\bar B}_{|\underline{b}]} + {\bar A}_{2[\underline{a}|}{}^k{}_k B_{|\underline{b}]} \right) = 0 
\end{align} 
Rep $\mathbf{(6,n(n-1)(n-2)/6)_0}$:
\begin{align}
\label{T44}
       A_{\underline{a} \underline{b} \underline{c}} {\bar B}_{ij} - \frac{1}{2} \epsilon_{ijkl} B^{kl} {\bar A}_{\underline{a} \underline{b} \underline{c}}   + 3 B_{[\underline{a}} {\bar A}_{\underline{b} \underline{c}] ij } - \frac{3}{2} \epsilon_{ijkl} {\bar B}_{[\underline{a}} A_{\underline{b} \underline{c}]}{}^{kl} = 0 
\end{align}
Rep $ \mathbf{(1,n(n-1)(n-2)(n-3)/24)_0}$:
\begin{align}
\label{T45}
     B_{[\underline{a}} {\bar A}_{\underline{b} \underline{c} \underline{d}]} - {\bar B}_{[\underline{a}} A_{\underline{b} \underline{c} \underline{d}]} = 0 
\end{align}

\subsection{From \eqref{q11}}
Rep $(\mathbf{(15\times15)}_A,\mathbf{1})_{\mathbf{0}}$:
\begin{align}
    \label{T46}
     & \frac{2}{9} \delta^j_i \left( A_2^{(lm)} {\bar A}_{2 (km)} - A_1^{lm} {\bar A}_{1 km} \right) - \frac{2}{9} \delta^l_k \left( A_2^{(jm)} {\bar A}_{2 (im)} - A_1^{jm} {\bar A}_{1 im} \right) \nonumber \\
    & - \frac{2}{9} \epsilon_{ikmn} \left( A_1^{jm} A_2^{(ln)} - A_2^{(jm)} A_1^{ln}  \right) - \frac{2}{9} \epsilon^{jlmn} \left(  {\bar A}_{1im} {\bar A}_{2 (kn)} - {\bar A}_{2 (im)} {\bar A}_{1 kn} \right) \nonumber \\ 
    & + A_{2 \underline{a} k}{}^j {\bar A}_2{}^{\underline{a}l}{}_i - A_{2 \underline{a} i}{}^l {\bar A}_2{}^{\underline{a}j}{}_k + \frac{1}{4} \delta^j_k \left( A_{2 \underline{a} i}{}^l  {\bar A}_2{}^{\underline{a}m}{}_m - A_{2 \underline{a} m}{}^m  {\bar A}_2{}^{\underline{a}l}{}_i \right) \nonumber \\ 
    & - \frac{1}{4} \delta^l_i \left( A_{2 \underline{a} k}{}^j  {\bar A}_2{}^{\underline{a}m}{}_m - A_{2 \underline{a} m}{}^m  {\bar A}_2{}^{\underline{a}j}{}_k \right) - \frac{4}{9} A_2^{[jl]} {\bar A}_{2 (ik)} - \frac{4}{9} A_2^{(jl)} {\bar A}_{2 [ik]}  \nonumber \\ 
    & - \frac{2}{9} \delta^j_i \left( A_2^{[lm]} {\bar A}_{2 (km)} + A_2^{(lm)} {\bar A}_{2 [km]}  \right) + \frac{2}{9} \delta^l_k \left( 
      A_2^{[jm]} {\bar A}_{2 (im)} + A_2^{(jm)} {\bar A}_{2 [im]}   \right)  \nonumber \\ 
      & + \frac{1}{9} \delta^l_i \left(   A_2^{[jm]} {\bar A}_{2 (km)} - A_2^{(jm)} {\bar A}_{2 [km]}   \right) - \frac{1}{9} \delta^j_k \left( A_2^{[lm]} {\bar A}_{2 (im)} - A_2^{(lm)} {\bar A}_{2 [im]}\right) \nonumber \\ 
      & + \frac{1}{9} \epsilon^{jlmn} \left(   {\bar A}_{2 mn}  {\bar A}_{1ik} - {\bar A}_{2 [im]} {\bar A}_{1 kn} -  {\bar A}_{2 [km]} {\bar A}_{1 in}\right) \nonumber \\ 
      & + \frac{1}{9} \epsilon_{ikmn} \left(  A_2^{mn} A_1^{jl} - A_2^{[jm]} A_1^{ln} - A_2^{[lm]} A_1^{jn}  \right) + \frac{1}{18} \delta^j_i \Big{(} \epsilon_{kmnp} A_2^{np} A_1^{lm} \nonumber \\ 
      & + \epsilon^{lmnp} {\bar A}_{2 np} {\bar A}_{1 km} \Big{)} - \frac{1}{18} \delta^l_k \left( \epsilon_{imnp} A_2^{np} A_1^{jm} + \epsilon^{jmnp} {\bar A}_{2 np} {\bar A}_{1 im} \right)  \\
      &  + \frac{2}{9} \delta^j_i A_2^{[lm]} {\bar A}_{2 [km]} - \frac{2}{9} \delta^l_k A_2^{[jm]} {\bar A}_{2 [im]} + \frac{2}{3} {\bar A}_{2 (ik)} B^{jl} + \frac{2}{3} A_2^{(jl)} {\bar B}_{ik} \nonumber \\
      & + \frac{1}{3} \delta^j_i \left( {\bar A}_{2 (km)} B^{lm} + A_2^{(lm)} {\bar B}_{km}   \right) - \frac{1}{3} \delta^l_k \left(  {\bar A}_{2 (im)} B^{jm} + A_2^{(jm)} {\bar B}_{im} \right) \nonumber \\ 
      & - \frac{1}{6} \delta^l_i \left(   {\bar A}_{2 (km)} B^{jm} - A_2^{(jm)} {\bar B}_{km}  \right) + \frac{1}{6} \delta^j_k \left(   {\bar A}_{2 (im)} B^{lm} - A_2^{(lm)} {\bar B}_{im} \right) \nonumber \\ 
      & - \frac{1}{6} \epsilon^{jlmn} \left(  {\bar A}_{1 ik} {\bar B}_{mn} + {\bar A}_{1im} {\bar B}_{kn} + {\bar A}_{1km} {\bar B}_{in}   \right) \nonumber \\ 
      & - \frac{1}{6} \epsilon_{ikmn} \left( A_1^{jl} B^{mn} + A_1^{jm} B^{ln} + A_1^{lm} B^{jn}  \right) - \frac{1}{12} \delta^j_i \Big{(} \epsilon_{kmnp} A_1^{lm} B^{np}  \nonumber \\ & + \epsilon^{lmnp} {\bar A}_{1 km} {\bar B}_{np}  \Big{)} + \frac{1}{12} \delta^l_k \left(  \epsilon_{imnp} A_1^{jm} B^{np} + \epsilon^{jmnp} {\bar A}_{1 im} {\bar B}_{np}  \right) \nonumber \\
      & - \frac{1}{3} \delta^j_i \left(  {\bar A}_{2 [km]} B^{lm} + A_2^{[lm]} {\bar B}_{km}  \right) + \frac{1}{3} \delta^l_k \left( {\bar A}_{2 [im]} B^{jm} + A_2^{[jm]} {\bar B}_{im} \right) \nonumber \\
      & - \frac{3}{2} \delta^j_i B^{lm} {\bar B}_{km} + \frac{3}{2} \delta^l_k B^{jm} {\bar B}_{im} = 0 \nonumber 
\end{align}
The tensor product $(\mathbf{15} \times \mathbf{15})_A$ of SU(4) decomposes as 
\begin{equation}
    (\mathbf{15} \times \mathbf{15})_A = \mathbf{15} + \mathbf{45} + \overline{\mathbf{45}}\,.
\end{equation}
The component of the quadratic constraint \eqref{T46} in the $\mathbf{15}$ of SU(4) follows from contracting \eqref{T46} with $\delta^k_l$, which yields
\begin{align}
    \label{T4615}
    \mathbf{(15,1)_0}: \qquad & \frac{8}{9} A_1^{jk} {\bar A}_{1ik} - \frac{8}{9} A_2^{(jk)} {\bar A}_{2 (ik)} - \frac{2}{9} \delta^j_i \left( A_1^{kl} {\bar A}_{1kl} - A_2^{(kl)} {\bar A}_{2 kl} \right) \nonumber \\ 
    & + A_{2 \underline{a} k}{}^j {\bar A}_{2}{}^{\underline{a}k}{}_i -   A_{2 \underline{a} i}{}^k {\bar A}_{2}{}^{\underline{a}j}{}_k + \frac{4}{9} A_2^{[jk]} {\bar A}_{2 (ik)} + \frac{4}{9} A_2^{(jk)} {\bar A}_{2 [ik]} \nonumber \\ 
    & - \frac{2}{9} \epsilon^{jklm} {\bar A}_{1ik} {\bar A}_{2 lm} - \frac{2}{9} \epsilon_{iklm} A_1^{jk} A_2^{lm} - \frac{8}{9} A_2^{[jk]} {\bar A}_{2 [ik]} + \frac{2}{9} \delta^j_i A_2^{[kl]} {\bar A}_{2 kl} \nonumber \\
    & - \frac{2}{3} {\bar A}_{2 (ik)} B^{jk} - \frac{2}{3} A_2^{(jk)} {\bar B}_{ik} + \frac{1}{3} \epsilon^{jklm} {\bar A}_{1ik} {\bar B}_{lm} + \frac{1}{3} \epsilon_{iklm} A_1^{jk} B^{lm}  \\ 
    &  + \frac{4}{3} {\bar A}_{2 [ik]} B^{jk}  + \frac{4}{3} A_2^{[jk]} {\bar B}_{ik} - \frac{1}{3} \delta^j_i \left( {\bar A}_{2 kl} B^{kl} + A_2^{kl} {\bar B}_{kl} \right) \nonumber \\
    & + 6 B^{jk} {\bar B}_{ik} - \frac{3}{2} \delta^j_i B^{kl} {\bar B}_{kl} = 0 \nonumber
\end{align}
Rep $\mathbf{(15\times6,n)_0}$:
\begin{align}
    \label{T47}
      & - \frac{2}{3} A_{2 \underline{a}[i}{}^m \epsilon_{j]kmn} A_1^{ln} - \frac{2}{3} {\bar A}_{2 \underline{a}}{}^m{}_{[i} \epsilon_{j]kmn} A_2^{(ln)} + \frac{2}{3} {\bar A}_{2 \underline{a}}{}^l{}_{[i} {\bar A}_{1 j]k} \nonumber \\ & - \frac{1}{3} {\bar A}_{2 (ik)} A_{2 \underline{a}j}{}^l + \frac{1}{3} {\bar A}_{2 (jk)} A_{2 \underline{a}i}{}^l - \frac{2}{3}
    {\bar A}_{2 \underline{a}}{}^m{}_{[i} \delta^l_{j]} {\bar A}_{1 km} - \frac{2}{3} A_{2 \underline{a} [i}{}^m \delta^l_{j]} {\bar A}_{2 (km)}   \nonumber \\ 
    & + {\bar A}_{\underline{a} 
    \underline{b} ij } \left(  A_2{}^{\underline{b}}{}_k{}^l - \frac{1}{4} \delta^l_k A_2{}^{\underline{b}}{}_m{}^m  \right) + \frac{1}{2} \epsilon_{ijmn} A_{\underline{a} \underline{b} }{}^{mn} \left( 
     {\bar A}_{2}{}^{\underline{b} l}{}_k - \frac{1}{4} \delta^l_k {\bar A}_{2}{}^{\underline{b} p}{}_p    \right) \nonumber \\ 
     & + \frac{1}{3} \epsilon_{ijkm} A_1^{lm} A_{2 \underline{a} n}{}^n - \frac{1}{3} {\bar A}_{2 [ij]} A_{2 \underline{a} k}{}^l - \frac{1}{3}  {\bar A}_{2 [ik]} A_{2 \underline{a} j}{}^l + \frac{1}{3}  {\bar A}_{2 [jk]} A_{2 \underline{a} i}{}^l \nonumber \\ 
     & - \frac{1}{6} \epsilon_{ijmn} A_2^{mn} {\bar A}_{2 \underline{a}}{}^l{}_k + \frac{1}{6} {\bar A}_{2 \underline{a}}{}^l{}_{[i} \epsilon_{j]kmn} A_2^{mn} + \frac{1}{3} {\bar A}_{2 \underline{a}}{}^m{}_{[i} \epsilon_{j]kmn} A_2^{[ln]} \nonumber \\
      & - \frac{1}{12} \epsilon_{ijkm} A_2^{[lm]} {\bar A}_{2 \underline{a}}{}^n{}_n - \frac{1}{3} \delta^l_i {\bar A}_{2 jk} A_{2 \underline{a} m}{}^m + \frac{1}{3} \delta^l_j {\bar A}_{2 ik} A_{2 \underline{a} m}{}^m  \nonumber \\ & + \frac{2}{3} A_{2 \underline{a} [i}{}^m \delta^l_{j]} {\bar A}_{2 [km]} + \frac{1}{12} \delta^l_{[i} \epsilon_{j]kmn} A_2^{mn} {\bar A}_{2 \underline{a}}{}^p{}_p + \frac{1}{6} {\bar A}_{2 \underline{a}}{}^m{}_{[i} \delta^l_{j]} \epsilon_{kmnp} A_2^{np} \nonumber \\ 
      & + \frac{1}{6} \delta^l_k \left( 
       {\bar A}_{2 [im]} A_{2 \underline{a} j}{}^m -   {\bar A}_{2 [jm]} A_{2 \underline{a} i}{}^m  \right) + \frac{1}{12} \delta^l_k \epsilon_{ijmn} A_2^{mn} {\bar A}_{2 \underline{a}}{}^p{}_p \nonumber \\& - \frac{1}{12} \delta^l_k {\bar A}_{2 [ij]} A_{2 \underline{a} m}{}^m  + \frac{1}{6} \epsilon_{ijkm} \left( A_1^{lm} B_{\underline{a}} - A_2^{(lm)} {\bar B}_{\underline{a}}  \right) - A_{2 \underline{a} [i}{}^l {\bar B}_{j]k} \\ 
       & + \frac{1}{2} A_{2 \underline{a} k}{}^l {\bar B}_{ij} + \frac{1}{4} \epsilon_{ijmn} {\bar A}_{2 \underline{a}}{}^l{}_k B^{mn} - \frac{1}{4} {\bar A}_{2 \underline{a}}{}^l{}_{[i} \epsilon_{j]kmn} B^{mn} \nonumber \\ & - \frac{1}{2} {\bar A}_{2 \underline{a}}{}^m{}_{[i} \epsilon_{j]kmn} B^{ln}  + \frac{1}{8} \epsilon_{ijkm} {\bar A}_{2 \underline{a}}{}^n{}_n B^{lm } + 
 \frac{1}{3} \delta^l_{[i} {\bar A}_{1 j]k} {\bar B}_{\underline{a}} \nonumber \\ & - \frac{1}{6} \delta^l_i {\bar A}_{2 (jk)} B_{\underline{a}} + \frac{1}{6} \delta^l_j {\bar A}_{2 (ik)} B_{\underline{a}} + \delta^l_{[i} {\bar B }_{j]k} A_{2 \underline{a} m}{}^m - A_{2 \underline{a} [i}{}^m \delta^l_{j]} {\bar B}_{km} \nonumber \\ & - \frac{1}{4} {\bar A}_{2 \underline{a} }{}^m{}_{[i}  \delta^l_{j]} \epsilon_{kmnp} B^{np} - \frac{1}{8} \delta^l_{[i} \epsilon_{j]kmn} {\bar A}_{2 \underline{a}}{}^p{}_p B^{mn} + \frac{1}{2} \delta^l_k A_{2 \underline{a} [i}{}^m {\bar B}_{j]m} \nonumber \\ & + \frac{1}{8} \delta^l_k A_{2 \underline{a} m}{}^m {\bar B}_{ij}  - \frac{1}{8} \delta^l_k \epsilon_{ijmn} {\bar A}_{2 \underline{a}}{}^p{}_p B^{mn} - \frac{1}{6} \delta^l_i {\bar A}_{2 [jk] } B_{\underline{a}} + \frac{1}{6} \delta^l_j {\bar A}_{2 [ik]} B_{\underline{a}} \nonumber \\ 
 & + \frac{1}{6} \delta^l_{[i} \epsilon_{j]kmn} A_2^{mn} {\bar B}_{\underline{a}} - \frac{1}{12} \delta^l_k {\bar A}_{2 [ij]} B_{\underline{a}} + \frac{1}{24} \delta^l_k \epsilon_{ijmn} A_2^{mn} {\bar B}_{\underline{a}} \nonumber \\ &  - \frac{3}{2} \delta^l_{[i} {\bar B}_{j] k} B_{\underline{a}} + \frac{3}{4} \delta^l_{[i} \epsilon_{j]kmn} B^{mn} {\bar B}_{\underline{a}} - \frac{3}{8} \delta^l_k {\bar B}_{ij} B_{\underline{a}}  + \frac{3}{16} \delta^l_k \epsilon_{ijmn} B^{mn} {\bar B}_{\underline{a}} = 0  \nonumber 
    \end{align}
   The tensor product $\mathbf{15\times6}$ of SU(4) decomposes according to  
\begin{equation}
    \mathbf{15} \times \mathbf{6} = \mathbf{6} + \mathbf{10} + \overline{\mathbf{10}} + \mathbf{64} \,.
\end{equation}
In order to specify the components of \eqref{T47} in the 
$\mathbf{10}$ and $\mathbf{6}$ representations of SU(4), we first contract \eqref{T47} with $\delta^j_l$. To obtain the $\mathbf{10}$ component, we symmetrize the resulting identity in $i$ and $k$, whereas to get the $\mathbf{6}$ component, we antisymmetrize in $i$ and $k$. The results are 
\begin{align}
    \label{T4710}
    \mathbf{(10,n)_0}: \qquad & - \frac{2}{3} {\bar A}_{2 \underline{a}}{}^j{}_{(i} {\bar A}_{1 k)j} - \frac{1}{3} {\bar A}_{2 (ij)} A_{2 \underline{a} k}{}^j - \frac{1}{3} {\bar A}_{2 (jk)} A_{2 \underline{a} i}{}^j - {\bar A}_{\underline{a} \underline{b} j (i} A_2{}^{\underline{b}}{}_{k)}{}^j \nonumber \\ 
    &  + \frac{1}{2} {\bar A}_2{}^{\underline{b} l}{}_{(i} \epsilon_{k)lmn} A_{\underline{a} \underline{b}}{}^{mn} - \frac{1}{3} {\bar A}_{1 ik} {\bar A}_{2 \underline{a}}{}^j{}_j + \frac{2}{3} {\bar A}_{2 (ik)} A_{2 \underline{a} j}{}^j  + \frac{1}{6} {\bar A}_{2 [ij]} A_{2 \underline{a} k}{}^j \nonumber \\ 
    & - \frac{1}{6} {\bar A}_{2 [jk]} A_{2 \underline{a} i}{}^j + \frac{1}{6} {\bar A}_{2 \underline{a}}{}^l{}_{(i} \epsilon_{k)lmn} A_2^{mn} - \frac{1}{2} A_{2 \underline{a} (i}{}^j {\bar B}_{k)j} - \frac{1}{4} {\bar A}_{2 \underline{a}}{}^l{}_{(i} \epsilon_{k)lmn} B^{mn} \\ 
    & - \frac{1}{2} {\bar A}_{1 ik} {\bar B}_{\underline{a}} + \frac{1}{2} {\bar A}_{2 (ik)} B_{\underline{a}} = 0 \nonumber
\end{align}
\begin{align}
    \label{T476}
    \mathbf{(6,n)_0}: \qquad & - \frac{1}{3} \epsilon_{ikmn} \left( A_1^{lm} A_{2 \underline{a} l}{}^n + A_2^{(lm)} {\bar A}_{2 \underline{a}}{}^n{}_l  \right) - \frac{2}{3} {\bar A}_{2 \underline{a}}{}^j{}_{[i} {\bar A}_{1 k]j} + \frac{1}{3} {\bar A}_{2 (ij)} A_{2 \underline{a} k}{}^j \nonumber \\ & - \frac{1}{3} {\bar A}_{2 (jk)} A_{2 \underline{a} i}{}^j - {\bar A}_{\underline{a} \underline{b} j [i} A_2{}^{\underline{b}}{}_{k]}{}^j - \frac{1}{4} {\bar A}_{\underline{a} \underline{b} ik} A_2{}^{\underline{b}}{}_j{}^j - \frac{1}{2} {\bar A}_2{}^{\underline{b} l}{}_{[i} \epsilon_{k]lmn} A_{\underline{a} \underline{b}}{}^{mn} \nonumber \\ 
    & - \frac{1}{8} \epsilon_{iklm} A_{\underline{a} \underline{b}}{}^{lm} {\bar A}_2{}^{\underline{b} n}{}_n - \frac{1}{3} {\bar A}_{2 [ij]} A_{2 \underline{a} k}{}^j - \frac{1}{3} {\bar A}_{2 [jk]} A_{2 \underline{a} i}{}^j + \frac{1}{2} {\bar A}_{2 \underline{a}}{}^l{}_{[i} \epsilon_{k]lmn} A_2^{mn}  \nonumber \\ 
    & - \frac{1}{6} \epsilon_{iklm} A_2^{[ln]} {\bar A}_{2 \underline{a}}{}^m{}_n + \frac{7}{12} {\bar A}_{2 [ik]} A_{2 \underline{a} j}{}^j - \frac{1}{24} \epsilon_{iklm} A_2^{lm} {\bar A}_{2 \underline{a}}{}^n{}_n - A_{2 \underline{a} [i}{}^j {\bar B}_{k]j} \\ 
    & - \frac{3}{4}  {\bar A}_{2 \underline{a}}{}^l{}_{[i} \epsilon_{k]lmn} B^{mn} - \frac{1}{4} \epsilon_{iklm} {\bar A}_{2 \underline{a}}{}^l{}_n B^{mn} - \frac{7}{8} A_{2 \underline{a} j}{}^j {\bar B}_{ik} + \frac{1}{16} \epsilon_{ikmn} {\bar A}_{2 \underline{a}}{}^l{}_l B^{mn}  \nonumber \\ 
    & + \frac{5}{12} {\bar A}_{2 [ik]} B_{\underline{a}} - \frac{5}{24} \epsilon_{iklm} A_2^{lm} {\bar B}_{\underline{a}} + \frac{15}{8} {\bar B}_{ik} B_{\underline{a}} - \frac{15}{16} \epsilon_{iklm} B^{lm} {\bar B}_{\underline{a}} = 0 \nonumber 
\end{align}
Rep
$(((\mathbf{6},\mathbf{n})\times(\mathbf{6},\mathbf{n}))_A)_{\mathbf{0}}$:
\begin{align}
    \label{T48}
     \qquad & - 2 A_{2 \underline{a} [i|}{}^m {\bar A}_{2 \underline{b}}{}^{[k}{}_m \delta^{l]}_{|j]} + 2 {\bar A}_{2 \underline{a}}{}^m{}_{[i|} A_{2 \underline{b} m}{}^{[k} \delta^{l]}_{|j]} - 4 A_{2 [\underline{a}[i|}{}^{[k} {\bar A}_{2 |\underline{b}]}{}^{l]}{}_{|j]}  \nonumber \\ & 
    + 2 A_{[\underline{a}}{}^{\underline{c} kl} {\bar A}_{\underline{b}] \underline{c} ij} - 4 \delta^{[k}_{[i|} A_{\underline{a} \underline{c}}{}^{l] m} {\bar A}_{\underline{b}}{}^{\underline{c}}{}_{|j]m} + \delta^{[k}_i \delta^{l]}_j A_{\underline{a} \underline{c}}{}^{mn} {\bar A}_{\underline{b}}{}^{\underline{c}}{}_{mn} \nonumber \\ 
    &  + 2 A_{2 \underline{a} m}{}^m {\bar A}_{2 \underline{b}}{}^{[k}{}_{[i} \delta^{l]}_{j]} + 2 {\bar A}_{2 \underline{b}}{}^m{}_m A_{2 \underline{a} [i}{}^{[k} \delta^{l]}_{j]} + \frac{2}{3} {A}_2^{[kl]} {\bar A}_{\underline{a} \underline{b} ij} \nonumber \\ 
    & - \frac{2}{3} {\bar A }_{2[ij]} A_{\underline{a} \underline{b}}{}^{kl}  - \frac{2}{3} \delta^k_{[i|} A_2^{[lm]} {\bar A}_{\underline{a} \underline{b} |j] m} + \frac{2}{3} \delta^l_{[i|} A_2^{[km]} {\bar A}_{\underline{a} \underline{b} |j] m} \nonumber \\ 
    &  + \frac{2}{3} \delta^{[k}_i {\bar A}_{2[jm]} A_{\underline{a} \underline{b}}{}^{l]m} -  \frac{2}{3} \delta^{[k}_j {\bar A}_{2[im]} A_{\underline{a} \underline{b}}{}^{l]m}  + \frac{1}{3} \delta^{[k}_i \delta^{l]}_j \big{(} A_2^{mn} {\bar A}_{\underline{a} \underline{b} mn} \nonumber \\ & - {\bar A}_{2 mn} A_{\underline{a} \underline{b}}{}^{mn} \big{)} - \frac{1}{9} \delta_{\underline{a} \underline{b}} \Big{(} \delta^k_i A_2^{[lm]} {\bar A}_{2 [jm]} - \delta^k_j A_2^{[lm]} {\bar A}_{2 [im]} \nonumber \\ 
    & - \delta^l_i A_2^{[km]} {\bar A}_{2 [jm]}  + \delta^l_j A_2^{[km]} {\bar A}_{2 [im]} \Big{)} + \frac{1}{9} \delta_{\underline{a} \underline{b}} \delta^{[k}_i \delta^{l]}_j A_2^{[mn]} {\bar A}_{2 mn}  \\ 
    & - \delta^{[k}_{i} \delta^{l]}_j A_{2 \underline{a} m}{}^m {\bar A}_{2 \underline{b}}{}^n{}_n + A_{\underline{a} \underline{b}}{}^{kl} {\bar B}_{ij} - {\bar A}_{\underline{a} \underline{b} ij} B^{kl} -2 \delta^{[k}_{[i} {\bar B}_{j]m} A_{\underline{a} \underline{b}}{}^{l]m}\nonumber \\ &  + 2 \delta^{[k}_{[i|} {\bar A}_{\underline{a} \underline{b} |j]m} B^{l]m}   + \frac{1}{2} \delta^{[k}_i \delta^{l]}_j \left( A_{\underline{a} \underline{b}}{}^{mn} {\bar B}_{mn} - {\bar A}_{\underline{a} \underline{b} mn} B^{mn}  \right) \nonumber \\&+ 2 B_{(\underline{a}} {\bar A}_{2 \underline{b})}{}^{[k}{}_{[i} \delta^{l]}_{j]}   + 2 {\bar B}_{(\underline{a}} A_{2 \underline{b}) [i}{}^{[k} \delta^{l]}_{j]} - \frac{1}{2} \delta^{[k}_i \delta^{l]}_j \Big{(} B_{\underline{a}} {\bar A}_{2 \underline{b}}{}^m{}_m \nonumber \\ 
    & + {\bar B}_{\underline{b}} A_{2 \underline{a} m}{}^m  \Big{)} + \frac{1}{3} \delta_{\underline{a} \underline{b}} \Big{(}    \delta^k_{[i} A_2^{[lm]} {\bar B}_{j]m} - \delta^l_{[i} A_2^{[km]} {\bar B}_{j]m}  + \delta^{[k}_i B^{l]m} {\bar A}_{2 [jm]} \nonumber \\ 
    & - \delta^{[k}_j B^{l]m} {\bar A}_{2 [im]} \Big{)} - \frac{1}{6} \delta_{\underline{a} \underline{b}} \delta^{[k}_i \delta^{l]}_j \left( 
     A_2^{mn} {\bar B}_{mn}  + {\bar A}_{2 mn} B^{mn}   \right) \nonumber \\
     &  + 3 \delta_{\underline{a} \underline{b}} \delta^{[k}_{[i} B^{l]m} {\bar B}_{j]m} - \frac{3}{4}   \delta_{\underline{a} \underline{b}} \delta^{[k}_i \delta^{l]}_j B^{mn} {\bar B}_{mn}  + \frac{3}{2} \delta^{[k}_i \delta^{l]}_j B_{[\underline{a}} {\bar B}_{\underline{b}]} = 0 \nonumber 
\end{align}
The tensor product $((\mathbf{6},\mathbf{n})\times(\mathbf{6},\mathbf{n}))_A$ of $\text{SU}(4) \times \text{SO}(n)$ decomposes as
\begin{align}
((\mathbf{6},\mathbf{n})\times(\mathbf{6},\mathbf{n}))_A = & \, \left(\mathbf{1},\mathbf{ n (n-1)/2} \right) + \left({\mathbf{20}}^{'},\mathbf{ n (n-1)/2}\right) \nonumber \\
& + \left( \mathbf{15}, \mathbf{ n (n+1)/2 -1}  \right) + \left( \mathbf{15},\mathbf{1}  \right).
\end{align} 
In order to specify the component of \eqref{T48} transforming in the (reducible) $\left( \mathbf{15}, \mathbf{ n (n+1) /2 }\right)$ representation of $\text{SU}(4) \times \text{SO}(n)$, we contract \eqref{T48} with $\delta^j_l$ and we then symmetrize the resulting equation in $\underline{a}$ and $\underline{b}$. We find
\begin{align}
\label{T4815}
    \mathbf{(15,n(n+1)/2)_0}: \qquad & - A_{2 (\underline{a}|i}{}^j {\bar A}_{2 |\underline{b})}{}^k{}_j + A_{2 (\underline{a}|j}{}^k {\bar A}_{2 |\underline{b})}{}^j{}_i - 2 A_{(\underline{a}}{}^{\underline{c} kj} {\bar A}_{\underline{b}) \underline{c} ij} \nonumber \\ & + \frac{1}{2} \delta^k_i A_{(\underline{a}}{}^{\underline{c} jl} {\bar A}_{\underline{b}) \underline{c} jl} + A_{2 (\underline{a}|j}{}^j {\bar A}_{2 |\underline{b})}{}^k{}_i   + A_{2 (\underline{a}|i}{}^k {\bar A}_{2 |\underline{b})}{}^j{}_j \nonumber \\ & - \frac{1}{2} \delta^k_i A_{2 (\underline{a}|j}{}^j {\bar A}_{2 |\underline{b})}{}^l{}_l - \frac{2}{9} \delta_{\underline{a} \underline{b}} \left( A_2^{[kj]} {\bar A}_{2 [ij]} - \frac{1}{4} \delta^k_i A_2^{[jl]} {\bar A}_{2 jl} \right)  \nonumber \\ 
    & + B_{(\underline{a}} {\bar A}_{2 \underline{b})}{}^k{}_i + {\bar B}_{(\underline{a}} A_{2 \underline{b})i}{}^k - \frac{1}{4} \delta^k_i \left( B_{(\underline{a}} {\bar A}_{2 \underline{b})}{}^j{}_j + {\bar B}_{(\underline{a}} A_{2 \underline{b})j}{}^j  \right) \\
    & + \frac{1}{3} \delta_{\underline{a} \underline{b}} \left( A_2^{[kj]} {\bar B}_{ij} + {\bar A}_{2 [ij]} B^{kj} \right) - \frac{1}{12} \delta_{\underline{a} \underline{b}} \delta^k_i \left( 
     A_2^{jl} {\bar B}_{jl} + {\bar A}_{2 jl} B^{jl}   \right) \nonumber \\ 
     & + \frac{3}{2} \delta_{\underline{a} \underline{b}} \left( { B}^{kj}  {\bar B}_{ij} - \frac{1}{4} \delta^k_i B^{jl} {\bar B}_{jl}  \right) =  0  \nonumber 
     \end{align}
On the other hand, the $(\mathbf{1},\mathbf{n(n-1)/2})_{\mathbf{0}}$ component of the quadratic constraint \eqref{T48} follows from contracting \eqref{T48} with $\delta^i_k \delta_l^j$, which gives
\begin{align}
    \label{T481}
    \mathbf{(1,n(n-1)/2)_0}: \qquad  & - 4 A_{2 [\underline{a}|i}{}^j {\bar A}_{2 |\underline{b}]}{}^i{}_j + 2 A_{[\underline{a}}{}^{\underline{c} ij} {\bar A}_{\underline{b}] \underline{c} ij} + \frac{2}{3} A_2^{ij} {\bar A}_{\underline{a} \underline{b} ij} - \frac{2}{3} {\bar A}_{2 ij} A_{\underline{a} \underline{b}}{}^{ij} \nonumber \\ 
    & - 2 A_{2 [\underline{a}|i}{}^i {\bar A}_{2 |\underline{b}]}{}^j{}_j + A_{\underline{a} \underline{b}}{}^{ij} {\bar B}_{ij} - {\bar A}_{\underline{a} \underline{b} ij} B^{ij} - 3 B_{[\underline{a}} {\bar A}_{2 \underline{b}]}{}^i{}_i \\
    & + 3 {\bar B}_{[\underline{a}} A_{2 \underline{b}]i}{}^i + 9 B_{[\underline{a}} {\bar B}_{\underline{b}]} = 0 \nonumber
\end{align}
Rep $\mathbf{(15,n(n-1)/2)_0}$:
\begin{align}
\label{T49}
     & \frac{2}{3} A_2^{(jk)} {\bar A}_{\underline{a} \underline{b} ik} + \frac{2}{3} {\bar A}_{2 (ik)} A_{\underline{a} \underline{b}}{}^{jk} - \frac{1}{3} \epsilon_{iklm} A_1^{jk} A_{\underline{a} \underline{b}}{}^{lm}     - \frac{1}{3} \epsilon^{jklm} {\bar A}_{1ik} {\bar A}_{\underline{a} \underline{b} lm} \nonumber \\ &  - A_{\underline{a} \underline{b} \underline{c}} \left( {\bar A}_2{}^{\underline{c} j}{}_i - \frac{1}{4} \delta^j_i {\bar A}_2{}^{\underline{c} k}{}_k   \right) - {\bar A}_{\underline{a} \underline{b} \underline{c}} \left( A_2{}^{\underline{c}}{}_i{}^j - \frac{1}{4} \delta^j_i A_2{}^{\underline{c}}{}_k{}^k  \right) \nonumber \\ &  - A_{2 [\underline{a}|k}{}^k {\bar A}_{2 |\underline{b}]}{}^j{}_i  +  A_{2 [\underline{a}|i}{}^j {\bar A}_{2 |\underline{b}]}{}^k{}_k   - \frac{2}{3} A_2^{[jk]} {\bar A}_{\underline{a} \underline
    {b} ik} - \frac{2}{3} {\bar A}_{2 [ik]} A_{\underline{a} \underline{b} }{}^{jk} \nonumber \\ & + \frac{1}{6} \delta^j_i \left(A_2^{kl} {\bar A}_{\underline{a} \underline{b} kl} + {\bar A}_{2 kl} A_{\underline{a} \underline{b}}{}^{kl} \right) \\
    & + A_{\underline{a} \underline{b}}{}^{jk} {\bar B}_{ik} + {\bar A}_{\underline{a} \underline{b} ik} B^{jk} - \frac{1}{4} \delta^j_i \left( A_{\underline{a} \underline{b}}{}^{kl} {\bar B}_{kl} + {\bar A}_{\underline{a} \underline{b} kl} B^{kl}  \right) \nonumber \\ 
    & - B_{[\underline{a}} \left(  {\bar A}_{2 \underline{b}]}{}^j{}_i - \frac{1}{4} {\bar A}_{2 \underline{b}]}{}^k{}_k  \delta^j_i \right) - {\bar B}_{[\underline{a}} \left(  A_{2 \underline{b}] i}{}^j - \frac{1}{4}  A_{2 \underline{b}] k}{}^k  \delta^j_i \right) = 0 \nonumber 
\end{align}
Rep $\mathbf{(6,n\times n(n-1)/2)_0}$:
\begin{align}
\label{T50}
      & - 2 {A}_{2 \underline{c} [i|}{}^k {\bar A}_{\underline{a} \underline{b} |j] k} - {\bar A}_{2 \underline{c}}{}^k{}_{[i} \epsilon_{j]klm} A_{\underline{a} \underline{b}}{}^{lm} - A_{\underline{a} \underline{b}}{}^{\underline{d}} {\bar A}_{\underline{c} \underline{d} ij}  + \frac{1}{2} \epsilon_{ijkl} {\bar A}_{\underline{a} \underline{b}}{}^{\underline{d}} A_{\underline{c} \underline{d}}{}^{kl} \nonumber \\ &  + A_{2 [\underline{a}| k}{}^k {\bar A}_{|\underline{b}] \underline{c} ij} - \frac{1}{2} \epsilon_{ijlm} {\bar A}_{2 [\underline{a}|}{}^k{}_k A_{|\underline{b}] \underline{c} }{}^{lm} - A_{2 \underline{c} k}{}^k {\bar A}_{\underline{a} \underline{b} ij} + \frac{1}{3} {\bar A}_{2 [ij]} A_{\underline{a} \underline{b} \underline{c}} - \frac{1}{6} \epsilon_{ijkl} A_2^{kl} {\bar A}_{\underline{a} \underline{b} \underline{c}} \nonumber \\ & - \frac{1}{3} \delta_{\underline{c} [\underline{a}} A_{2 \underline{b}]k}{}^k {\bar A}_{2 [ij]} + \frac{1}{6} \epsilon_{ijlm} \delta_{\underline{c} [\underline{a}} {\bar A}_{2 \underline{b}]}{}^k{}_k A_2^{lm} + B_{[\underline{a}} {\bar A}_{\underline{b}] \underline{c} ij} \nonumber \\
     & - \frac{1}{2} A_{\underline{a} \underline{b} \underline{c}} {\bar B}_{ij} + \frac{1}{4} \epsilon_{ijkl} B^{kl} {\bar A}_{\underline{a} \underline{b} \underline{c}}   - \frac{1}{2} \epsilon_{ijkl} {\bar B}_{[\underline{a}} A_{\underline{b}] \underline{c}}{}^{kl}  - \frac{1}{2} B_{\underline{c}} {\bar A}_{\underline{a} \underline{b} ij} + \frac{1}{4} \epsilon_{ijkl} {\bar B}_{\underline{c}} A_{\underline{a} \underline{b}}{}^{kl} \\   
    & + \frac{1}{2} \delta_{\underline{c} [\underline{a}} A_{2 \underline{b}] k}{}^k {\bar B}_{ij} - \frac{1}{4} \epsilon_{ijlm} \delta_{\underline{c} [\underline{a}} {\bar A}_{2 \underline{b}]}{}^k{}_k B^{lm}\nonumber \\ &  - \frac{1}{3} \delta_{\underline{c} [\underline{a}} B_{\underline{b}]} {\bar A}_{2 [ij]} + \frac{1}{6} \epsilon_{ijkl} \delta_{\underline{c} [\underline{a}} {\bar B}_{\underline{b}]} A_2^{kl} - \frac{3}{2} \delta_{\underline{c} [\underline{a}} B_{\underline{b}]} {\bar B}_{ij} + \frac{3}{4} \epsilon_{ijkl}  \delta_{\underline{c} [\underline{a}} {\bar B}_{\underline{b}]} B^{kl}  = 0 \nonumber 
\end{align}
Rep $\mathbf{(1,\left(  n (n-1) /2 \times n (n-1)/2 \right)_A)_0} $:
\begin{align}
\label{T51}
     & - A_{\underline{a} \underline{b}}{}^{ij} {\bar A}_{\underline{c} \underline{d} ij} + A_{\underline{c} \underline{d}}{}^{ij} {\bar A}_{\underline{a} \underline{b} ij} + A_{\underline{a} \underline{b} \underline{e}} {\bar A}_{\underline{c} \underline{d}}{}^{\underline{e}}  - A_{\underline{c} \underline{d} \underline{e}} {\bar A}_{\underline{a} \underline{b}}{}^{\underline{e}} - {\bar A}_{\underline{c} \underline{d} [\underline{a}} A_{2 \underline{b}]i}{}^i + A_{\underline{c} \underline{d} [\underline{a}} {\bar A}_{2 \underline{b}]}{}^i{}_i \nonumber \\ & + {\bar A}_{\underline{a} \underline{b} [\underline{c}} A_{2 \underline{d}]i}{}^i - A_{\underline{a} \underline{b} [\underline{c}} {\bar A}_{2 \underline{d}]}{}^i{}_i + \delta_{[\underline{a} [\underline{c}} {\bar A}_{2 \underline{d}]}{}^i{}_i A_{2 |\underline{b}]j}{}^j   -  \delta_{[\underline{a} [\underline{c}} A_{2 \underline{d}]i}{}^i {\bar A}_{2 |\underline{b}]}{}^j{}_j \nonumber \\ 
    & - {\bar A}_{\underline{c} \underline{d} [\underline{a}} B_{\underline{b}]} + A_{\underline{c} \underline{d} [\underline{a}} {\bar B}_{\underline{b}]}  + {\bar A}_{\underline{a} \underline{b} [\underline{c}} B_{\underline{d}]} - A_{\underline{a} \underline{b} [\underline{c}} {\bar B}_{\underline{d}]} \\ 
    &+ \delta_{[\underline{a} [\underline{c}} \big{(} {\bar B}_{\underline{d]}} A_{2 \underline{b}]i}{}^i  - B_{\underline{d}]} {\bar A}_{2 \underline{b}]}{}^i{}_i - A_{2 \underline{d}]i}{}^i {\bar B}_{|\underline{b}]}  + {\bar A}_{2 \underline{d}]}{}^i{}_i B_{|\underline{b}]}  - 3 {\bar B}_{\underline{d}]} B_{\underline{b}]} + 3 B_{\underline{d}]} {\bar B}_{\underline{b}]} \big{)} = 0 \nonumber
\end{align}
For $B^{ij}=B^{\underline{a}}=0$, the $T$-identities given in this appendix consistently reduce to those of the standard gauged four-dimensional ${\cal N}=4$ matter-coupled supergravity, which are provided in \cite{DallAgata:2023ahj}.


\begin{thebibliography}{100}

\bibitem{Das:1977uy}
A.~K.~Das,
``SO(4) Invariant Extended Supergravity,''
Phys. Rev. D \textbf{15} (1977), 2805

\bibitem{Cremmer:1977tc}
E.~Cremmer and J.~Scherk, ``Algebraic Simplifications in Supergravity Theories,''
Nucl. Phys. B \textbf{127} (1977), 259-268

\bibitem{Cremmer:1977tt}
E.~Cremmer, J.~Scherk and S.~Ferrara,
``SU(4) Invariant Supergravity Theory,''
Phys. Lett. B \textbf{74} (1978), 61-64

\bibitem{Freedman:1978ra}
D.~Z.~Freedman and J.~H.~Schwarz,
``N=4 Supergravity Theory with Local SU(2) x SU(2) Invariance,''
Nucl. Phys. B \textbf{137} (1978), 333-339

\bibitem{deRoo:1984zyh} M.~de Roo,
``Matter Coupling in N=4 Supergravity,''
Nucl. Phys. B \textbf{255} (1985), 515-531

\bibitem{Bergshoeff:1985ms} 
E.~Bergshoeff, I.~G.~Koh and E.~Sezgin,
``Coupling of Yang-Mills to N=4, D=4 Supergravity,''
Phys. Lett. B \textbf{155} (1985), 71

\bibitem{deRoo:1985np}
M.~de Roo,
``GAUGED N=4 MATTER COUPLINGS,''
Phys. Lett. B \textbf{156} (1985), 331-334

\bibitem{deRoo:1985jh} M.~de Roo and P.~Wagemans,
``Gauged Matter Coupling in $N=4$ Supergravity,''
Nucl. Phys. B \textbf{262} (1985), 644

\bibitem{Perret:1987nk} R.~E.~C.~Perret, ``GEOMETRIC N=4 SUPERGRAVITY,''
Class. Quant. Grav. \textbf{5} (1988), 1109

\bibitem{Perret:1988jq} R.~E.~C.~Perret,
``GEOMETRIC STRUCTURE OF N=4 MATTER COUPLED SUPERGRAVITY,''
Class. Quant. Grav. \textbf{5} (1988), 1115

\bibitem{Frey:2002hf} A.~R.~Frey and J.~Polchinski,
``N=3 warped compactifications,''
Phys. Rev. D \textbf{65} (2002), 126009
[arXiv:hep-th/0201029 [hep-th]].

\bibitem{Kachru:2002he} S.~Kachru, M.~B.~Schulz and S.~Trivedi,
``Moduli stabilization from fluxes in a simple IIB orientifold,''
JHEP \textbf{10} (2003), 007 [arXiv:hep-th/0201028 [hep-th]].

\bibitem{DAuria:2002qje} 
R.~D'Auria, S.~Ferrara and S.~Vaula,
``N=4 gauged supergravity and a IIB orientifold with fluxes,''
New J. Phys. \textbf{4} (2002), 71
[arXiv:hep-th/0206241 [hep-th]].

\bibitem{DAuria:2003nhg} R.~D'Auria, S.~Ferrara, F.~Gargiulo, M.~Trigiante and S.~Vaula,
``$N=4$ supergravity Lagrangian for type IIB on $T^6 / \mathbb{Z}_2$ in presence of fluxes and $D3$-branes,''
JHEP \textbf{06} (2003), 045
[arXiv:hep-th/0303049 [hep-th]].

\bibitem{Berg:2003ri}
M.~Berg, M.~Haack and B.~Kors,
``An Orientifold with fluxes and branes via T duality'',
Nucl. Phys. B \textbf{669} (2003), 3-56
[arXiv:hep-th/0305183 [hep-th]].

\bibitem{Angelantonj:2003rq}
C.~Angelantonj, S.~Ferrara and M.~Trigiante,
``New D = 4 gauged supergravities from N=4 orientifolds with fluxes'',
JHEP \textbf{10} (2003), 015
[arXiv:hep-th/0306185 [hep-th]].

\bibitem{Angelantonj:2003up}
C.~Angelantonj, S.~Ferrara and M.~Trigiante,
``Unusual gauged supergravities from type IIA and type IIB orientifolds,''
Phys. Lett. B \textbf{582} (2004), 263-269 [arXiv:hep-th/0310136 [hep-th]].

\bibitem{Villadoro:2004ci} G.~Villadoro and F.~Zwirner, ``The Minimal N=4 no-scale model from generalized dimensional reduction,'' JHEP \textbf{07} (2004), 055 [arXiv:hep-th/0406185 [hep-th]].

\bibitem{Derendinger:2004jn} J.~P.~Derendinger, C.~Kounnas, P.~M.~Petropoulos and F.~Zwirner,
``Superpotentials in IIA compactifications with general fluxes,'' Nucl. Phys. B \textbf{715} (2005), 211-233 [arXiv:hep-th/0411276 [hep-th]].

\bibitem{Villadoro:2005cu} G.~Villadoro and F.~Zwirner, ``N=1 effective potential from dual type-IIA D6/O6 orientifolds with general fluxes,'' JHEP \textbf{06} (2005), 047 
[arXiv:hep-th/0503169 [hep-th]].

\bibitem{DallAgata:2009wsi} G.~Dall'Agata, G.~Villadoro and F.~Zwirner, ``Type-IIA flux compactifications and N=4 gauged supergravities,'' JHEP \textbf{08} (2009), 018 [arXiv:0906.0370 [hep-th]].


\bibitem{Schon:2006kz} 
 J.~Schon and M.~Weidner,
``Gauged N=4 supergravities,''
JHEP \textbf{05} (2006), 034 [arXiv:hep-th/0602024 [hep-th]].

\bibitem{Cordaro:1998tx}
F.~Cordaro, P.~Fre, L.~Gualtieri, P.~Termonia and M.~Trigiante,
``N=8 gaugings revisited: An Exhaustive classification,''
Nucl. Phys. B \textbf{532} (1998), 245-279 [arXiv:hep-th/9804056 [hep-th]].

\bibitem{Nicolai:2000sc} H.~Nicolai and H.~Samtleben,
``Maximal gauged supergravity in three-dimensions,''
Phys. Rev. Lett. \textbf{86} (2001), 1686-1689
[arXiv:hep-th/0010076 [hep-th]].

\bibitem{Nicolai:2001sv} H.~Nicolai and H.~Samtleben,
``Compact and noncompact gauged maximal supergravities in three-dimensions,''
JHEP \textbf{04} (2001), 022
[arXiv:hep-th/0103032 [hep-th]].

\bibitem{deWit:2002vt} B.~de Wit, H.~Samtleben and M.~Trigiante,
``On Lagrangians and gaugings of maximal supergravities,''
Nucl. Phys. B \textbf{655} (2003), 93-126
[arXiv:hep-th/0212239 [hep-th]].

\bibitem{deWit:2004nw} 
B.~de Wit, H.~Samtleben and M.~Trigiante,
``The Maximal D=5 supergravities,''
Nucl. Phys. B \textbf{716} (2005), 215-247 [arXiv:hep-th/0412173 [hep-th]].

\bibitem{deWit:2005ub} B.~de Wit, H.~Samtleben and M.~Trigiante,
``Magnetic charges in local field theory,''
JHEP \textbf{09} (2005), 016
[arXiv:hep-th/0507289 [hep-th]].

\bibitem{deWit:2007kvg} B.~de Wit, H.~Samtleben and M.~Trigiante,
``The Maximal D=4 supergravities,''
JHEP \textbf{06} (2007), 049
[arXiv:0705.2101 [hep-th]].

\bibitem{Samtleben:2008pe} H.~Samtleben,
``Lectures on Gauged Supergravity and Flux Compactifications,''
Class. Quant. Grav. \textbf{25} (2008), 214002
[arXiv:0808.4076 [hep-th]].



\bibitem{Trigiante:2016mnt} 
M.~Trigiante, ``Gauged Supergravities,''
Phys. Rept. \textbf{680} (2017), 1-175
[arXiv:1609.09745 [hep-th]].

\bibitem{DallAgata:2021uvl} 
G.~Dall\textquoteright{}Agata and M.~Zagermann,
``Supergravity: From First Principles to Modern Applications,''
Lect. Notes Phys. \textbf{991} (2021), 1-263


\bibitem{DallAgata:2023ahj}
G.~Dall'Agata, N.~Liatsos, R.~Noris and M.~Trigiante,
``Gauged D = 4 $ \mathcal{N} $ = 4 supergravity,''
JHEP \textbf{09} (2023), 071
[arXiv:2305.04015 [hep-th]].

\bibitem{Howe:1997qt}
P.~S.~Howe, N.~D.~Lambert and P.~C.~West,
``A New massive type IIA supergravity from compactification,''
Phys. Lett. B \textbf{416} (1998), 303-308
[arXiv:hep-th/9707139 [hep-th]].

\bibitem{Lavrinenko:1997qa}
I.~V.~Lavrinenko, H.~Lu and C.~N.~Pope,
``Fiber bundles and generalized dimensional reduction,''
Class. Quant. Grav. \textbf{15} (1998), 2239-2256
[arXiv:hep-th/9710243 [hep-th]].



\bibitem{Scherk:1979zr}
J.~Scherk and J.~H.~Schwarz,
``How to Get Masses from Extra Dimensions,''
Nucl. Phys. B \textbf{153} (1979), 61-88


\bibitem{Romans:1985tz}
L.~J.~Romans,
``Massive N=2a Supergravity in Ten-Dimensions,''
Phys. Lett. B \textbf{169} (1986), 374







 \bibitem{Bergshoeff:2002nv}
E.~Bergshoeff, T.~de Wit, U.~Gran, R.~Linares and D.~Roest,
``(Non)Abelian gauged supergravities in nine-dimensions,''
JHEP \textbf{10} (2002), 061
[arXiv:hep-th/0209205 [hep-th]].



\bibitem{Kerimo:2003am}
J.~Kerimo and H.~Lu,
``New D = 6, N=(1,1) gauged supergravity with supersymmetric $\text{(Minkowski)}_4 \times S^2$ vacuum,''
Phys. Lett. B \textbf{576} (2003), 219-226
[arXiv:hep-th/0307222 [hep-th]].

\bibitem{Kerimo:2004md}
J.~Kerimo, J.~T.~Liu, H.~Lu and C.~N.~Pope,
``Variant N = (1,1) supergravity and $\text{(Minkowski)}_4 \times S^2$ vacua,''
Class. Quant. Grav. \textbf{21} (2004), 3287-3300
[arXiv:hep-th/0401001 [hep-th]].


\bibitem{LeDiffon:2008sh}
A.~Le Diffon and H.~Samtleben,
``Supergravities without an Action: Gauging the Trombone,''
Nucl. Phys. B \textbf{811} (2009), 1-35
[arXiv:0809.5180 [hep-th]].

\bibitem{LeDiffon:2011wt}
A.~Le Diffon, H.~Samtleben and M.~Trigiante,
``N=8 Supergravity with Local Scaling Symmetry,''
JHEP \textbf{04} (2011), 079
[arXiv:1103.2785 [hep-th]].

\bibitem{Prins:2013yra}
H.~J.~Prins,
``Gauging the half-maximal trombone in 4D,'' Master's Thesis, University of Groningen

\bibitem{deWit:2005hv}  
B.~de Wit and H.~Samtleben,
``Gauged maximal supergravities and hierarchies of nonabelian vector-tensor systems,''
Fortsch. Phys. \textbf{53} (2005), 442-449 [arXiv:hep-th/0501243 [hep-th]].

\bibitem{Awada:1985ep}
M.~Awada and P.~K.~Townsend,
``$N=4$ Maxwell-Einstein Supergravity in Five-dimensions and Its SU(2) Gauging,''
Nucl. Phys. B \textbf{255} (1985), 617-632

\bibitem{DallAgata:2001wgl}
G.~Dall'Agata, C.~Herrmann and M.~Zagermann,
``General matter coupled N=4 gauged supergravity in five-dimensions,''
Nucl. Phys. B \textbf{612} (2001), 123-150
[arXiv:hep-th/0103106 [hep-th]].


\bibitem{Castellani:1991eu} L.~Castellani, R.~D'~Auria and P.~Fr\'{e}, ``Supergravity and Superstrings: A Geometric Perspective," World Scientific (1991), Vol. 1,2,3


\bibitem{deWit:1975veh} B.~de Wit and D.~Z.~Freedman,
``On Combined Supersymmetric and Gauge Invariant Field Theories,''
Phys. Rev. D \textbf{12} (1975), 2286



\bibitem{Jackiw:1978ar} R.~Jackiw,
``GAUGE COVARIANT CONFORMAL TRANSFORMATIONS,''
Phys. Rev. Lett. \textbf{41} (1978), 1635

\bibitem{Freedman:2012zz} D.~Z.~Freedman and A.~Van Proeyen,
``Supergravity,''
Cambridge Univ. Press, 2012


\bibitem{deVroome:2007unr} M.~de Vroome and B.~de Wit,
``Lagrangians with electric and magnetic charges of N=2 supersymmetric gauge theories,''
JHEP \textbf{08} (2007), 064 [arXiv:0707.2717 [hep-th]].







\end{thebibliography}
\end{document}